\documentclass[journal]{IEEEtran}
\usepackage{amsmath}
\usepackage{amssymb}
\usepackage{amsthm}
\usepackage{cite}
\usepackage{color}
\usepackage{xcolor}
\usepackage{booktabs}
\usepackage{array}
\usepackage{epsfig,latexsym}
\usepackage{float}
\usepackage{fancyhdr}
\usepackage{graphicx}
\usepackage{indentfirst}
\usepackage{mdwmath}
\usepackage{mdwtab}
\usepackage{subfigure}
\usepackage{setspace}
\usepackage{times}
\usepackage{url}
\usepackage{multirow}
\usepackage{algorithm}
\usepackage{algorithmic}
\usepackage{verbatim}
\usepackage{tabularx} 
\usepackage{makecell}

\newtheorem{theorem}{Theorem}

\newtheorem{lemma}{Lemma}

\newtheorem{corollary}{Corollary}

\begin{document}

\title{Pinching Antenna Systems (PASS): Principles, Positioning and Opportunities}
\author{Tianwei Hou,~\IEEEmembership{Member,~IEEE}, Yaoyu Zhang, Jinhua Wang, Guangyu Li, Zongyi Li, Minghao Jin, 

        Anna Li,~\IEEEmembership{Member,~IEEE,}
        and Mugen Peng,~\IEEEmembership{Fellow,~IEEE}

\thanks{This work is supported in part by the Beijing Natural Science Foundation L232041. (Corresponding author: Tianwei Hou)}
\thanks{Tianwei Hou and Mugen Peng are with the Beijing Key Laboratory of Convergent Communications and Networking Technologies in LEO Satellite Systems, and also with the State Key Laboratory of Networking and Switching Technology, Beijing University of Posts and Telecommunications, Beijing 100876, China (email: htw@bupt.edu.cn; pmg@bupt.edu.cn).}
\thanks{Yaoyu Zhang, Jinhua Wang, Guangyu Li and Zongyi Li are with the School of Electronic and Information Engineering, Beijing Jiaotong University, Beijing 100044, China (email: 24110070@bjtu.edu.cn; 25110041@bjtu.edu.cn; 26110103@bjtu.edu.cn; 26110108@bjtu.edu.cn).}
\thanks{Minghao Jin and Anna Li are with the School of Computing and Communications, Lancaster University, Lancaster LA1 4WA, U.K. (e-mail: m.jin6@lancaster.ac.uk; a.li16@lancaster.ac.uk).}

}

\maketitle

\begin{abstract}
Pinching-antenna systems (PASS) have garnered significant attention owing to their ability to enhance the capacity and coverage of wireless networks by guiding electromagnetic waves through waveguides before radiation. As such, PASS are regarded as promising enabling technologies for sixth-generation (6G) communication networks.
In this paper, we present a comprehensive survey of the state of the art in PASS, covering operating principles, positioning methodologies, performance evaluation, resource allocation, and the application of artificial intelligence (AI) to PASS-enhanced wireless networks.
We first introduce the fundamental principles of PASS from both materials and physics perspectives. Building on this foundation, we present positioning methodologies for both the user side and the base station side. We then evaluate the communication performance of PASS in both uplink and downlink scenarios under different blockage models. Furthermore, we systematically review existing designs for PASS-enhanced wireless networks from a performance optimization perspective. Finally, we survey research efforts that leverage AI to address challenges in dynamic and complex scenarios. 
\end{abstract}



\section{Introduction}

\begin{figure*}[!t]
  \centering
  \includegraphics[width=0.9\textwidth]{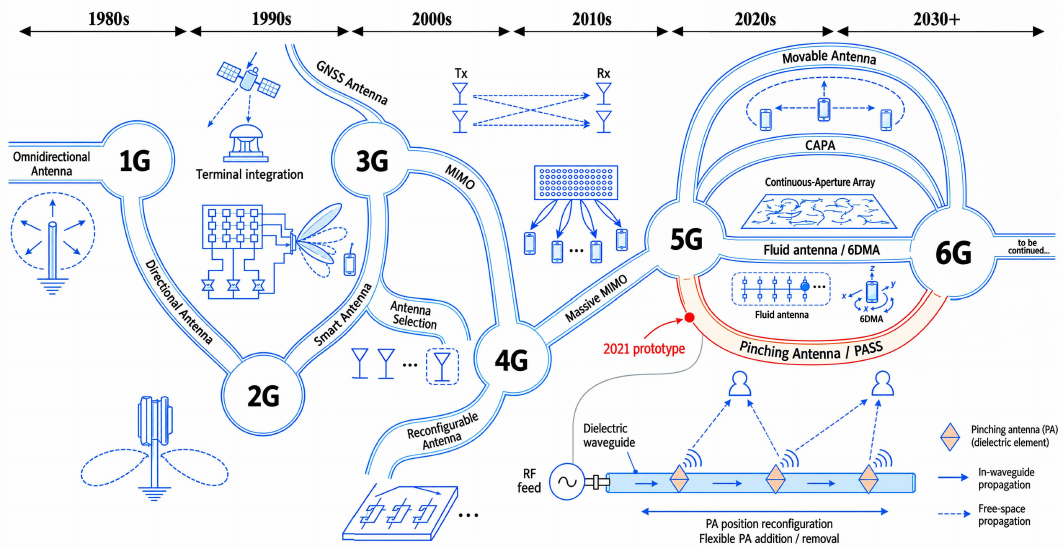}
  \caption{Evolution of Antenna Technologies from 1G to 6G: Emerging Flexible Antenna Architectures for 6G Wireless Systems. }
  \label{fig:1Gto5G}
\end{figure*}

The growing demand for ubiquitous connectivity and ultra-reliable communication in next-generation networks is driving the need for fundamentally new wireless architectures. As the distance between base station (BS) or access point (AP) and users continues to decrease, conventional fixed antenna deployments are no longer able to fully exploit the spatial degrees of freedom (DoF) required to meet such stringent requirements. In this context, novel antenna systems are emerging as promising enablers for sixth-generation (6G) networks, offering unprecedented capabilities to dynamically adapt to enclosed environments, varying channel conditions, user mobility, and diverse service requirements~\cite{11433651, 10858129, Zhu2026MovableReconfigurable}.

In mobile communication systems, the evolution of antennas is of fundamental importance. From the first generation (1G) to 6G, advances in antenna technology have continuously reshaped communication paradigms, including multiple access, beamforming, and resource allocation, among others~\cite{zhang20236g, 11219348, liu2026surveypinchingantennasystemspass}. Since the advent of the first-generation analog cellular networks, each breakthrough in antenna technology has not only enhanced radio frequency performance but also profoundly transformed the architecture and operational principles of communication systems.

In the 1G era, omnidirectional coverage and cellular broadcasting laid the foundation for mobile connectivity~\cite{rappaport2002book}. With the introduction of the second generation, directional antennas and sectorization were adopted to improve spectral efficiency (SE)~\cite{mouly1992gsm}. In the third-generation (3G) era, smart antennas began to support beam switching and interference suppression, marking the initial exploitation of spatial information~\cite{liberti1999smart}. The fourth-generation (4G) era witnessed the large-scale adoption of multiple-input multiple-output (MIMO) technology, elevating the spatial dimension into a key resource for capacity enhancement~\cite{6736761, MIMO_poi_1}. In the fifth generation (5G), the deep integration of massive antenna arrays and hybrid beamforming in millimeter-wave bands has transformed the antenna from a mere electromagnetic transducer into a core enabler of network topology~\cite{7593259, 7448962}. For example, as the antenna aperture increases, near-field effects become more pronounced, thereby opening up new paradigms for wireless communications~\cite{cui2023nearfield, an2023beamfocusing}. However, over the past decades, antenna systems have largely acted as fixed infrastructure.

In recent years, the application of flexible antenna systems has emerged as a potential solution for 6G. Unlike conventional 3G or 4G networks, where the distance between the BS and the user remains at the kilometer level, 6G is expected to provide ``last-meter'' or even ``last-wavelength'' services. As illustrated in Fig.~\ref{fig:1Gto5G}, several emerging techniques, such as fluid antenna systems (FAS), movable antenna systems (MAS), and pinching-antenna systems (PASS), are envisioned to serve as new flexible infrastructures. By employing mechanisms such as liquid-metal antennas, radio frequency (RF)-switching networks, and mechanically movable structures, FAS enables an antenna element to dynamically access different spatial positions within a sub-wavelength region, thereby exploiting spatial variations of the wireless channel~\cite{9264694, wong2023faspart1, new2025fastutorial}. Similarly, by employing a motor, a movable antenna can adjust its position, azimuth, and elevation angle in six dimensions, thereby reshaping the electromagnetic radiation pattern of the antenna~\cite{10286328, Shao2026SixDMA, Zheng2026RATutorial}. It is worth noting that both FAS and MAS mainly operate on small-scale fading channels, while the exploration of large-scale flexibility remains largely unaddressed.

By transmitting electromagnetic waves through a waveguide and radiating them in close proximity to the user, PASS can mitigate the impact of large-scale fading, thereby enabling ``last-meter'' services. Unlike free-space propagation, where path loss exhibits a logarithmic relationship with distance, the path loss in a waveguide exhibits a linear relationship with distance. Therefore, rather than conventional cellular networks, PASS is more suitable for enclosed spaces with pronounced multipath effects, such as indoor environments, trenches, underground facilities, and subways. The key advantages of PASS are summarized as follows.
\begin{itemize}
  \item Ubiquitous communication: unlike the dense deployment strategies traditionally used in underground spaces, PASS is capable of independently covering an area of approximately one hundred meters through waveguides, thereby providing ubiquitous communications for underground trenches and similar environments~\cite{southworth1950principles, southworth1936general, barrow1936hollow}.
  \item Easy to deploy: in PASS, compared with conventional communication systems, the only required modification is the antenna architecture, where a dielectric waveguide can be directly substituted. Moreover, the cost of PASS is significantly lower than that of phased array antenna or leaky coaxial cable (LCX), which provides PASS with a substantial practical advantage~\cite{wang2001periodicslots, delogne1975leakytheory, zhang2001leakyfeeder}.
  \item PASS is capable of reconfiguring the wireless propagation environment by reallocating the positions of pinching antennas (PAs). Line-of-sight (LoS) links between BSs and users can be transformed from direct BS-user links to BS-waveguide-user links. Owing to the intelligent deployment of PAs and the low path loss of waveguides, significant throughput enhancement can be achieved when the LoS link between the BS and users is obstructed by obstacles~\cite{11533561, 11178241, wang2026blockageeffectpinchingantennasystems}.
  \item Applicable for NG networks: due to the property of dielectric waveguides, the communication bandwidth can reach one-third of the carrier frequency, which can provide support for the smooth evolution toward 6G. Additionally, PASS-enhanced wireless networks are compatible with the standards and hardware of existing wireless networks~\cite{11169486, liu2026passtutorial, 11434944}.
\end{itemize}

Although PASS has attracted increasing attention, comprehensive magazine and tutorial coverage remains limited. Existing works have primarily focused on the fundamental principles, system architectures, optimization methods, and communication-oriented applications of PASS. In particular, early magazine articles introduced the basic principles of PASS and distinguished PAs from conventional fixed-position antennas by highlighting their ability to establish favorable LoS links~\cite{yang2025pinching, 11169486}. Subsequent studies further investigated PASS transmission architectures and communication schemes, including single- and multi-waveguide configurations, MIMO, non-orthogonal multiple access (NOMA), and other representative 6G-oriented applications~\cite{11169486, liu2026passtutorial}. From a network optimization perspective, \cite{R3} summarized PA placement, beamforming and resource allocation strategies, while \cite{R4} further introduced wireless-fed PASS, which extends the conventional wired-feeding architecture by enabling waveguide excitation via relay-assisted wireless links. Tutorial papers subsequently provided more formal and systematic treatments of PASS, covering propagation and channel modeling, system design and optimization, and diverse physical implementations~\cite{liu2026passtutorial,11434944}. More recent surveys have broadened the scope toward intelligent design, multifunctional services, physical-layer security (PLS), and broader integration with emerging 6G technologies \cite{liu2026surveypinchingantennasystemspass, illi2026pinchingantennasystemspass, R9}. Meanwhile, \cite{11433651} examined PASS alongside FAS, MAS, and reconfigurable intelligent surface (RIS), providing a cross-paradigm comparison of reconfigurable antenna technologies.

Different from existing communication-oriented and optimization-oriented reviews, our survey studies PASS from the waveguide and geometry-centered perspective toward integrated positioning, communication and navigation. We first focus on the fundamental design of different waveguide structures, including the material properties, geometrical configurations, modal characteristics and attenuation behavior. Building on this physical foundation, we further show that accurate positioning is a prerequisite for efficient PASS communications: user-location information are needed to guide PA activation, waveguide selection and beamforming design. Meanwhile, the communication signals transmitted and received from PAs naturally provide received signal strength indicator (RSSI), phase, angle, and delay observations for ranging, positioning and navigation. Therefore, our PASS forms an intrinsic positioning-communication-navigation closed loop, where uplink positioning provides location information to enhance downlink communication performance, reliable communication supports navigation service, and the resulting communication feedback further facilitates more accurate positioning.

Against the above observations, our main contributions are as follows.
\begin{itemize}
  \item We provide a comprehensive overview of the waveguide foundation of PASS, covering the historical evolution, the influence of materials and geometries on electromagnetic performance, and its architectural distinction from other flexible communication systems. A systematic comparison of the representative waveguide structures is conducted, examining the effects of materials and geometries on modal behavior, field confinement, attenuation, and external‑field accessibility. Finally, a unified channel model and quantitative comparison are established to distinguish PASS from LCX and other flexible architectures, highlighting its superior frequency scalability and negligible coupling loss.
  \item We review the existing PASS-based positioning methods from two perspectives: BS-side and user-side positioning. We summarize single-waveguide and multi-waveguide architectures under both LoS and multipath conditions. Furthermore, representative algorithms based on RSSI, phase, and delay measurements are investigated. By comparing their observables, advantages, and limitations, this review highlights several key research directions, including multi-user scheduling, robust LoS extraction, and lightweight algorithms for PASS-based positioning.
  \item We present a systematic performance analysis of PASS, which covers both downlink and uplink transmissions, LoS blockage, and mixed LoS and non-line-of-sight (NLoS) channel models. The mechanism by which the reconfigurability of PASS yields performance gains over conventional antenna systems is examined. LoS blockage modeling is then categorized into probabilistic and geometry-aware approaches, which differ in tractability and physical interpretability.
  \item We survey resource allocation for PASS, covering PA positioning, transmit beamforming, and energy efficiency (EE) and quality-of-service (QoS)-aware optimization. PA position optimization is investigated under continuous, discrete, and practical constraints, evolving from idealized to robust multi-user designs.
  \item We survey the synergistic integration of AI and PASS, covering deep learning (DL), reinforcement learning (RL), federated learning (FL), meta-learning, and large language model (LLM). Design principles and solution paradigms across different AI methodologies are systematically compared. We also identify key challenges and promising future directions, offering a structured reference for AI-driven PASS optimization in next-generation wireless networks.
\end{itemize}

\begin{figure}[t]
	\centering
	\includegraphics[width= 3 in]{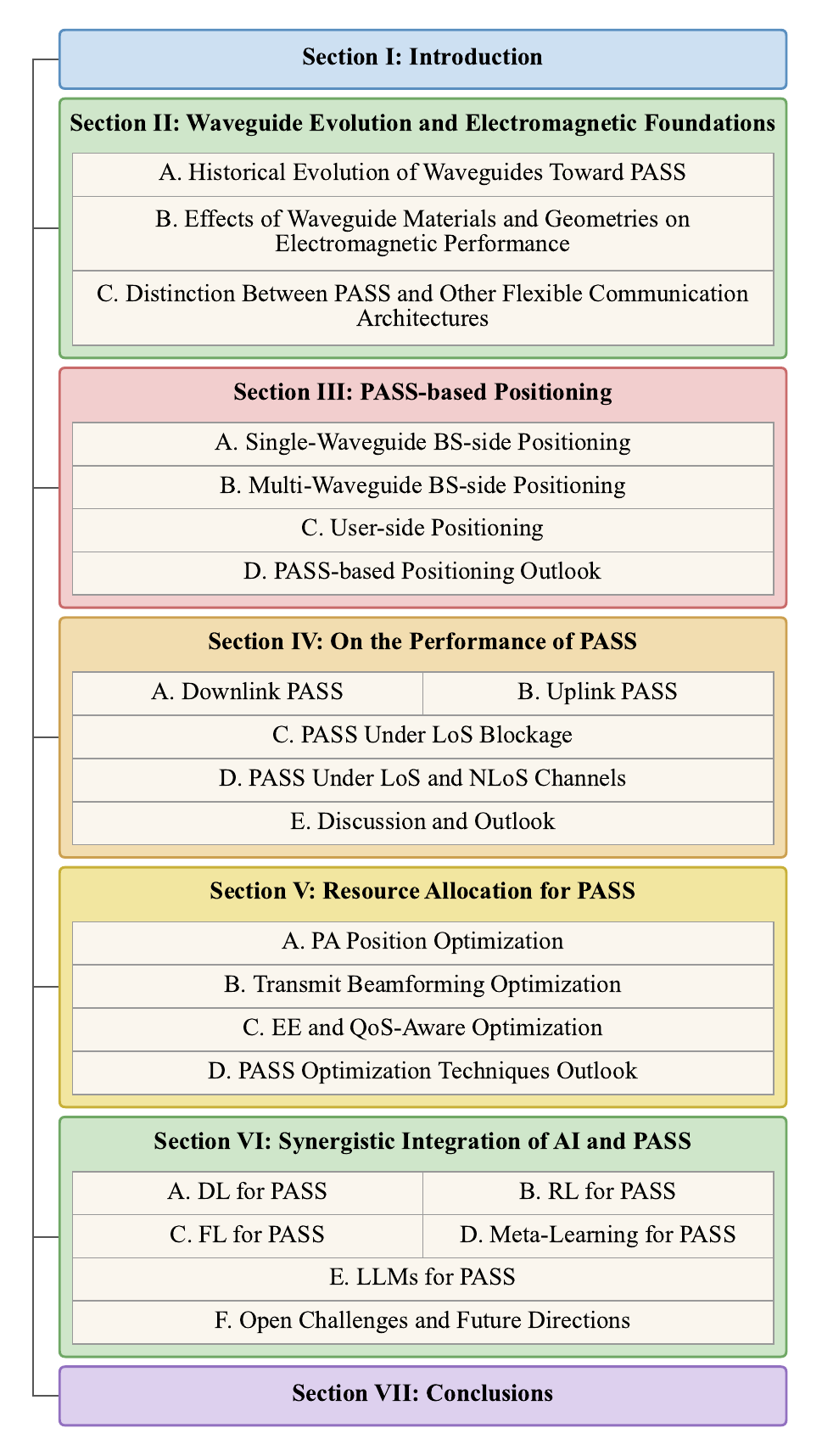} 
	\caption{Organization of the present paper.}
	\label{fig:orgnization}
\end{figure}


As illustrated in Fig.~\ref{fig:orgnization}, this paper is structured as follows.
Section II elaborates on the fundamental operating principles of PASS-enhanced wireless networks. Section III focuses on the positioning method of PASS-assisted systems. In Section IV, the performance analysis of PASS in both LoS and NLoS are investigated. The latest research on the resource allocation are discussed in Section V. The synergistic integration of AI and PASS is surveyed in Section VI. Finally, Section VII concludes this survey.


\section{Waveguide Evolution and Electromagnetic Foundations of PASS}  

\subsection{Historical Evolution of Waveguides Toward PASS}

\begin{figure*}[!t]
    \centering
    \includegraphics[width=0.8\textwidth]{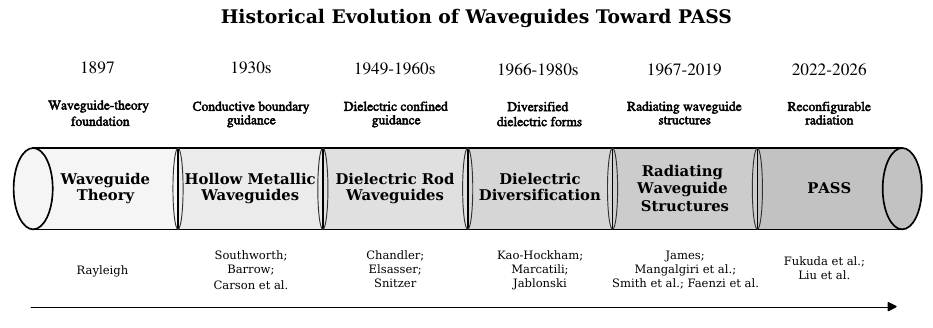}
    \caption{Historical evolution of waveguides toward PASS.}
    \label{fig:Historical evolution}
\end{figure*}

As shown in Fig.~\ref{fig:Historical evolution}, the development of waveguides in electromagnetic engineering follows a progressive evolution from guided transmission to controllable radiation. Hollow metallic waveguides enable the confinement and transmission of high-frequency electromagnetic waves through artificial structures, while dielectric waveguides subsequently provide greater structural flexibility and different field-confinement mechanisms. With the continued development of dielectric waveguides, they are no longer regarded solely as passive transmission media, since their guided fields can interact with the surrounding space to support radiation under suitable conditions. The combination of guided transmission and radiation capability eventually provides the physical basis for PASS, in which dielectric waveguides act as low-loss transmission media and support reconfigurable radiating elements for electromagnetic-wave emission.

The origin of waveguide theory can be traced back to Rayleigh’s 1897 paper, \emph{On the Passage of Electric Waves through Tubes, or the Vibrations of Dielectric Cylinders} \cite{rayleigh1897passage}. In this work, Rayleigh investigated the propagation of electric waves in tubes and dielectric cylindrical structures, establishing an early theoretical foundation for subsequent waveguide research. More importantly, the study already indicated two fundamental routes for guiding electromagnetic waves: one based on conductive boundaries and the other based on dielectric confinement. These two mechanisms later evolved into the two major branches of waveguide development, namely metallic waveguides and dielectric waveguides. Rayleigh’s work thus extended the understanding of electromagnetic propagation from free space radiation to structurally confined transmission, showing that electromagnetic energy can be guided by properly designed physical structures \cite{packard1984origin}.

Building on the early theoretical foundations, the transition from conceptual analysis to practical engineering began in the 1930s with the development of hollow metallic waveguides. Southworth’s work \cite{southworth1936general} provided one of the earliest systematic experimental investigations of wave propagation in metallic guiding structures and demonstrated their feasibility for high-frequency transmission. Barrow’s work \cite{barrow1936hollow} further examined the electromagnetic transmission in hollow metallic tubes and clarified their propagation characteristics in high-frequency regimes. In parallel, Carson~\textit{et al.}~ \cite{carson1936mathematical} established a mathematical framework for guided-wave propagation in metallic waveguides. The studies in \cite{southworth1936general, barrow1936hollow, carson1936mathematical} showed that, when the operating frequency was sufficiently high, conventional wire-based transmission structures became less suitable, whereas hollow metallic waveguides could support efficient guided transmission above cutoff. As a result, hollow metallic waveguides became important low-loss transmission structures with strong field confinement, high shielding capability, and high power handling capacity \cite{southworth1950principles}. Despite these advantages, hollow metallic waveguides are inherently based on conductive boundaries and usually require rigid, closed metallic enclosures. The structural feature limits their applicability in scenarios that require lightweight deployment, geometrical flexibility, or direct interaction between the guided field and the surrounding environment \cite{monticone2015leakywave}.

To overcome the structural limitations of hollow metallic waveguides, subsequent research turned to dielectric waveguides. Chandler’s work \cite{chandler1949rod} investigated dielectric rod waveguides, showing that electromagnetic waves could be guided by dielectric structures without metallic boundaries. Further analyzing attenuation in dielectric circular rods, Elsasser's work \cite{elsasser1949attenuation} provided an early characterization of propagation loss in dielectric guiding structures. Thus, the corresponding electromagnetic foundation shifted from conductive boundary guided propagation to dielectric confinement guided propagation, where the guided field was maintained by permittivity contrast and modal balance between the waveguide and the surrounding space \cite{snitzer1961cylindrical}. Compared with hollow metallic waveguides, dielectric rod waveguides provided lower structural rigidity and greater field accessibility, facilitating stronger interaction with the external environment \cite{king1957losses}. This transition from metallic to dielectric guidance laid the groundwork for waveguide structures in which guided transmission and radiation could coexist \cite{kelebekler2021leaky}.

The development of dielectric waveguides subsequently expanded across different frequency bands, materials, and geometrical forms. Kao and Hockham’s work \cite{kao1966fibre} established the foundation of optical-fiber guidance, demonstrating the potential of dielectric confinement for low-loss transmission. Marcatili’s work \cite{marcatili1969rectangular} showed, through a systematic analysis of rectangular dielectric waveguides, that dielectric guidance could be realized in geometries beyond cylindrical rods. Jablonski’s work \cite{jablonski1978circular} extended dielectric waveguides to millimeter-wave regimes by characterizing the attenuation of circular dielectric waveguides at millimeter wavelengths. Yoneyama and Nishida’s work \cite{yoneyama1983nrd} developed nonradiative dielectric waveguide components for high-frequency circuit applications. Malekabadi~\textit{et al.}~ \cite{malekabadi2014ribbon} demonstrated that high-resistivity silicon ribbon waveguides can achieve single-mode low-loss propagation at F/G bands. Atakaramians~\textit{et al.}~\cite{atakaramians2013terahertz} further summarized the material and structural diversity of dielectric waveguides for very-high-frequency and terahertz applications. Collectively, the studies in \cite{kao1966fibre, marcatili1969rectangular, jablonski1978circular, yoneyama1983nrd, malekabadi2014ribbon, atakaramians2013terahertz} showed that dielectric waveguides were not limited to a specific material or cross-sectional shape, with their propagation properties tunable via permittivity, loss tangent, geometry, and structural openness.

Later studies showed that dielectric waveguides could support not only guided transmission but also radiation. Unlike hollow metallic waveguides, whose electromagnetic fields were strongly confined within closed conductive boundaries, dielectric waveguides usually exhibited partially open field distributions, where part of the guided field extended beyond the dielectric body into the surrounding medium. The electromagnetic feature allowed the guided field to couple more readily with external structures or local perturbations, thereby enabling radiation under suitable structural conditions. James's work \cite{james1967cylindrical} showed that dielectric rod structures could support guided transmission while also functioning as effective radiators. Callahan's work \cite{callahan1968moths} further demonstrated that dielectric waveguides could be designed according to antenna principles to shape and control their radiation characteristics. The studies in \cite{james1967cylindrical} and \cite{callahan1968moths} indicated that dielectric waveguides could evolve from pure transmission media into waveguide-based radiating structures, in which the waveguide itself participated directly in radiation generation.

Following the transition, subsequent research increasingly addressed the systematic exploitation of the coexistence of guided transmission and radiation in a more application-oriented manner. Mangalgiri~\textit{et al.}~ \cite{mangalgiri2017nanorod} showed that open dielectric structures could exhibit leaky-mode behavior and stronger outward field coupling under suitable conditions, thereby providing further physical insight into the interaction of guided energy with the surrounding space. Smith~\textit{et al.}~\cite{smith2017waveguidefed} developed waveguide-fed radiating architectures, in which guided waves are intentionally coupled to distributed radiating elements to realize controllable aperture radiation. Faenzi~\textit{et al.}~\cite{faenzi2019metasurface} extended the line of development to surface-wave-based and metasurface-enabled radiating structures, which expanded the flexibility of guided-wave radiation and provided more diverse physical implementations for integrating transmission and radiation on the same platform. These developments provided the conceptual and technical bridge from conventional radiating waveguide structures to more flexible and reconfigurable architectures.

The guided-radiative concept was developed in the context of PASS. Fukuda~\textit{et al.}~\cite{suzuki2022pinching} introduced the basic concept of PASS by using dielectric waveguides as transmission media and small dielectric particles attached along the waveguides as radiating elements. In this framework, these radiating elements, referred to as PAs, couple guided electromagnetic energy out of the dielectric waveguide and radiate it into free space at selected locations. Therefore, the waveguide no longer serves only as a signal transmission path, but also acts as a spatial platform that supports localized radiation. Building on the basic concept, Liu~\textit{et al.}~\cite{liu2026passtutorial} further developed PASS in the context of wireless communications, presenting it as a waveguide-based flexible-antenna architecture. Specifically, the study discussed several communication-oriented architectures and operation modes for PASS, covering signal models, hardware and power radiation models, PA activation methods, single-user and multi-user transmission structures, as well as practical protocols such as waveguide switching, waveguide division, and waveguide multiplexing.

\subsection{Effects of Waveguide Materials and Geometries on Electromagnetic Performance}


\begin{table*}[t!]
	\centering
	\caption{Electromagnetic characteristics of representative waveguide structures.}
	\label{tab:waveguide_em_summary}
	\renewcommand{\arraystretch}{1.15}
	\footnotesize
	\begin{tabularx}{\textwidth}{
			>{\raggedright\arraybackslash}p{2.5cm}
			>{\centering\arraybackslash}p{1.6cm}
			>{\raggedright\arraybackslash}p{2.5cm}
			>{\raggedright\arraybackslash}p{2.3cm}
			>{\raggedright\arraybackslash}X
			>{\centering\arraybackslash}p{1.8cm}}
		\toprule
		\textbf{Waveguide type} &
		\textbf{Axial loss} &
		\textbf{Loss origin} &
		\textbf{Modes} &
		\textbf{Transverse field} &
		\textbf{External interaction} \\
		\midrule
		
		Circular hollow metallic &
		Moderate &
		Conductor, filling medium &
		TE/TM &
		Bessel-type distribution in hollow region &
		Weak \\
		
		Rectangular hollow metallic &
		Moderate &
		Conductor, filling medium &
		TE/TM &
		Sinusoidal distribution in hollow region &
		Weak \\
		
		Dielectric rod &
		Moderate &
		Dielectric loss, leakage &
		Hybrid guided modes &
		Core oscillation with external evanescence &
		Moderate \\
		
		Rectangular dielectric &
		Moderate &
		Dielectric loss, leakage &
		Hybrid/Marcatili modes &
		Separable core field with external evanescence &
		Moderate \\
		
		Optical fiber &
		Low &
		Material, scattering, bending &
		LP/hybrid modes &
		Core-cladding confinement &
		Weak \\
		
		Dielectric ribbon &
		High &
		Dielectric loss, scattering, leakage &
		Ribbon guided modes &
		Broad-surface evanescence &
		Strong \\
		
		\bottomrule
	\end{tabularx}
\end{table*}

Under the time-harmonic assumption, electromagnetic fields in a uniform waveguide can be analyzed via separation of variables. The total field is decomposed into a transverse distribution and an axial propagation factor \cite{southworth1936general}. Specifically, the longitudinal component satisfies the transverse Helmholtz equation \cite{carson1936mathematical}. The eigen-solutions of the equation are determined by the boundary conditions at the waveguide walls or dielectric interfaces. The eigenvalue $k_{\mathrm{c}}$ characterizes the transverse field variation. In metallic waveguides, 
the eigenvalue $k_{\mathrm{c}}$ equals the discrete cutoff wavenumber, whereas in dielectric waveguides, it represents the transverse phase constant. The transverse field components are then derived from Maxwell’s curl relations. All field components share an axial propagation factor $e^{-(\alpha + j\beta)z}$, where $\alpha$ denotes the amplitude attenuation coefficient, and $\beta$ denotes the phase constant.

Although the analysis method is general, waveguide materials and geometries strongly affect modal behavior, axial attenuation, field confinement, and external-field accessibility. The following comparison covers representative waveguide structures from the perspectives discussed above. The main characteristics of the structures are summarized in Table~\ref{tab:waveguide_em_summary}.

\subsubsection{Circular Hollow Metallic Waveguides}

Circular hollow metallic waveguides are among the earliest guided-wave structures for high-frequency electromagnetic transmission. The fundamental configuration is a hollow cylindrical region bounded by a conducting wall. That wall provides both field confinement and axial guidance for the waves \cite{southworth1936general}.

\begin{figure}[t]
	\centering
	\includegraphics[width=0.4\textwidth]{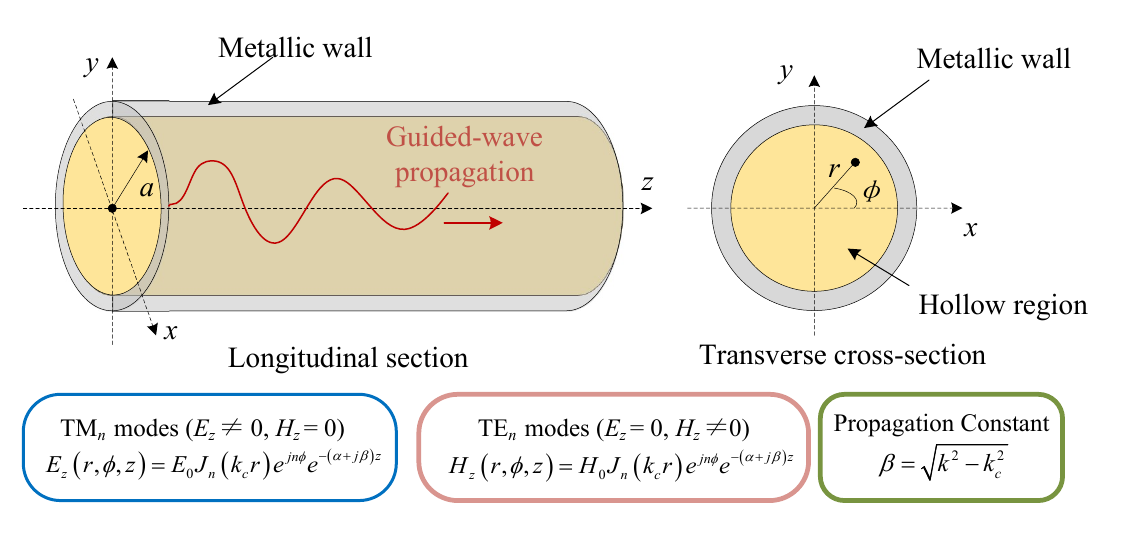} 
	\caption{Circular hollow metallic waveguide model.} 
	\label{Circular Hollow Metallic Waveguide} 
\end{figure}

As illustrated in Fig.~\ref{Circular Hollow Metallic Waveguide}, the structure is naturally described in a cylindrical coordinate system $(r,\phi,z)$, where wave propagation occurs along the $z$-axis. Owing to the cylindrical symmetry, the longitudinal field components are expressed in terms of Bessel functions. The perfectly conducting wall at $r = a$ imposes a boundary condition that leads to quantization of the transverse eigenvalues and discrete cutoff wavenumbers for transverse electric (TE) and transverse magnetic (TM) modes. Fig.~\ref{Circular Hollow Metallic Waveguide} provides the detailed field expressions and eigenvalue equations. For hollow metallic waveguides, the phase propagation constant is given by $\beta = \sqrt{\kappa^2 - \kappa_{\mathrm{c}}^2}$. Guided propagation requires the operating frequency to exceed the cutoff frequency, i.e., $\kappa > \kappa_{\mathrm{c}}$.


Circular hollow metallic waveguides generally provide strong modal confinement, high shielding capability, and high power-handling capacity. However, the continuous metallic boundary creates a rigid, fully enclosed structure, severely restricting geometrical flexibility and direct interaction between the guided field and the external space. PASS requires efficient near-field power extraction by PAs, which the enclosed field distribution cannot support. Therefore, circular hollow metallic waveguides are not suitable as the PASS waveguide. The structure may still serve as a shielded feeder or a high-power transmission section.

\subsubsection{Rectangular Hollow Metallic Waveguides}

Rectangular hollow metallic waveguides are defined by two transverse dimensions, namely the broad wall dimension $a$ and the narrow wall dimension $b$. Conducting walls enclose a hollow rectangular region, confining the electromagnetic fields. The fields propagate along the longitudinal direction. The waveguide model is governed by rectangular boundary conditions, which yield sinusoidal field distributions and geometry-dependent cutoff characteristics \cite{marcuvitz1951waveguide, pozar2021microwave}.

\begin{figure}[t]
	\centering
	\includegraphics[width=0.4\textwidth]{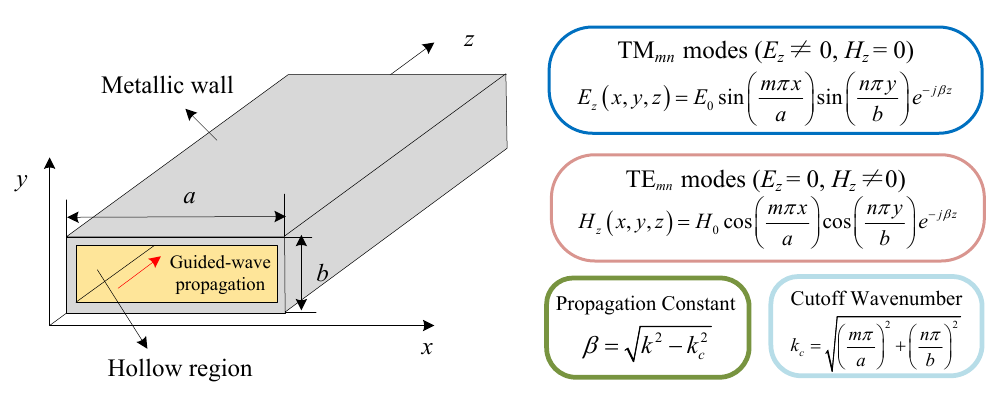} 
	\caption{Rectangular hollow metallic waveguide model.} 
	\label{Rectangular Hollow Metallic Waveguide} 
\end{figure}

Fig.~\ref{Rectangular Hollow Metallic Waveguide} shows the Cartesian coordinate system $(x,y,z)$ commonly used for the waveguides. The waveguide occupies the region $0<x<a$ and $0<y<b$, with propagation along the $z$-axis. Perfectly conducting walls at $x=0,a$ and $y=0,b$ force the tangential electric field to vanish. The boundary condition restricts the transverse field variation to sinusoidal functions. The eigenvalues are quantized into discrete cutoff wavenumbers determined by the dimensions $a, b$ and the mode indices. Detailed field expressions and eigenvalue equations appear in Fig.~\ref{Rectangular Hollow Metallic Waveguide}. As in the circular case, the propagation constant is $\beta = \sqrt{\kappa^2 - \kappa_{\mathrm{c}}^2}$. Axial loss mechanisms resemble those of circular hollow metallic waveguides. The total loss includes both dielectric and conductor contributions.

Rectangular hollow metallic waveguides offer clearer dominant-mode operation, more stable field polarization, and easier integration with planar microwave structures. Similar to circular hollow metallic waveguides, the closed metallic wall structure ensures efficient transmission and low radiation leakage. However, the structure cannot support efficient power extraction by PAs in PASS. Therefore, rectangular hollow metallic waveguides are not suitable as the PASS waveguide.

\subsubsection{Dielectric Rod Waveguides}

Dielectric rod waveguides are open dielectric guiding structures. Unlike hollow metallic waveguides, dielectric rod waveguides do not require closed metallic walls. Guided propagation relies on the modal balance between the field inside the rod and the evanescent field in the surrounding medium \cite{chandler1949rod, elsasser1949attenuation}.

\begin{figure}[t]
	\centering
	\includegraphics[width=0.4\textwidth]{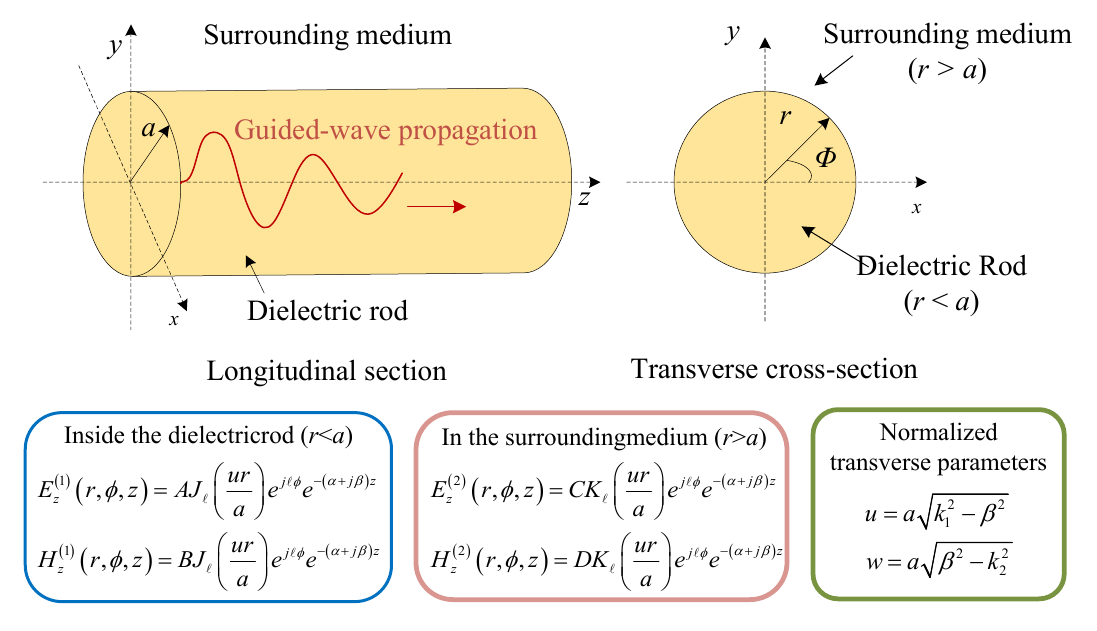} 
	\caption{Dielectric rod waveguide model.}
	\label{Dielectric Rod Waveguide}
\end{figure}

As shown in Fig.~\ref{Dielectric Rod Waveguide}, the structure can be modeled as a circular dielectric cylinder of permittivity $\varepsilon_1$, and permeability $\mu_1$, surrounded by an infinite medium with $\varepsilon_2$ and $\mu_2$. Guided propagation typically requires $\varepsilon_1 > \varepsilon_2$. A cylindrical coordinate system is adopted, with wave propagation along the $z$-axis. No conducting boundary exists at $r=a$. Instead, the tangential electric and magnetic field components must be continuous across the dielectric interface \cite{snitzer1961cylindrical}.
For a bound guided mode, the phase propagation constant $\beta$ satisfies $k_2 < \beta < k_1$, ensuring an oscillatory field inside the rod and evanescent decay in the surrounding medium. Inside the dielectric rod, the longitudinal fields are expressed in terms of Bessel functions $J_\ell(\cdot)$. In the exterior region, modified Bessel functions guarantee radial decay. Enforcing the continuity conditions at $r=a$ yields a characteristic equation that determines the admissible values of $\beta$. Detailed field expressions and the eigenvalue equation appear in Fig.~\ref{Dielectric Rod Waveguide}.

In conclusion, the guided fields are confined not by metallic walls, but by dielectric contrast and modal continuity at an open interface. Consequently, dielectric rod waveguides support low-loss transmission while maintaining an externally accessible evanescent field. The open field distribution enhances interaction with external radiating elements, thereby enabling near-field coupling with PAs, which is desirable for PASS. Since sensitivity to external perturbations, bending and surface contamination remains a drawback, mechanical support and alignment may be required in practical deployment. Overall, dielectric rod waveguides are suitable for PASS when low-loss guidance and controllable external coupling are both required.

\subsubsection{Rectangular Dielectric Waveguides}

Rectangular dielectric waveguides extend dielectric guiding to non-circular cross sections. Unlike circular rods, rectangular waveguides are defined by two independent transverse dimensions, width $a$ and height $b$. The electromagnetic response depends on both the permittivity contrast and the ratio $a/b$. The resulting geometrical tunability complicates the analysis \cite{marcatili1969rectangular, goell1969circular}.

\begin{figure}[t]
	\centering
	\includegraphics[width=0.4\textwidth]{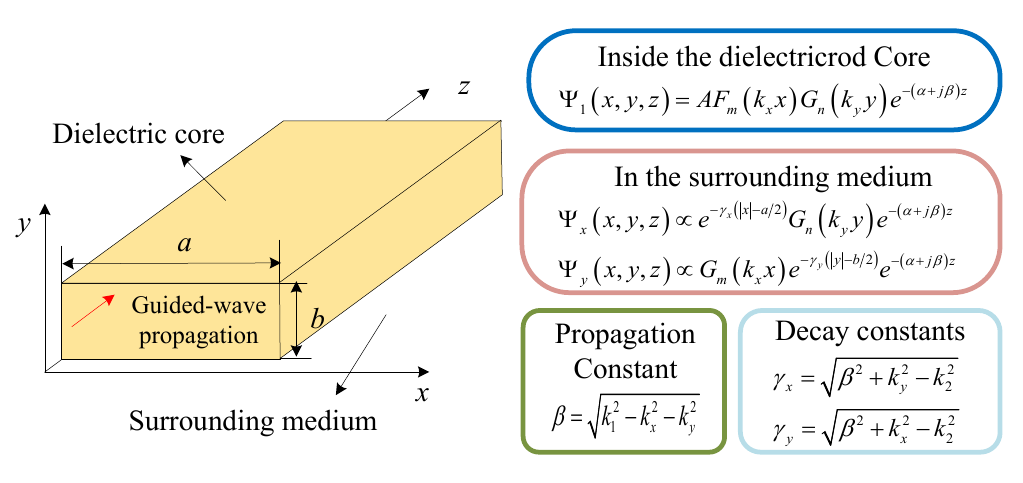} 
	\caption{Rectangular dielectric waveguide model.}
	\label{Rectangular Dielectric Waveguide}
\end{figure}

As shown in Fig.~\ref{Rectangular Dielectric Waveguide}, the waveguide consists of a dielectric core with permittivity $\varepsilon_1$ and permeability $\mu_1$, surrounded by an infinite medium of lower permittivity $\varepsilon_2$ and permeability $\mu_2$. The core occupies the region $0<x<a$, $0<y<b$, with wave propagation along the $z$-axis. No conducting walls exist at the transverse boundaries. Instead, the tangential fields must be continuous across the dielectric interfaces. Consequently, the guided field extends outside the core as an evanescent tail.

The rectangular geometry lacks rotational symmetry. The presence of four dielectric corners adds further complexity. A widely used simplification is the Marcatili approximation \cite{marcatili1969rectangular}, where the weak fields in the four corner regions are neglected. Inside the core, the longitudinal field components take a separable form. The transverse dependence is described by sine and cosine functions, with parity set by the mode orders. In the exterior side regions, the field decays exponentially away from the core surfaces. The corresponding transverse wavenumbers and decay constants are linked to the propagation constant $\beta$ through the continuity conditions at the dielectric interfaces. Detailed field expressions, parameter definitions, and the resulting eigenvalue equations appear in Fig.~\ref{Rectangular Dielectric Waveguide}.

Electromagnetic properties of rectangular dielectric waveguides depend on dielectric contrast, dimensions, and aspect ratio. Larger permittivity contrast or core size leads to stronger modal confinement. Weaker confinement extends the guided field into the surrounding medium, thereby increasing evanescent field accessibility. The planar open geometry supports low-loss propagation and localized PA coupling. Integration with flat surfaces and radiating elements is convenient. Rotational asymmetry and corner-field complexity tolerances complicate practical implementation. Excessive openness may increase leakage and scattering loss. Therefore, rectangular dielectric waveguides are promising for PASS in planar or integrated deployments. The confinement accessibility tradeoff requires careful engineering~\cite{hou2019rectangularrod}.

\subsubsection{Optical Fiber Waveguides}

Optical fibers are cylindrical dielectric waveguides with an additional cladding layer. A typical step-index fiber comprises a high-permittivity core surrounded by a lower-permittivity cladding. The cladding stabilizes the guided mode while confining most of the field within the fiber, thereby reducing direct exposure to the external environment \cite{snitzer1961cylindrical, kao1966fibre}.

\begin{figure}[t]
	\centering
	\includegraphics[width=0.4\textwidth]{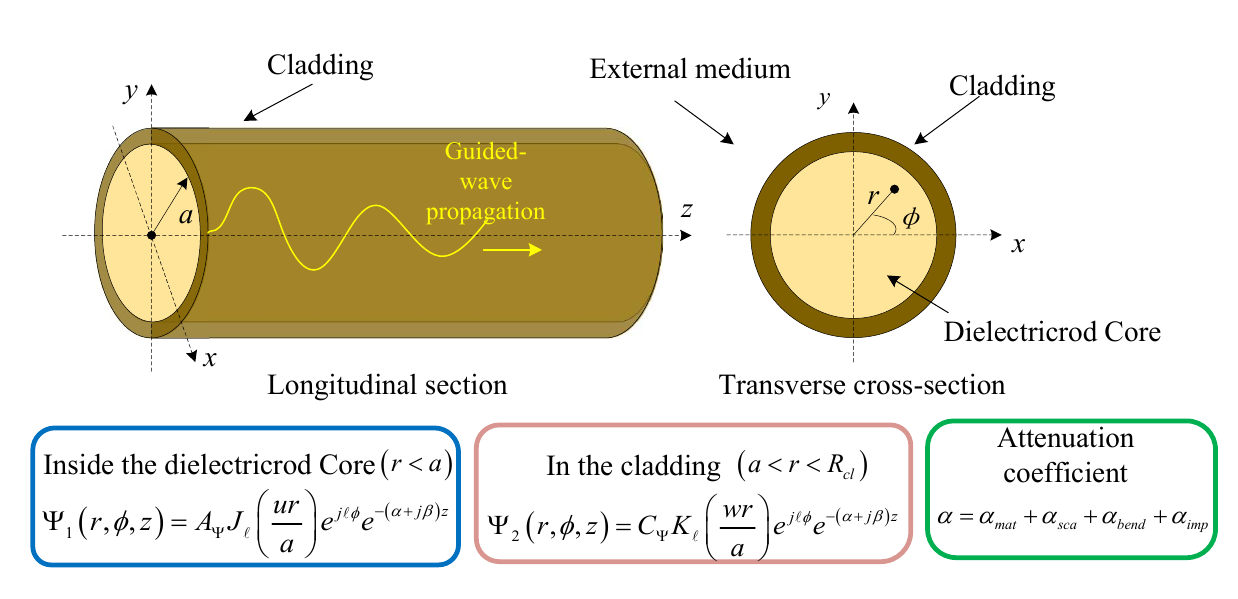} 
	\caption{Optical fiber waveguide model.}
	\label{Optical Fiber Waveguide}
\end{figure}

Similar to dielectric rod waveguides, the guided fields are expressed using Bessel functions in the core and modified Bessel functions in the cladding. Attenuation in practical fibers arises from several independent mechanisms, including material absorption, Rayleigh scattering, bend-induced radiation, and fabrication imperfections. A low-loss base material alone cannot guarantee ultralow total attenuation, because bending and surface defects must also be carefully controlled. Detailed field expressions are provided in Fig.~\ref{Optical Fiber Waveguide}.

Optical fibers avoid conductor loss by eliminating conducting boundaries, thus enabling stable low-loss signal transmission. However, the cladding confines the evanescent field inside the fiber, thereby significantly weakening direct interaction with the external environment. PASS requires efficient near-field power extraction by movable PAs. The cladding prevents such extraction, thereby limiting controllable radiation. Bending-induced radiation and fabrication imperfections further reduce suitability. Therefore, optical fibers are not suitable as the PASS waveguide. The fiber can serve as a stable signal-delivery link in PASS-enabled networks.

\subsubsection{Dielectric Ribbon Waveguides}

Dielectric ribbon waveguides are rectangular dielectric waveguides with an extremely high width-to-thickness ratio. The thin, wide cross section creates broad dielectric-air interfaces, which provide highly accessible evanescent fields. Since no cladding is present, a portion of the guided field extends beyond the dielectric body, enabling effective external coupling over a spatially extended region \cite{malekabadi2014ribbon, atakaramians2013terahertz}. At the same time, the exposed interfaces make the guided mode sensitive to surface roughness, local perturbations, and radiation leakage.

\begin{figure}[t]
	\centering
    \includegraphics[width=0.4\textwidth]{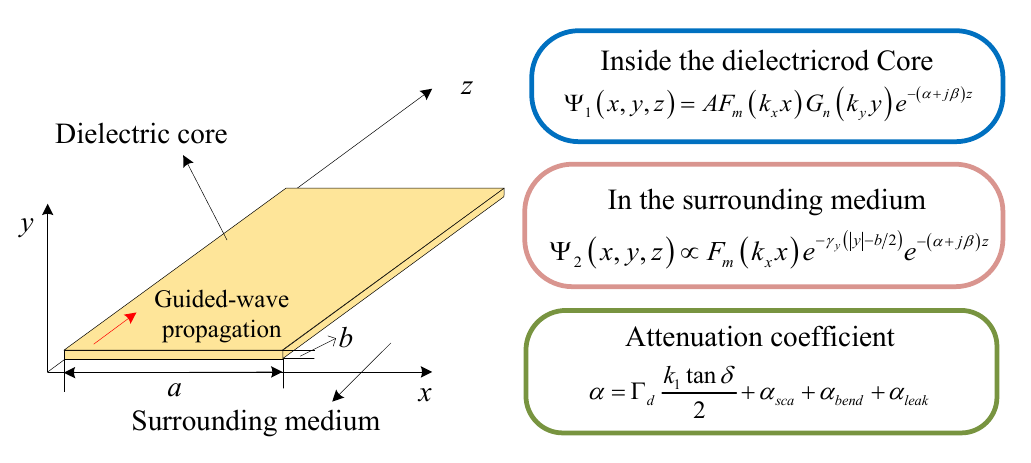} 
	\caption{Dielectric ribbon waveguide model.}
	\label{Dielectric Ribbon Waveguide}
\end{figure}

Inside the waveguide, the field distribution takes a separable form with sinusoidal functions along the transverse directions. Outside, the evanescent field decays exponentially from the broad surfaces and the edges. The corresponding transverse wavenumbers and decay constants are linked to the propagation constant through the continuity conditions at the dielectric interfaces. Full field expressions and the resulting characteristic equations are summarized in Fig.~\ref{Dielectric Ribbon Waveguide}.

Total attenuation is influenced by several factors. Dielectric loss is weighted by the fraction of modal power inside the ribbon. Additional contributions arise from scattering at surface irregularities, bend-induced radiation, and leakage into the surrounding space. The interplay among these mechanisms creates a design trade-off. Stronger field confinement reduces leakage and scattering loss. However, it limits external accessibility. Weaker confinement enhances radiation and coupling, but it raises attenuation.

Dielectric ribbon waveguides provide a tunable balance between low-loss guidance and external-field interaction. The planar open geometry creates broad dielectric-air interfaces, which support simultaneous waveguiding and controlled radiation. PAs can be positioned along the ribbon for near-field power extraction. However, surface roughness, local perturbations, and bending leakage create difficulties for practical deployment. Therefore, dielectric ribbon waveguides show strong potential as PASS waveguides, but further investigation remains necessary.

\subsection{Distinction Between PASS and Other Flexible Communication Architectures}

Emerging flexible communication architectures aim to reshape wireless propagation or adjust the spatial relationship between antennas and users. Representative examples include RISs, FASs and MAS \cite{huang2019reconfigurable, 9264694, 10286328}. RISs reconfigure the wavefront via programmable surfaces, whereas FASs and MAS enhance spatial flexibility by adjusting the effective antenna position. PASS follow a different implementation principle, where dielectric waveguides are used as signal-guiding media and PAs serve as reconfigurable radiation points along the waveguides~\cite{wang2025passmodeling, xu2026powerradiation}. Therefore, PASS enable waveguide-based radiation-point reconfiguration, jointly involving guided signal transport and free-space radiation.

Among existing architectures, the LCX is the closest counterpart to PASS. Both systems combine guided signal transport with radiation into the surrounding space~\cite{wang2001periodicslots, hou2017capacitylcx}. However, the radiation mechanisms and system costs are fundamentally different. PASS localize radiation through detachable or movable PAs placed on a dielectric waveguide. In contrast, LCX radiates through pre-designed slots or apertures distributed along the outer conductor of a coaxial cable. To highlight these differences, a unified mathematical model is presented below, followed by a quantitative comparison of the main characteristics.

\begin{table*}[t]
	\centering
	\caption{Comparison of key loss parameters between LCX and PASS}
	\label{tab:lcx_pass_compare}\footnotesize
	\begin{tabular}{l c c}
		\toprule
		Parameter & LCX ${1 \mathord{\left/
				{\vphantom {1 2}} \right.
				\kern-\nulldelimiterspace} 2}$ & PASS \\
		\midrule
		Typical attenuation at 2.4 GHz (dB/m) & 0.062 & 0.08 \\
		Coupling loss (dB) & 56--77 & \(\ll 1\) (negligible) \\
		Total link loss for 100 m at 2.4 GHz (dB) & 62.2--83.2 & 8.0 \\
		Typical operating frequency range & 0.1--3 GHz & 2.4--28 GHz \\
		Primary loss sources & Conductor loss, dielectric loss, \textbf{slot coupling} & \textbf{Dielectric absorption}, scattering \\
		\bottomrule
	\end{tabular}
\end{table*}

\subsubsection{Unified Model}

A unified channel representation is first established to enable a direct mathematical comparison between PASS and LCX. Let $\mathbf{p}^{\mathrm{RE}}_{m,n}$ denote the position of the $n$-th radiating element on the $m$-th transmission line. For LCX, the slot position is fixed after manufacturing. For PASS, the PA can be relocated along the waveguide. Let $\ell_{m,n}$ be the propagation distance from the feeding point to that element along the line. Let $d_{k,m,n} = \|\mathbf{p}^{\mathrm{u}}_k - \mathbf{p}^{\mathrm{RE}}_{m,n}\|$ be the distance from that element to user $k$. 

Under a LoS dominant condition, the baseband channel coefficient from the transmission-line input to user $k$ via the $(m,n)$-th element can be written in the following unified form \cite{wang2025passmodeling, wang2001adjustablelcx}:
\begin{equation}
	h^{\mathrm{CH}}_{k,m,n} = 
	\xi_{k,m,n} e^{-(\alpha_m + j\beta_m)\ell_{m,n}} \frac{c}{4\pi f_{\mathrm{c}} d_{k,m,n}} e^{-j\frac{2\pi}{\lambda}d_{k,m,n}},
	\label{eq:unified_channel}
\end{equation}
where $\xi_{k,m,n}$ is the effective radiation factor, $\alpha_m$ and $\beta_m$ are the attenuation and phase constants of the $m$-th line, $f_{\mathrm{c}}$ is the carrier frequency, $c$ is the speed of light, and $\lambda$ is the free-space wavelength. Two fundamental distinctions between the two architectures are encapsulated in the factor $\xi_{k,m,n}$ and in the frequency dependence of $\alpha_m$.

The factor $\xi_{k,m,n}$ captures the radiation extraction mechanism. For PASS, each PA is deployed on the dielectric waveguide to extract the guided power through proximity coupling. A portion of the guided electromagnetic energy is coupled out through near-field interaction, while the remaining power continues to propagate along the waveguide. The extracted power is determined by the PA structure, the coupling distance, and the dielectric properties of the waveguide. The coupling factor for PASS is given by $\xi_{k,m,n} = \sqrt{\eta_{m,n}}$, where $\eta_{m,n}$ is the coupling efficiency of PAs. The efficiency is engineered to be close to unity, corresponding to a coupling loss less than 1 dB. Since the PA can be designed with a nearly omnidirectional or controllable radiation pattern, the coupling factor is essentially independent of the user direction.

For LCX, radiation emanates from fixed apertures prefabricated on the outer conductor. The aperture geometry and the slot distribution along the cable determine the radiation characteristics. Accordingly, the effective radiation factor for LCX is given by \cite{wang2016modelinglcx, wang2025generalizedlcxpass}:
\begin{equation}
	\xi_{k,m,n} = \sqrt{\rho_{m,n} G_{m,n}(\theta_{k,m,n}, \phi_{k,m,n})},
	\label{eq:lcx_xi}
\end{equation}
where $\rho_{m,n}$ is the slot radiation efficiency, and $G_{m,n}$ is the slot gain toward the user at the elevation angle $\theta_{k,m,n}$ and azimuth angle $\phi_{k,m,n}$. In typical LCX designs, the factor corresponds to a coupling loss of tens of decibels, often ranging from 56 dB to 77 dB. The gain $G_{m,n}$ exhibits strong angular dependence, meaning the received power varies significantly with the user position relative to the slot orientation.

The propagation constants $\alpha_m$ and $\beta_m$ also reflect fundamental differences in waveguiding physics. LCX is based on coaxial cable structures with periodic slots. Above a few gigahertz, the attenuation $\alpha_m$ rises rapidly due to increased conductor loss and reduced radiation efficiency of the slots. At millimeter-wave frequencies, the loss becomes prohibitive. In contrast, PASS employ dielectric waveguides. The attenuation $\alpha_m$ is governed primarily by material absorption and scattering, which increase with frequency at a much slower rate. 

\subsubsection{Comparison}
The above models reveal a shared guided-radiative principle, but the two systems differ in how radiation is generated and reconfigured. In PASS, the dielectric waveguide serves as a low-loss transmission path. The PA acts as an external, detachable radiator that extracts guided power through controlled near-field coupling. This design allows flexible PA placement, activation, and movement along the waveguide, enabling user-centric reconfiguration at modest incremental hardware cost. In LCX, radiation is an inherent property of the cable structure, with slots or apertures prefabricated on the outer conductor for a specified frequency band and coverage pattern. While LCX provides stable continuous coverage in tunnels or corridors, its radiation positions and coupling characteristics are fixed upon manufacturing, offering limited reconfigurability without significant hardware modifications.

The frequency scalability of the two systems also diverges sharply. At lower frequencies, such as 2.4 GHz, both technologies exhibit comparable propagation attenuation. However, the total link budget is dominated by fundamentally different mechanisms. LCX requires the signal to overcome both cable attenuation and substantial slot coupling loss, which can reach tens of dB over 100 m, severely limiting EE. By contrast, PASS incur negligible coupling loss, since each PA extracts only a small fraction of guided power while the remainder propagates along the waveguide. At millimeter-wave and terahertz frequencies, LCX becomes practically unusable. Its slot-based radiation mechanism grows increasingly inefficient, yielding theoretical propagation attenuation of tens of dB/m. PASS can maintain low propagation attenuation into the hundreds of gigahertz by leveraging advanced dielectric materials and geometries, such as low-loss polymers, ceramic ribbons, or porous dielectrics. Consequently, PASS outperform LCX across the entire frequency spectrum. It reduces coupling loss at lower frequencies, while preserving scalability at higher frequencies through material flexibility that LCX inherently lacks. The distinctions are quantitatively summarized in Table~\ref{tab:lcx_pass_compare}.


\section{PASS-based Positioning}

PASS guides electromagnetic waves along low-loss dielectric waveguides before radiation, which improves wireless coverage and capacity. Unlike conventional fixed antenna arrays or emerging flexible architectures that typically operate within wavelength scale, PASS uniquely achieves meter-scale spatial reconfigurability, allowing radiating PAs to be deployed in close proximity to the users~\cite{liu2026passtutorial,yang2025pinching}.

Despite these advances, existing reconfigurable antenna techniques mainly reshape the channel within a wavelength-scale region. By contrast, PASS enables meter-scale reconfigurability along dielectric waveguides~\cite{ding2025flexible, suzuki2022pinching, 11169486}. However, a fundamental question raises: \emph{where should the PAs be moved or activated?} Addressing this question requires the BS to first acquire reliable user position information. The significance of BS-side positioning in PASS can be summarized as follows:
\begin{itemize}
    \item Since PASS improves the propagation condition by activating or relocating PAs close to users, reliable user location information is needed to determine which PAs should be selected or where the PAs should moved.
    \item The movable PAs can also serve as dynamic anchor nodes for localization, providing geometry-dependent measurements that are unavailable in conventional fixed-anchor systems.
    \item The estimated user positions can further support waveguide selection, beamforming, and communication optimization.
\end{itemize}

The new positioning paradigm also introduces novel challenges, such as in-waveguide signal attenuation, inter-PA signal superposition, and the uncertainty of dynamic anchor positions. This section presents a comprehensive tutorial on BS-side positioning for PASS, covering the positioning methodologies from single-waveguide to multi-waveguide architectures and from pure LoS propagation to complex multipath environments~\cite{zhang2025pinchingantennasystemspassbasedindoor, PASS_poi_4, PASS_poi_2, PASS_poi_40, PASS_poi_3, PASS_poi_7}.

\subsection{Single-Waveguide BS-side Positioning}

Single-waveguide PASS is the most fundamental architecture for BS-side positioning. In this setup, an AP to one dielectric waveguide deployed along a ceiling edge, wall edge, or corridor boundary, and multiple PAs are activated sequentially or in groups along the waveguide to receive uplink pilots from the users. Since the PA positions are known to the AP, each PA can be regarded as an anchor. The key objective is to estimate the user position from measurements associated with these PA anchors. This architecture is attractive for early deployment because it requires only one waveguide, which can operate with low RF-chain complexity, exploiting the meter-scale reconfigurability of PAs.

\begin{figure}[t!]
\centering
\includegraphics[width = 2.5 in]{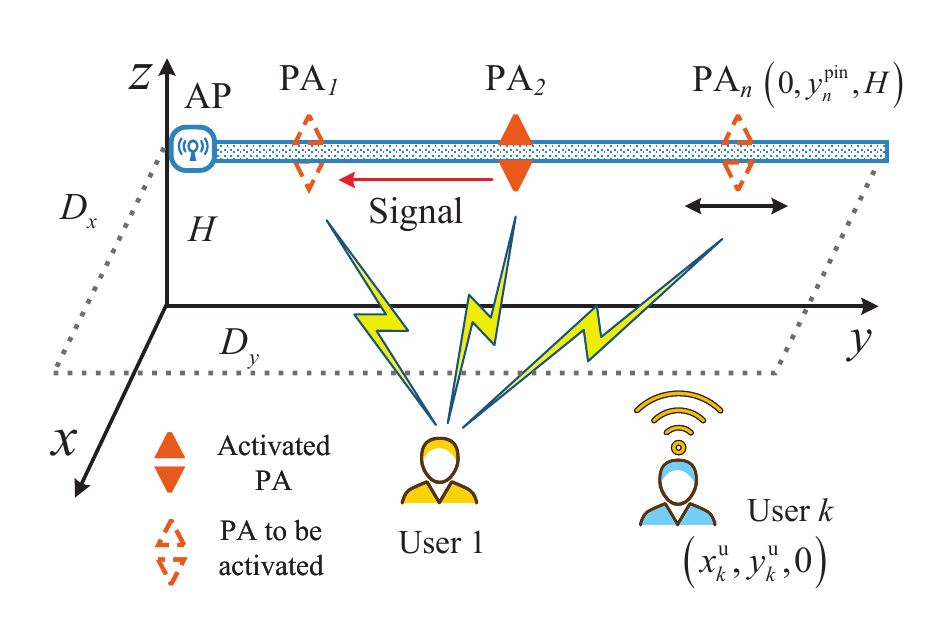}
\caption{Single-waveguide positioning system model.}
\label{system_model}
\end{figure}

As illustrated in Fig.~\ref{system_model}, consider a single-waveguide PASS deployed in a room of size $D_x \times D_y \times H $. The AP is located at the corner of ceiling, denoted by $\mathbf{p}^{\mathrm{AP}}=(0,0,H)$. The waveguide extends along the $y$-axis at height $H$, and the $n$-th PA is located at $\mathbf{p}^{\mathrm{Pin}}_n=(0,y^{\mathrm{Pin}}_n,H)$, where $y^{\mathrm{Pin}}_n \in [0,D_y]$. The PAs are activated following a time division multiple access (TDMA) scheme, i.e., only a single PA along the waveguide is active during each time slot, and the coordinates of each PA are known. The $k$-th single-antenna user is located on the ground plane at $\mathbf{p}^{\mathrm{u}}_k=(x^{\mathrm{u}}_k,y^{\mathrm{u}}_k,0)$. Thus, the distance between the $k$-th user and the $n$-th PA can be written as:
\begin{equation}\label{user-PA distance}
d_{k,n} = \left\| {{\mathbf{p}^{\mathrm{u}}_k} - {\mathbf{p}^{\mathrm{Pin}}_n}} \right\| = \sqrt{{x^{\mathrm{u}}_k}^2+(y^{\mathrm{u}}_k-y^{\mathrm{Pin}}_n)^2+H^2}.
\end{equation}

The uplink signal first propagates from the user to the activated PA through free space, and then propagates along the dielectric waveguide to the AP. Under the LoS assumption, the received signal through the $n$-th PA can be written as:
\begin{equation}\label{rx_signal}
y_{k,n}
=
\underbrace{
\frac{c}{4\pi f_{\mathrm{c}} d_{k,n}}
e^{-j\frac{2\pi}{\lambda}d_{k,n}}
e^{-(\alpha+j\beta)y^{\mathrm{Pin}}_n}
}_{h^{\mathrm{CH}}_{k,n}}
\sqrt{P^{\mathrm{tx}}_k}s_k+n^{\mathrm{AWGN}}_{k,n},
\end{equation}
where $s_k$ is the pilot symbols and $n^{\mathrm{AWGN}}_{k,n}$ is additive white Gaussian noise (AWGN). 

Based on the amount of signal information exploited from \eqref{rx_signal}, existing single-waveguide positioning methods can be broadly classified into three categories: RSSI-based positioning, phase-aware localization, and orthogonal matching pursuit (OMP) geometry-consistent localization.

%
%
%
%
%
%

\subsubsection{RSSI-Based localization}

RSSI-based localization is the simplest approach for single-waveguide PASS positioning. The first RSSI-based PASS positioning work proposes a single-waveguide uplink localization model and develops a PASS-specific weighted least squares (WLS) algorithm tailored to the one-dimensional PA geometry~\cite{zhang2025pinchingantennasystemspassbasedindoor}. By taking the received power of \eqref{rx_signal}, the PA-user distance can be inferred from the amplitude attenuation. Specifically, the estimated distance from the $k$-th user to the $n$-th PA is given by:
\begin{equation}\label{eq:sw_rssi_range}
\hat d_{k,n}
=
\frac{c e^{-\alpha y^{\mathrm{Pin}}_n}}{4\pi f_{\mathrm{c}}}
\sqrt{\frac{P^{\mathrm{tx}}_k}{\hat P^{\mathrm{rx}}_{k,n}}},
\end{equation}
where $\hat P^{\mathrm{rx}}_{k,n}$ denotes the measured received power. 

By combining \eqref{user-PA distance} and \eqref{eq:sw_rssi_range}, a set of distance equations can be constructed. In practical ranging, different PA observations have unequal reliability due to the joint effects of free-space path loss, waveguide attenuation, and nonlinear power-to-distance conversion. To address this issue, the WLS estimator in~\cite{zhang2025pinchingantennasystemspassbasedindoor} constructs a PASS-aware weight based on the inverse of the observation-error variance. In this way, unreliable measurements are down-weighted, while high-quality observations contribute more to the final estimate. Furthermore, since all PA centers lie on the same waveguide, the localization problem has a special one-dimensional anchor geometry. The WLS algorithm may generate two symmetric candidate points, and the infeasible candidate can be removed using the known room boundary or feasible user region.

\begin{figure}[htbp]
\centering
\subfigure[High-SNR conditions.]{\includegraphics[width =0.4\textwidth]{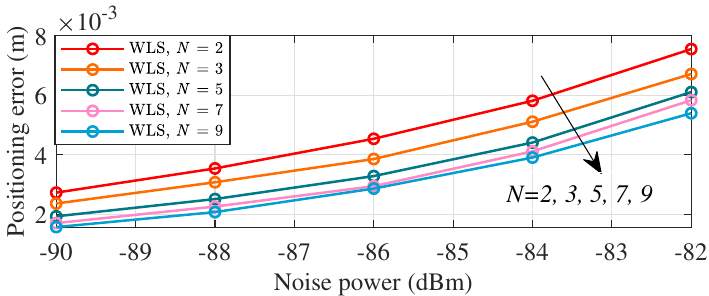} \label{SW_RSSI1}}
\subfigure[Low-SNR conditions.]{\includegraphics[width =0.4\textwidth]{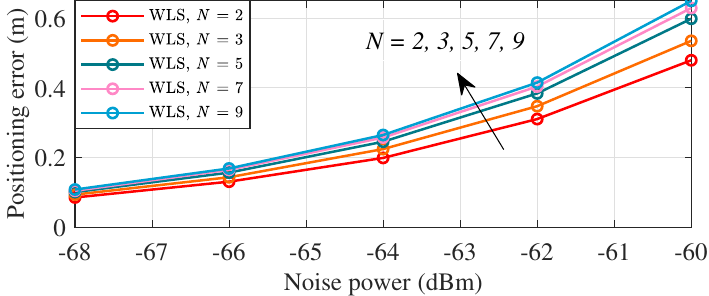} \label{SW_RSSI2}}
\caption{Positioning errors of RSSI-based WLS algorithms under different numbers of PAs.}
\label{positioning error distribution}
\end{figure}

The numerical results validate the feasibility of RSSI-based positioning in PASS. As demonstrated in Fig.~\ref{positioning error distribution}, the PASS-based WLS estimator achieves reliable positioning accuracy across various noise power. Furthermore, the findings reveal an important design trade-off regarding the number of activated PAs: while more PAs enhances geometric diversity under high signal-to-noise ratio (SNR) conditions, it may introduce unreliable measurements under low-SNR conditions. Therefore, an adaptive PA selection strategy should be employed in practical PASS-based localization.

From a system-design viewpoint, RSSI-based positioning is suitable for low-cost coarse localization, especially in LoS-dominant environments. Its main limitation lies in the sensitivity to transmit power calibration, path loss modeling, and environmental dynamics. Therefore, RSSI-based WLS is a lightweight positioning baseline or initialization method, which can be further refined by phase-aware or delay-aware localization when higher accuracy is required.

\subsubsection{Phase-Aware Localization}

Although RSSI based positioning is attractive due to its simplicity and low hardware cost, amplitude only measurements provide relatively coarse range information and fail to exploit the fine grained geometric cues carried by signal phase. Therefore, a natural refinement is to exploit the full complex baseband observation rather than only the received amplitude. Based on the signal model in \eqref{rx_signal}, the likelihood of the received vector $\mathbf{y}_k=[y_{k,1},\ldots,y_{k,N}]^T$ can be formulated as a function of the unknown user position $\mathbf{p}^{\mathrm{u}}_k$. Then, the Fisher information is contributed by both the amplitude variation and the phase rotation. The resulting Fisher information matrix (FIM) admits a two factor interpretation: geometric sensitivity determined by the PA user relative geometry, and amplitude phase sensitivity determined by the distance dependent variation of the received complex signal.

Feng et al.~\cite{PASS_poi_4} investigated phase-aware PASS localization utilizing the complex received signal model, as shown in \eqref{rx_signal}. Based on the complex received signal model, the FIM, Cramér-Rao lower bound (CRLB), and the position error bound are derived, which reveal the joint impact of geometric, amplitude and phase on positioning accuracy. The study further develops a ML estimator that jointly uses amplitude and phase information. Since the likelihood function is non-convex with respect to the user location, a two-stage solution is adopted: coarse grid search is first used to find reliable initial points, and Levenberg-Marquardt refinement is then applied for local optimization. When the phase synchronization and calibration are available, the same sequential PA measurements contain richer geometric information.

Nevertheless, phase-aware positioning introduces practical challenges. Phase wrapping, imperfect synchronization, waveguide-parameter calibration errors, and user motion during PA scanning can degrade performance. Moreover, the ML problem is non-convex, so initialization is important. A practical implementation may therefore use RSSI-based WLS to provide a coarse estimate and then apply phase-aware ML refinement within a smaller candidate region.

\subsubsection{OMP-Based Geometry-Consistent Localization}

Beyond RSSI-based and phase-based localization, another line of work formulates PASS positioning as a sparse geometric parameter estimation problem. This approach exploits the near field structure of PASS channels, where each candidate user or scatterer location corresponds to a deterministic spherical wave response across the activated PAs. Then, the localization can therefore be achieved by constructing a geometry-dependent dictionary and applying sparse recovery methods, such as OMP algorithm. 

A recent user localization and channel estimation framework proposes an OMP-based geometry-consistent localization (OMP-GCL) algorithm for PASS~\cite{PASS_poi_2}. In the single-waveguide structure, multiple PA subarrays are placed along the same waveguide and activated in different time slots. Each subarray obtains compressed channel measurements by randomly activating its PAs. By representing near-field channel vector over a distance-parameterized angular dictionary, the localization problem becomes a sparse recovery problem. The OMP algorithm first estimates the direction-related parameter from each PA group. Then, the estimates from different PA groups are fused through a geometry-consistency step to infer the user and scatterer location.

%
%

\subsection{Multi-Waveguide BS-side Positioning}

In a single-waveguide structure, all activated PA groups remain constrained to a one-dimensional aperture. As a result, the available observation directions are limited to a half-space, which may cause angle or mirror ambiguity. Multi-waveguide deployment can provide full-space observations, improve geometric diversity, and reduce such ambiguity. Existing studies on muti-waveguide PASS positioning can be broadly categorized based on the propagation environment and the type of signal information utilized. In LoS environments, research primarily focuses on power-based localization across different multi-waveguide architectures. In multipath environments, the key challenge lies in extracting reliable geometric information from channel observations.

%
%
%
%
%
%
%
%

\subsubsection{LoS Environments}

\begin{figure}[t!]
\centering
\includegraphics[width =0.4\textwidth]{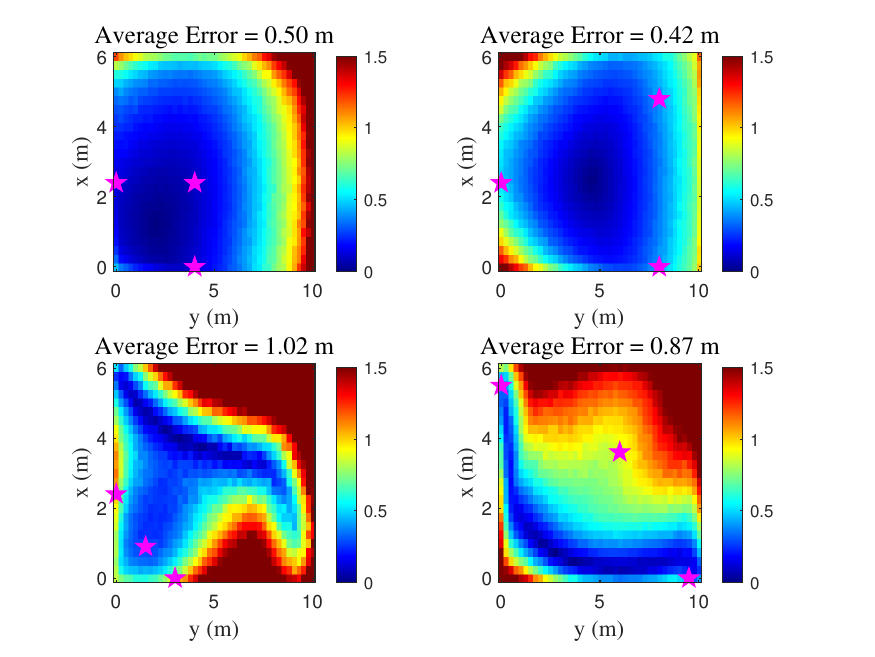}
\caption{RSSI-based positioning error in MWSP scenario, where the noise power is set to -50 dBm.}
\label{MWSP}
\end{figure}

A LoS-based multi-waveguide positioning framework is developed in~\cite{PASS_poi_40}, where two architectures are considered: multi-waveguide single-PA (MWSP) and multi-waveguide multi-PA (MWMP). For the MWSP architecture, only one PA is activated on each waveguide, which contributes one RSSI observation. The distance between the user and PA can be obtained using the same RSSI-to-range principle as in \eqref{eq:sw_rssi_range}. The key difference from single waveguide RSSI positioning is that the PA anchors are no longer collinear. Since the PAs are distributed across different waveguides, their measurements can be obtained in parallel rather than through sequential TDMA based activation, which enables real time localization with reduced measurement latency. Moreover, the range equations associated with different waveguides can be linearized and solved using the least square (LS) method. Therefore, MWSP can be regarded as a natural and interpretable extension of RSSI-based LS positioning from single-waveguide to multi-waveguide scenarios. The performance analysis and simulation results show that MWSP accuracy is jointly determined by ranging quality, anchor geometry, and user location, as illustrated in Fig. \ref{MWSP}. When the selected PAs span a larger effective area or provide more diverse observation directions, the geometric dilution of precision (GDoP) becomes smaller, leading to reduced positioning error. Nevertheless, the analysis of MWSP also exhibits a specific phenomenon: even when the anchor geometry is not globally optimal, users with balanced distances to multiple PAs may still achieve favorable accuracy, because the range errors are more evenly propagated through the linearized equations. Hence, RSSI-based LS algorithm can serve as a low complexity and interpretable baseline for LoS dominant MWSP positioning.

MWMP considers a more complicated activation structure, where multiple PAs on each waveguide are simultaneously active. This configuration is attractive for downlink communication because that simultaneous multi-PA activation can enhance signal strength and support flexible PA deployment. However, the BS no longer observes an independent RSSI measurement associated with each PA. Instead, the received power is determined by the superposition of multiple PA on the same waveguide. Consequently, the observation becomes a nonlinear function of the user-PA position, PA spacing $\Delta d$ and waveguide distribution. Let $\Delta d = \lambda/\sqrt{\varepsilon_{\mathrm r}}$, the received power from the $m$-th waveguide can be written as~\cite{PASS_poi_40}:
\begin{equation}\label{the received power}
\begin{aligned}
{P^{\mathrm{rx}}_{k,m}} &= {P^{\mathrm{tx}}}{\left( {\frac{{c{e^{ - \alpha \sqrt {{x^{\mathrm{Pin}}_m}^2 + {y^{\mathrm{Pin}}_m}^2} }}}}{{4\pi {f_{\mathrm{c}}}{d_{k,m}}}}\frac{{\sin \left( {\frac{{N{\psi _m}}}{2}} \right)}}{{\sin \left( {\frac{{{\psi _m}}}{2}} \right)}}} \right)^2} + {\sigma _m^2},\\
{\psi _m} &= \frac{{2\pi \Delta d\cos {\theta^{\mathrm{Pin}}_m}}}{\lambda },
\end{aligned}
\end{equation}
where $x^{\mathrm{Pin}}_m$ and $y^{\mathrm{Pin}}_m$ represent the coordinate of the PA array center of the $m$-th waveguide, and ${d_{k,m}}$ denotes the distance between the $k$-th user and the array center. ${\theta^{\mathrm{Pin}}_m}$ denotes the angle between signal and the array center, with ${\theta^{\mathrm{Pin}}_m} \in \left( {0,\pi } \right)$. ${\sigma _m^2}$ represents the noise power.

To handle this superposed observation, Zhang et al.~\cite{PASS_poi_40} adopts a grid-search-based positioning strategy. Specifically, for each candidate user location, the BS calculates the predicted received power according to the closed-form expression in \eqref{the received power}, and then selects the location by maximum likelihood strategy. This method directly uses the predicted received power as a spatial fingerprint generated by the MWMP geometry. Nevertheless, the spatial fingerprinting approach introduces a trade-off between received power and spatial ambiguity. Although activating multiple PAs can increase the received signal power, the superposed power map may become more ambiguous, making it more difficult to distinguish the true user location. Moreover, non-parallel waveguides provide complementary spatial signatures, which helps break the symmetry in individual waveguide observations. Thus, the accuracy of MWMP grid search based positioning is jointly affected by the waveguide distribution, the number of activated PAs, and the grid resolution. Overall, MWMP is more suitable when positioning is integrated with multi PA communication, whereas MWSP is preferable when explicit geometric measurements and low complexity are prioritized.



\subsubsection{Multipath Environments}

LoS-based positioning reveals the fundamental geometric principles intrinsic to the PASS architecture. However, practical indoor environments usually contain walls, furniture and moving obstacles, which generate NLoS components. In such cases, RSSI or superposed power may become unreliable because the received signal no longer maps uniquely to the LoS distance. The impact of NLoS motivates the research of PASS positioning in multipath environments, where the geometric information for localization must be extracted from signals containing LoS and NLoS components. 
Existing positioning methods for PASS in multipath environments can be broadly divided into two categories. The first category is \emph{joint channel estimation and localization} (JCEL), where localization is embedded into the channel estimation procedure. Representative methods include OMP-GCL, Orthogonal frequency-division multiplexing (OFDM)-domain OMP delay extraction, and belief propagation variational inference (BP-VI). The second category is \emph{direct solution}, where LoS delays are first extracted from multipath observations and then directly solve for the user position using geometric ranging equations. Representative methods include matrix pencil (MP)-based ranging and its low-complexity Rank-1 approximation.

\paragraph{Joint Channel Estimation and Localization}

The core principle of JCEL is to jointly estimate the channel and user location within a unified framework, which is achieved by formulating the problem as sparse parameter recovery, where location related geometric information is embedded in the channel parameters~\cite{PASS_poi_4,PASS_poi_3}. The OMP-GCL framework proposed in \cite{PASS_poi_4} groups PAs into subarrays, with each acquiring compressed channel measurements via random activation. By deploying these subarrays across multiple waveguides, this framework gathers channel observations from diverse directions, thereby alleviating the half-space ambiguity inherent to single-waveguide setups. Consequently, compared to its single-waveguide counterpart, the multi-waveguide PASS exploits full-space observations to achieve superior localization and channel reconstruction performance. 

Furthermore, the study in~\cite{PASS_poi_3} extends the OMP-based methodology to wideband OFDM systems. While OMP-GCL focuses on near-field spatial signatures, the method in~\cite{PASS_poi_3} exploits the delay-domain sparsity of wideband PASS channels. In such systems, different propagation paths introduce distinct delays, which manifest as phase rotations across subcarriers. By discretizing the delay domain into a dictionary, this method first employs OMP to extract the dominant delay components from the frequency-domain channel observations. Since these extracted delays are inherently unordered, a delay-anchor matching procedure is introduced to associate each delay with the corresponding PA. Subsequently, the user's geometric position is iteratively determined by solving the nonlinear ranging equations derived from these matched delays.

To mitigate the grid mismatch inherent to OMP-based delay discretization, Liu et al.~\cite{PASS_poi_3} proposed the BP-VI approach. Unlike OMP-based methods, which search over a predefined delay grid, the BP-VI framework models the path delays and channel coefficients as continuous latent variables. In this framework, the delay extraction is formulated on a factor graph, where belief propagation efficiently resolves the linear Gaussian mixing, while variational inference approximates the highly nonlinear dependence between continuous delays and subcarrier-domain channel responses. The hybrid inference structure enables the receiver to extract multipath delays with arbitrary resolution. The extracted delays are first associated with their corresponding PAs via a delay-anchor matching procedure, followed by a Newton refinement step that directly solves for the user's position using geometric constraints. 

The main advantage of BP-VI lies in its off-grid delay estimation capability. When the true delays are off the discretized grid, BP-VI provides more accurate delay estimation than grid-based OMP. However, BP-VI also increases computational complexity because it involves iterative probabilistic message passing and variational updates. As a result, BP-VI is more suitable for accuracy-oriented JCEL.

\paragraph{Direct Solution}

In contrast to the JCEL approaches discussed above, direct solution decouple the localization process from channel estimation. Zhang~\textit{et al.}~\cite{PASS_poi_7} estimate the LoS distances directly from multipath observations, and then solve the user's position via geometric ranging equations. This design provides a more direct connection between delay estimation and position calculation, making it highly efficient when the main objective is user localization rather than full channel reconstruction.

For a fixed PA, the channel observation can be regarded as a sum of delay-dependent complex exponentials, which can be written as:
\begin{equation}\label{channel observation}
\begin{aligned}
{h^{\mathrm{CH}}_s} &= \sum\limits_{\ell = 1}^L {\underbrace {{h^{\mathrm{path}}_\ell}{e^{ - j2\pi {f_{\mathrm{c}}}{\tau _\ell}}}}_{a^{\mathrm{path}}_\ell}} {\left( {\underbrace {{e^{ - j2\pi \Delta f{\tau _\ell}}}}_{{z_\ell }}} \right)^s} + n^{\mathrm{AWGN}}_s\\
 &= \sum\limits_{\ell = 1}^L {{a^{\mathrm{path}}_\ell}{{\left( {{z_\ell }} \right)}^s}}  + n^{\mathrm{AWGN}}_s,
\end{aligned}
\end{equation}
where $\ell$ and $s$ denote the indices of path and subcarrier, respectively. $\tau_{\ell}$ is the path delay, ${h^{\mathrm{path}}_\ell}$ is the path gain coefficient, and $n^{\mathrm{AWGN}}_{s}$ denotes the AWGN. ${f_{\mathrm{c}}}$ and $\Delta f$ are the carrier frequency and subcarrier spacing, respectively. This expression reveals the basic mechanism of delay extraction: estimating the poles $z_{\ell}$ is equivalent to estimating the multipath delays $\tau_\ell$.

For the MP-based ranging algorithm, the channel samples are first organized into two shifted Hankel matrices. Based on the shift-invariance of the exponential sequence in \eqref{channel observation}, the signal poles can be extracted from the corresponding MP. The phases of the valid poles are then converted into delay estimates, and the shortest valid delay is selected as the LoS delay. In this way, MP-based ranging explicitly separates the LoS component from NLoS components. However, the method requires pseudo-inverse computation and eigenvalue decomposition, resulting in high computational complexity.

To reduce the computational burden, the Rank-1-based algorithm provides a lightweight alternative. In mmWave indoor scenarios, the LoS path often contributes the strongest component to the channel observation. Therefore, the Hankel matrix can be approximated by the dominant rank-one component. Then, two adjacent submatrices obtained by shifting one column differ only by a constant phase-rotation factor, which is exactly the LoS pole. By avoiding full multipath decomposition, pseudo-inverse computation and eigenvalue decomposition, Rank-1-based ranging reduces the computational complexity. Nevertheless, residual NLoS components may cause deterministic ranging and positioning bias.


Given the LoS distances estimated by either MP-based or Rank-1 ranging, the user position is then obtained by fusing the multi-waveguide estimated distances through a two-stage weighted nonlinear least-squares (WNLS) estimator. The first stage provides an initial position estimate through linearized ranging equations, while the second stage refines the estimate by minimizing the weighted range residuals. These two direct ranging solutions embody a trade-off between ranging accuracy and computational efficiency. By resolving individual propagation paths, the MP-based approach achieves superior positioning accuracy at the cost of matrix decomposition complexity. Conversely, the Rank-1-based approach offers a computationally lightweight alternative by extracting only the dominant LoS component, at the expense of an error floor caused by unresolved multipath bias. Consequently, the MP-based estimator is well-suited for high-precision multipath scenarios, whereas the Rank-1 estimator is preferable for resource-constrained applications.

\subsection{User-side Positioning}

BS-side positioning focuses on centralized location management and network optimization, typically relying on uplink pilot transmission and centralized computing~\cite{zhang2025pinchingantennasystemspassbasedindoor, PASS_poi_4, PASS_poi_2, PASS_poi_40, PASS_poi_3}. 
Nevertheless, BS-side positioning requires continuous uplink signaling and precise coordination with known anchor positions, which may not be feasible in large-scale or resource-constrained deployments. 
User-side positioning empowers mobile terminals to perform self-localization using only downlink broadcast signals. 
It offers a distributed, scalable, and measurement-localized alternative~\cite{trodition_indoor_1, trodition_indoor_2, wifi_1, wifi_2, BLE, UWB_1, UWB_2}. 
For PASS, the user-side approach holds particular significance. 
It fully exploits the inherent ``deploy-and-use'' advantage of reconfigurable waveguide infrastructures~\cite{ding2025flexible, suzuki2022pinching, 11169486}. 
End-users can obtain positioning services immediately after system deployment. 
They do not need site surveys, manual coordinate calibration, or continuous network-side broadcasting of anchor coordinates.

Compared with BS-side positioning, the user-side paradigm has three fundamental differences. First, the signal direction is reversed: positioning is based on downlink broadcast signals rather than uplink pilots. Second, the PA coordinates are unknown to the user terminal, so PAs cannot be directly treated as conventional fixed anchor nodes. Third, user terminals typically have limited computational resources and hardware capabilities, which imposes stricter constraints on algorithm complexity.

Existing user-side positioning methods for PASS can be classified by both the number of waveguides and the type of signal information exploited. The following sections elaborate on single-waveguide and multi-waveguide architectures, covering LoS environments as well as preliminary multipath-resistant approaches.

\subsubsection{Single-Waveguide User-side Positioning}

Single-waveguide user-side positioning is the most fundamental architecture for terminal self-localization. 
It is especially suitable for long corridor scenarios such as underground parking lots, factory aisles and building corridors. 
The geometric setup and the cascaded channel gain are identical to those defined in the BS-side single-waveguide model. 
The core distinction is that the PA coordinates $\mathbf{p}^{\mathrm{Pin}}_n$ are not available at the user side, which prevents direct application of conventional trilateration.

The core challenge of single-waveguide user-side positioning lies in the coupling of two unknown distances: the AP-PA in-waveguide distance and the PA-user free-space distance. A single type of measurement cannot separate the two components. A natural solution is to exploit two heterogeneous measurements that carry different distance dependencies. Li~\textit{et al.}~\cite{li2026pinching} proposed an anchor-line navigation framework.
%
The user can obtain the PA position and pseudorange by using RSSI and ToA measurements via the Lambert W function (LWF). 
With estimated PA positions and pseudoranges from multiple PAs, the user can solve their position by using WLS. 

A key finding is the existence of a breakpoint distance $d_{\mathrm{0}} = \lambda/(\pi \tan\delta)$, within which the solution exists and is unique. Beyond the breakpoint, the strong nonlinearity of the LWF amplifies measurement errors sharply, leading to rapid performance degradation. 
To suppress the nonlinear error amplification near the LWF branch point, an optimal weight matrix is derived based on the Gauss-Markov theorem, which down-weights low-reliability measurements from far-end PAs.

\begin{figure}[t!]
\centering
\includegraphics[width = 3 in]{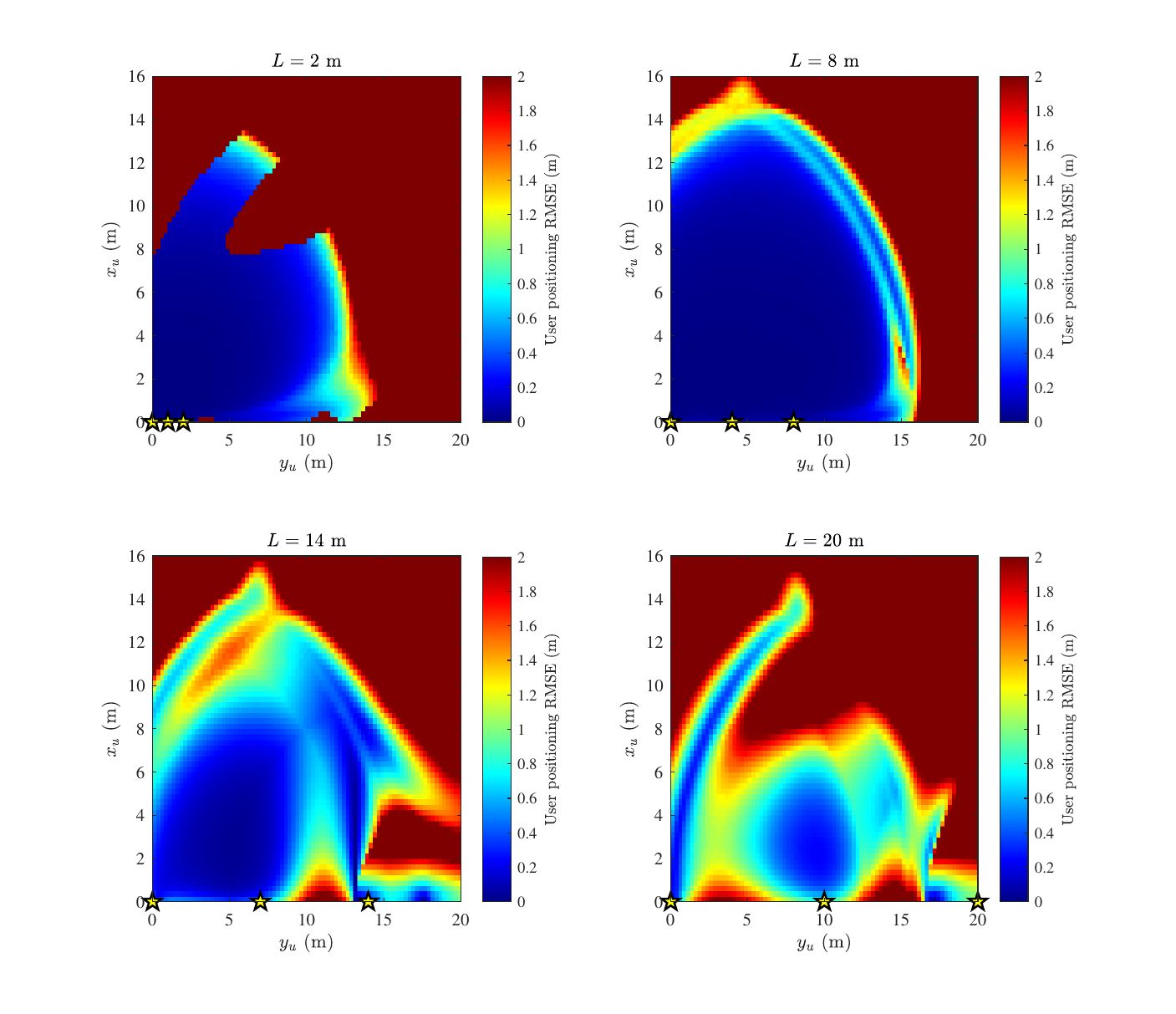}
\caption{User positioning error versus user position $[x^{\mathrm{u}}_k,y^{\mathrm{u}}_k]$ for varying number of PAs $N$ and waveguide lengths $L_{\mathrm{WG}}$.}
\label{lzy_Heatmap}
\end{figure}

Simulation results in Fig. \ref{lzy_Heatmap} verify that the anchor-line framework achieves centimeter-level user positioning accuracy within the working range. However, Fig. \ref{lzy_Heatmap} also reveals that the extent of this working range is constrained by the spacing between the PAs employed for navigation. Overall, anchor-line positioning remains the most mature single‑waveguide user‑side solution, featuring plug‑and‑play deployment and closed‑form solvability, yet its main limitations are the breakpoint distance constraint and the reliance on both power and timing measurements.

\subsubsection{Multi-Waveguide User-side Positioning}

Single-waveguide positioning offers a simple plug-and-play solution, but its inherent one-dimensional collinear geometry fundamentally limits spatial diversity. To evaluate network-level performance in scenarios where multiple waveguides are deployed in a random or semi-random fashion, He~\textit{et al.}~\cite{PASS_poi_5} proposed a stochastic geometry framework, where the spatial distribution of waveguides is modeled as a Poisson line process (PLP), while the pre-deployed PA positions along each waveguide follow a one-dimensional homogeneous Poisson point process (HPPP). 
Two analytical results are central to this work. The first is a closed-form expression for target localizability under a dominant-interference approximation, which reveals the fundamental trade-off between signal-to-interference-plus-noise ratio (SINR) requirements and localization feasibility. The second provides a compact approximation for the RSSI-based CRLB, establishing a direct link among waveguide density, PA density, channel parameters, and probabilistic localization accuracy. The results collectively characterize the theoretical limits of PASS-based positioning in dense random deployments.

The analysis uncovers a critical design trade-off: increasing the number of waveguides does not always improve positioning accuracy. When the number of waveguides is small, the additional geometric diversity dominates and accuracy improves. Beyond an optimal number, however, interference from extra waveguides becomes the dominant factor, leading to performance degradation. Compared with conventional uniform linear arrays, PASS-based user localization achieves a significantly lower CRLB, demonstrating the fundamental advantage of shortened LoS distances and enhanced spatial diversity.

\subsection{PASS-based Positioning Outlook}

Although existing PASS positioning has established useful algorithmic principles, yet several practical issues remain to be addressed in future research:

\begin{itemize}

    \item \textit{Geometric diversity and positioning range:} Single-waveguide deployments provide limited geometric diversity, which may lead to breakpoint constraints, nonlinear error amplification and mirror ambiguities. Multi-waveguide architectures and more robust methods are needed to extend the reliable positioning range and improve identifiability.

    \item \textit{Robust positioning in multipath environments:} Strong reflections in corridors, tunnels, and indoor environments can severely interfere with LoS or first-path estimation. Future methods should exploit joint delay, amplitude, phase, and spatial consistency across multiple PAs while maintaining manageable computational complexity.

    \item \textit{Multi-user positioning with limited resources:} Practical deployments must support multiple users under constrained PA activations, pilots and RF chains. Future work should jointly optimize PA scheduling and ranging/positioning resource allocation to reduce overhead while maintaining localization reliability.

    \item \textit{Joint positioning and dynamic PASS reconfiguration:} PA activation, waveguide selection, and beamforming may change dynamically according to communication requirements. Future systems should enable positioning algorithms to rapidly adapt to such reconfigurations and tightly couple localization with communication control, while minimizing additional signaling and latency.

\end{itemize}

\section{On the Performance of PASS}

Based on the established system and channel models, this section provides a comprehensive performance analysis of PASS. We focus on the mechanism by which the architectural flexibility of PASS, particularly the waveguide-enabled large-scale reconfigurability of PAs, translates into tangible performance gains. We elaborate on this topic from multiple perspectives, covering both downlink and uplink transmissions, as well as scenarios with LoS blockage and NLoS conditions.

\subsection{Downlink PASS}

\begin{figure}[t!]
  \centering
  \includegraphics[width= 2.5 in]{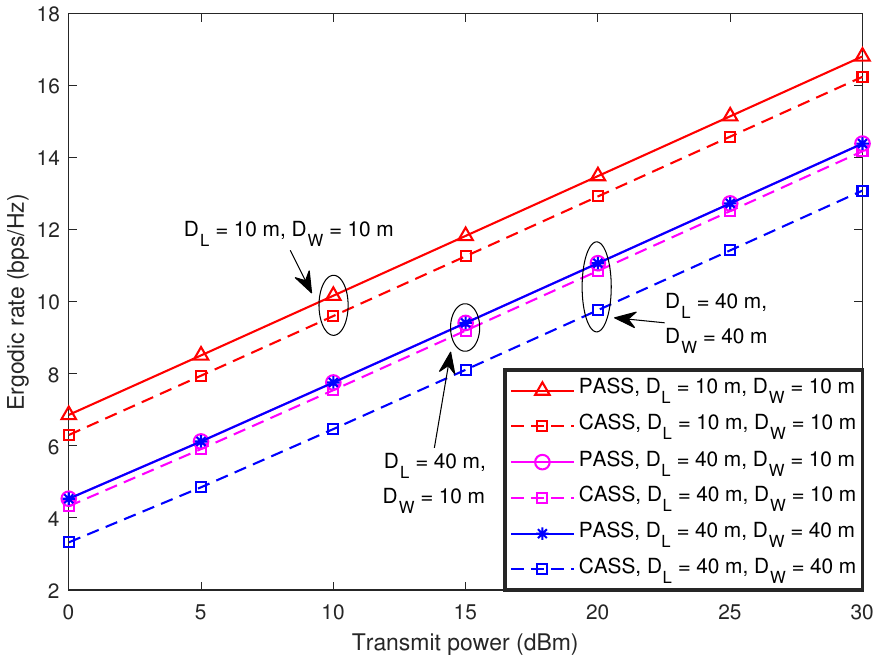}
  \caption{Ergodic rate comparison between single-PA PASS and CASS with a fixed antenna located at $\left( 0, 0, 3 \right)$ over a $D_x \times D_y$ service area.}
  \label{fig:PASS_outperform_CASS}
\end{figure}

Downlink analysis has been one of the earliest and most fundamental research directions in PASS, as it provides direct insight into the effect of PA deployment on the channel conditions. Ding~\textit{et al.}~\cite{ding2025flexible} initially established the system models and prospects of PASS. Two distinctive features of PASS compared to other flexible-antenna technologies are showed: the large-scale reconfigurability of waveguide-enabled antennas, and the capacity to support LoS link~\cite{suzuki2022pinching, example2025}. Therefore, downlink PASS analysis centers on three questions: whether PASS outperforms conventional antenna systems (CASS), what gives rise to that gain, and what limits it.

To validate the performance of PASS, Fig.~\ref{fig:PASS_outperform_CASS} illustrates the ergodic rate performance of PASS and CASS as a function of the transmit power under different area sizes. UEs are assumed to uniformly distributed in the service area. The carrier frequency is set to $28$ GHz. The effective refractive index of a dielectric waveguide is set to 1.4. It is observed that PASS consistently outperforms CASS across all considered settings. The advantage of PASS can be demonstrated through the variation in area size. Specifically, the ergodic rate of PASS is sensitive to the area width, since the activated PA can be repositioned along the waveguide to match the user location. In contrast, for CASS, enlarging the area in either length or width degrades the ergodic rate due to the increased average propagation distance. Furthermore, the performance of PASS is constrained by various factors, such as resource budget, transmit power, user distribution, and area size, where the transmit power is critical as the ergodic rate scales almost linearly with it.

Existing studies have gradually evolved from basic single-PA to multi-PA for multiple access and from single-waveguide to multi-waveguide. In the basic single-PA case, the main focus is to characterize the fundamental metrics of PASS, including outage probability (OP), average rate, and the optimal PA location over the service region. Tyrovolas~\textit{et al.}~\cite{10976621} investigated the fundamental performance analysis of a single-PA PASS, which shows that even a single PA can substantially improve downlink performance by reducing large-scale fading. Chen~\textit{et al.}~\cite{11540399} revealed a fundamental tradeoff between in-waveguide attenuation and free-space path loss in determining the optimal location of a single PA.
In the multi-PA case, the performance gain is no longer determined only by the nearest PA-user distance, but also by inter-PA spacing, phase coherence, power allocation, and in-waveguide attenuation. Xu~\textit{et al.}~\cite{10896748} studied the downlink performance of PASS with multiple PAs, which also provided strategies for multi-PA activation. Wang~\textit{et al.}~\cite{wang2025passmodeling} developed the physical coupler models for multiple PAs along the same waveguide, including equal power model and proportional power model. To explore a motor-based realization of PASS, Hou~\textit{et al.}~\cite{11434513} investigated a downlink PASS with PAs clustered on an array. Xu~\textit{et al.}~\cite{11457366} explicitly incorporated the in-waveguide attenuation into PASS modeling and analysis, where the determination standard of optimal PA position was concluded by categorically discussing the tradeoff between free-space path loss and in-waveguide attenuation.

In multiuser scenarios, the array gain enabled by flexible PA deployment must be effectively distributed among users through proper multiple access design. Orthogonal Multiple Access (OMA) avoids inter-user interference by allocating orthogonal resources, but its performance is constrained by resource partitioning. NOMA improves SE through power-domain multiplexing, but introduces inter-user interference and successive interference cancellation (SIC)-related constraints. Therefore, the comparison between OMA and NOMA provides an important perspective for understanding the translation of the spatial flexibility of PASS into fairness-oriented or rate-oriented multi-user gains.

Cheng~\textit{et al.}~\cite{M4} analyzed the performance of PASS-assisted multiple access under both OMA and NOMA, where a single PA serves a near user and a far user, simultaneously. They revealed that OMA was more beneficial to the strong user, whereas NOMA improved the weak user through power-domain multiplexing, thereby providing better fairness. Subsequently, Cheng~\textit{et al.}~\cite{cheng2026singlewaveguidemultiplepinchingantennasystemsoma} expanded the performance comparison of PASS-assisted multiple access to multi-PA settings, which revealed that the superiority of OMA or NOMA in downlink PASS depends on parameters such as the number of users and power allocation, rather than being absolute. By adopting a representative antenna approximation, Yue~\textit{et al.}~\cite{yue2026performanceanalysispinchingantenna} classified the considered users into near and far nodes, which indicated that blockage effects and LoS/NLoS propagation were critical factors in PASS.

However, the array gain of PASS is not unlimited. An excessive number of PAs reduces the power allocated to each PA and introduces weak contributions from far-away radiating points, leading to a non-monotonic relationship between the number of PAs and the achievable array gain. Ouyang~\textit{et al.}~\cite{10981775} analyzed the array gain of PASS, which demonstrated the existence of an optimal number of PAs and an optimal inter-antenna spacing that maximized the array gain. Then, Ouyang~\textit{et al.}~\cite{11358830} also analyzed the capacity limits of multi-user for multi-PA cases, as well as the inner and outer bounds, which provided a theoretical baseline for evaluating multiple access schemes. Qian~\textit{et al.}~\cite{11391537} compared two representative deployment strategies under a fixed number of PAs: multiple PAs on a single waveguide, and multiple waveguides, each with a single PA. The results revealed that, while the multi-waveguide deployment achieved a higher spectrum efficiency, it resulted in a low EE.

Overall, downlink performance analysis of PASS could further develop advanced channel models that jointly characterize electromagnetic modeling, propagation loss, coupling behavior, PA activation, and multi-user interference.

\subsection{Uplink PASS}

Efficient and reliable uplink communications are critical for many practical applications, such as indoor positioning and internet of things data reporting. Despite the dominant focus on downlink design, uplink transmission is of comparable importance in PASS-enabled networks. The uplink PASS introduces distinct system design issues that cannot be directly inferred from the downlink case, including resource allocation, signal detection, and interference management.

Specifically, Tegos~\textit{et al.}~\cite{tegos2025minimumdatarateuplink} investigated the uplink design of multi-user PASS for the first time. To balance throughput and fairness, by jointly optimizing PA positions and resource allocation, the minimum achievable data rate was maximized. Hou~\textit{et al.}~\cite{11195810} revealed the regular distribution of optimal PA positions within a single waveguide for uplink PASS. Specifically, the optimal positions for multiple PAs exhibit an asymmetric, non-uniform distribution. Zhang~\textit{et al.}~\cite{11175694} extended the study on uplink MISO-PASS to a NOMA-based multi-waveguide setting, where one PA was activated on each waveguide, and addressed the corresponding sum-rate maximization problem. To further improve the sum-rate and EE of uplink PASS, respectively, Zeng~\textit{et al.}~\cite{11215676, 11131179} studied the problem of power allocation and PA positioning. Moreover, Bereyhi~\textit{et al.}~\cite{11414143} explored the potential of PASS for joint uplink and downlink transmission in MIMO settings, where the simulation results demonstrate that PASS could achieve a greater performance gain in MIMO scenarios than conventional MIMO techniques.

\begin{figure}[t!]
  \centering
  \begin{minipage}{0.24\textwidth}
    \centering
    \includegraphics[width = \linewidth]{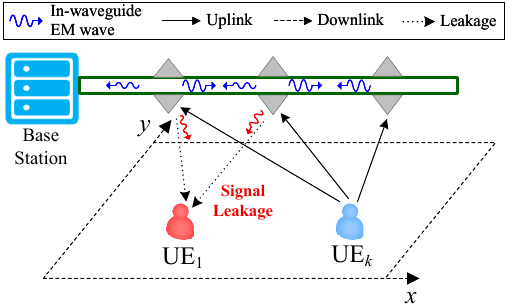}
    \centerline{\footnotesize(a)}
  \end{minipage}
  \begin{minipage}{0.24\textwidth}
    \centering
    \includegraphics[width = \linewidth]{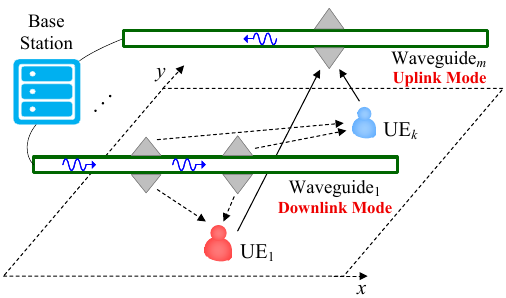}
    \centerline{\footnotesize(b)}    
  \end{minipage}
  \caption{Illustrations of uplink PASS: (a) Signal leakage in multi-PA deployments; (b) A multi-waveguide design of asynchronous uplink and downlink transmission.}
  \label{fig:uplink_PASS}
\end{figure}

However, previous studies have idealized the signal delivery from each PA to the BS, thereby neglecting a significant practical challenge in multi-PA deployments. As illustrated in Fig.~\ref{fig:uplink_PASS}(a), when multiple PAs are simultaneously activated on the same waveguide for uplink transmission, a fundamental issue named inter-antenna radiation (IAR) effect exists in uplink PASS, which manifests as signal leakage~\cite{liu2026passtutorial}. Specifically, a UE-emitted uplink signal enters the waveguide through one PA, but the direction in which electromagnetic waves propagate in the waveguide is uncertain, which may propagate towards either the feed point or the termination load of the waveguide. During such uncertain in-waveguide propagation process, part of the signal may be unintentionally re-radiated through other activated PAs on the same waveguide. The IAR effect not only reduces the effective signal power delivered to the BS, but also causes unintended interference to other UEs. Therefore, the IAR effect makes the signal model mathematically intractable and complicates the design of multi-PA uplink PASS.

Existing solutions mainly avoid the IAR effect by enforcing single-PA activation within each waveguide. Specifically, one solution is to constrain each waveguide to operate in either uplink or downlink mode only, while allowing only one active PA on an uplink waveguide. As illustrated in Fig.~\ref{fig:uplink_PASS}(b), asynchronous uplink/downlink transmission can be implemented by assigning different waveguides to different transmission modes, which is compatible with existing PASS architectures but requires multi-waveguide deployment and mode scheduling.
The other solution is to adopt a segmented waveguide-enabled PASS (SWANs) architecture~\cite{11348983}, where each segment is equipped with an independent feed point and is not physically interconnected to the others. The SWANs architecture shortens the in-waveguide propagation distance and thus mitigates in-waveguide attenuation, but it requires more feed points, leading to higher RF-chain complexity and deployment cost. Both solutions further require UE-aware waveguide or segment activation and mode scheduling according to the spatial distribution and traffic demands of UEs.



\subsection{PASS Under LoS Blockage}

\begin{figure}[t!]
  \centering
  \begin{minipage}{0.24\textwidth}
    \centering
    \includegraphics[width = \linewidth]{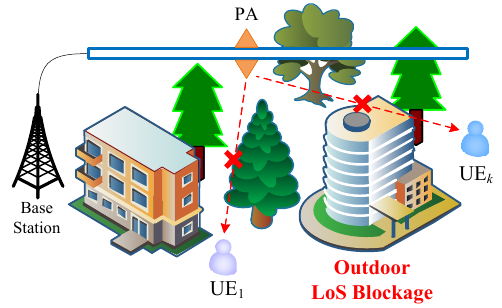}
    \centerline{\footnotesize(a)}
  \end{minipage}
  \begin{minipage}{0.24\textwidth}
    \centering
    \includegraphics[width = \linewidth]{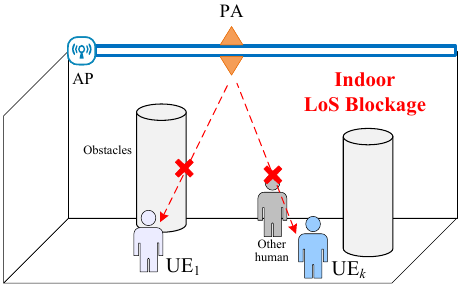}
    \centerline{\footnotesize(b)}    
  \end{minipage}
  \caption{Illustrations of PASS with LoS blockage: (a) In outdoor environments; (b) Indoors.}
  \label{fig:PASS_with_blockage}
\end{figure}

\begin{figure}[t!]
  \centering
  \includegraphics[width= 2.5 in]{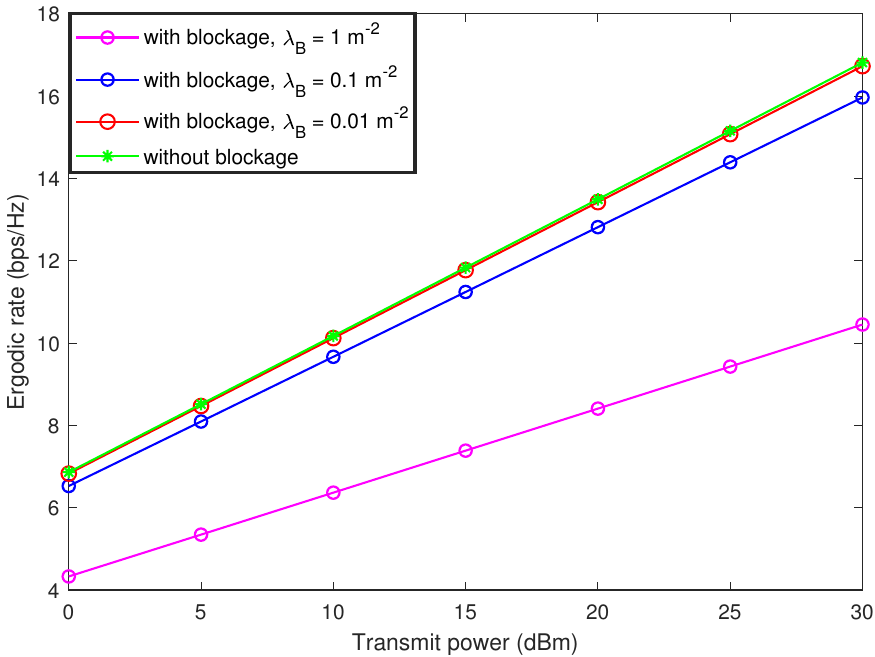}
  \caption{Ergodic rate of PASS with random LoS blockage and without LoS blockage in a $10 \times 10$ m$^2$ area, where $\lambda^{\mathrm{BL}}$ denotes the density of obstacles.}
  \label{fig:PASS_general_and_blockage}
\end{figure}

PASS is distinguished by its capability to flexibly establish favourable LoS links between PAs and UEs~\cite{11169486}. However, LoS blockage may impede the realization of this advantage. As illustrated in Fig.~\ref{fig:PASS_with_blockage}(a) and Fig.~\ref{fig:PASS_with_blockage}(b), in practical environments, LoS links are vulnerable to blockage induced by surrounding objects, such as buildings, vehicles, foliage in outdoor environments, and walls, columns, and human bodies indoors. LoS blockage can significantly reduce the achievable gain and practical deployment performance, particularly at high frequencies such as millimeter-wave (mmWave) bands, where weak diffraction and limited penetration make wireless links more susceptible to blockage~\cite{7593259, 9512383}.
Fig.~\ref{fig:PASS_general_and_blockage} illustrates that the performance of PASS is significantly impacted by LoS blockage. Although the ergodic rate increases almost linearly with transmit power, the rate gap between blocked and unblocked cases indicates that increasing the transmit power cannot fully compensate for the loss caused by LoS blockage. Moreover, the performance degradation becomes more pronounced as the obstacle density increases, which highlights the sensitivity of PASS to obstacle-rich environments. Therefore, it is essential to conduct a systematic investigation of blockage effects for PASS design, optimization, and practical deployment.

Existing studies on PASS under LoS blockage have mainly followed two research routes: probabilistic blockage models and geometry-aware blockage models. 

\emph{1) Probabilistic LoS blockage model:} the probabilistic LoS blockage model is based on the LoS probability, which is a global mathematical representation of the existence of a LoS path. For instance, the LoS probability is typically expressed as a simple exponential function related to the distance between a PA and a UE, which is expressed as follows~\cite{7448962, 3gpp.38.901}:
\begin{equation}\label{LoS_probability}
\mathbb{P}_{\mathrm{LoS}} = \exp \left( { - \beta^{\mathrm{BL}} {\left\| {\mathbf{p} ^{\mathrm{Pin}} - \mathbf{p} } \right\|}^{\alpha^{\mathrm{BL}}}} \right),
\end{equation}
where $\mathbf{p} ^{\mathrm{Pin}}$ and $\mathbf{p}$ denote the locations of a PA and a UE, respectively. $\beta^{\mathrm{BL}}$ denotes the LoS blockage exponent, which mathematically quantify the severity of LoS blockage, and $\alpha^{\mathrm{BL}}$ is a numerical fitting of the environmental density, which is usually set to $1$ or to $2$ for dense environments.

The tractability of probabilistic blockage models makes them convenient for initial analysis of PASS. Ding and Poor~\cite{11036558} first highlighted the dual effect of LoS blockage. While LoS blockage might degrade performance by blocking the desired LoS path, it could also be leveraged in multiuser PASS to suppress co-channel interference. Specifically, in PASS equipped with multiple waveguides, obstacles can be utilized to block undesired interference LoS paths by properly adjusting PA positions, thereby enhancing overall performance. Under a low-loss waveguide setting, Xu~\textit{et al.}~\cite{11533561} demonstrated that the in-waveguide attenuation had a negligible impact compared to blockage under meter-level propagation, particularly in densely obstructed environments. Li~\textit{et al.}~\cite{11516156} and Hanif~\textit{et al.}~\cite{11475477} exploited probabilistic blockage models to investigate power minimization and multicasting in PASS, respectively. Sun~\textit{et al.}~\cite{11315149} extended the PASS deployment to the multi-cell network, where spatially distributed waveguides are connected to different BSs, and both LoS blockage and inter-cell interference are taken into consideration. By exploiting LoS blockage as a multiple-access resource rather than treating it as a purely detrimental effect, Ding~\textit{et al.}~\cite{ding2025environmentdivisionmultipleaccess} proposed a technique termed environment division multiple access (EDMA) for PASS, which laid a foundation for environment-aware PASS design.

While probabilistic blockage models offer analytical simplicity and facilitate the tractable evaluation of system-level average performance, they usually describe blockage through empirical distance-dependent LoS probability functions rather than explicit obstacle geometry, which limits their physical interpretability in practical blockage conditions. For instance, both $\beta^{\mathrm{BL}}$ and $\alpha^{\mathrm{BL}}$ in~\eqref{LoS_probability} mainly control the statistical fitting of LoS availability with distance, but their association with specific physical properties of obstacles, such as locations, sizes, or shapes, is unclear. Moreover, since the LoS state of a specific PA-UE link is essentially binary, i.e., either blocked or unblocked, probabilistic blockage models are not well suited for link-level performance characterization or PA activation design.

\emph{2) Geometry-aware blockage model:} geometry-aware blockage models explicitly incorporate obstacle distributions, which provide a more physically interpretable description of blockage effects caused by real-world objects. Thus, geometry-aware blockage models are relevant for practical PASS deployment.

To exploit LoS blockage for interference suppression, Wang~\textit{et al.}~\cite{11178241} initially studied the problem of joint waveguide and PA association, where the size and position of obstacles were perfectly known. Building on the cylindrical blockage model, Xie~\textit{et al.}~\cite{xie2026pinchingantennasblockageawareenvironments} investigated both discrete and continuous PA placement designs, thereby treating blockage-aware LoS transitions as an integral part of PASS modeling and optimization. By adopting a cube-shaped blockage model for PASS, Zhao~\textit{et al.}~\cite{zhao2026robustsecureblockageawarepinching} derived the criteria for determining LoS blockage with the assistance of cubic envelopes. Xu~\textit{et al.}~\cite{xu2026environmentawarenetworkleveldesigngeneralizedII} developed a geometry-aware model corresponding to the channel model for obstacle-rich environments. The service area was discretized into a uniform grid, so that geometry-aware blockage states and geometry-dependent channel quantities could be precomputed, which facilitated tractable network-level optimization of PA activation under explicit obstacle layouts.

Nevertheless, geometry-aware blockage models still face several challenges. On the one hand, geometry-aware blockage models must balance modeling fidelity and mathematical tractability, as realistic environments contain obstacles of diverse shapes, sizes, and spatial distributions. On the other hand, most existing studies rely on detailed a priori maps of obstacle locations and geometries, which limits their universality to deployment scenarios where such information is unavailable. Moreover, the LoS/NLoS state may alternate rapidly due to dynamic obstacles or users, which causes the effective channel to switch discontinuously between distinct fading regimes. The LoS/NLoS state jumps render the associated optimization problems, e.g., PA positioning or beamforming design, non-convex and non-smooth, which poses significant challenges for conventional gradient-based solvers.

Beyond the two blockage models mentioned above, Wang~\textit{et al.}~\cite{wang2026blockageeffectpinchingantennasystems} recently proposed a stochastic geometry-based blockage model that retained the analytical tractability of probabilistic models while incorporating geometric information, such as the distribution and dimensions of obstacles. The randomness of obstacles and the LoS blockage criterion naturally yielded closed-form expressions of LoS probability, which preserved both physical interpretability and analytical tractability. Therefore, the stochastic geometry-based blockage model serves as an initial attempt to bridge purely probabilistic and geometry-aware approaches, which offers a principled and tunable trade-off between modeling fidelity and mathematical feasibility that deserves further investigation.

\subsection{PASS Under LoS and NLoS Channels}

\begin{figure}[t!]
  \centering
  \includegraphics[width= 2.2 in]{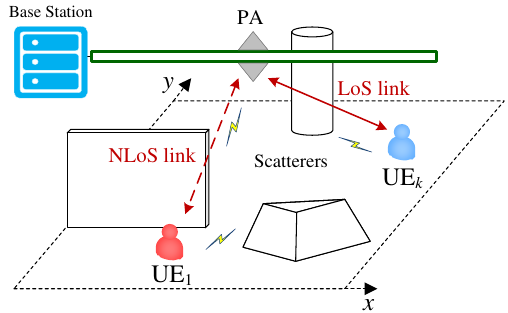}
  \caption{Illustration of PASS under LoS and NLoS channels.}
  \label{fig:PASS_under_LoS_and_NLoS_channels}
\end{figure}

Owing to the meter-scale activation range of PAs, the power strength of LoS propagation created by PASS is usually a hundred times stronger than that of NLoS scattering~\cite{yang2025pinching}. Therefore, the LoS-dominated channel setting is widely adopted, which is well justified in open or obstacle-sparse environments. The free-space path loss governs the large-scale fading of the LoS channel. The small-scale fading is limited to the phase shift of signals propagating in free-space environments. However, as illustrated in Fig.~\ref{fig:PASS_under_LoS_and_NLoS_channels}, LoS paths may be partially or fully blocked in practical complex environments, and then the performance is governed by scattered NLoS propagation. This phenomenon is particularly pronounced in scenarios with dense obstacles or rich multipath scattering, such as indoor spaces, underground utility corridors, and power cable tunnels. Hence, it is of great importance to jointly consider LoS and NLoS components for the accurate modeling, performance evaluation, and design of PASS.

The conventional Rician fading model, which prescribes the K-factor as an independent parameter~\cite{11526989}, is insufficient for PASS, where the PA-UE distance varies with PA repositioning. Therefore, the PA-UE distance is incorporated into the channel model by characterizing the LoS and NLoS components separately. In the presence of both in-waveguide attenuation and LoS blockage, the effective channel gain comprises both the in-waveguide and spatial channel gains, which can be expressed as follows~\cite{wang2026performancepinchingantennasystemspass}:
\begin{equation}\label{effective channel gain}
h = h^{\mathrm{WG}}\left( \zeta h^{\mathrm{LoS}} + h^{\mathrm{NLoS}} \right),
\end{equation}
where $h^{\mathrm{WG}}$ denotes the in-waveguide channel gain, $h^{\mathrm{LoS}}$ denotes the LoS channel component, $h^{\mathrm{NLoS}}$ denotes the NLoS channel component, and $\zeta \in \{0,1\}$ is a LoS indicator that represents the presence ($\zeta = 1$) or absence ($\zeta = 0$) of the LoS component. Note that when $\zeta = 0$, the spatial channel gain is characterized by a NLoS channel with Rayleigh distribution. Conversely, when $\zeta = 1$, the spatial channel gain is characterized by a LoS-dominated channel with Rician distribution.
The NLoS component is modeled as the superposition of multiple independent scattering clusters. As each cluster follows a zero-mean complex Gaussian distribution, their sum also follows such a distribution. Consequently, the NLoS channel component satisfies $h^{\mathrm{NLoS}} \sim \mathcal{CN}\left( 0,\frac{\mu^{\mathrm{NLoS}}}{d_{k,n}^{\alpha^{\mathrm{NLoS}}}} \right)$.


\begin{figure}[t!]
  \centering
  \begin{minipage}[b]{0.48\linewidth}
    \centering
    \includegraphics[width=\linewidth]{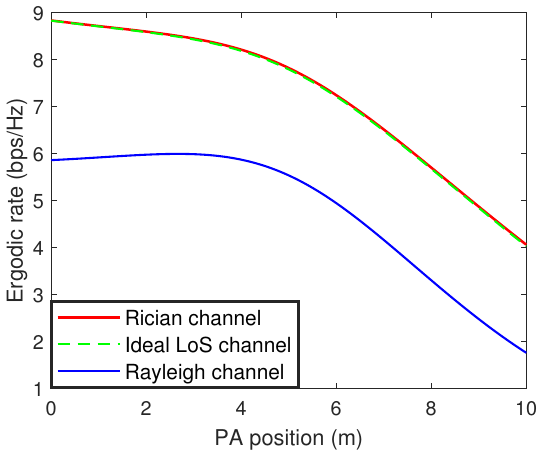}
    \centerline{\footnotesize(a)}
  \end{minipage}
  \hfill
  \begin{minipage}[b]{0.48\linewidth}
    \centering
    \includegraphics[width=\linewidth]{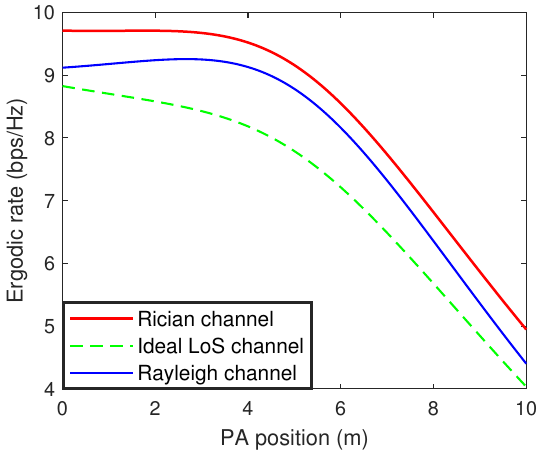}
    \centerline{\footnotesize(b)}
  \end{minipage}
  \caption{Ergodic rate versus PA position with different average NLoS power in SISO-PASS: (a) $-30$ dBm; and (b) $-20$ dBm.}
  \label{fig:PASS_Rician_Rayleigh}
\end{figure}

Fig.~\ref{fig:PASS_Rician_Rayleigh}(a) and Fig.~\ref{fig:PASS_Rician_Rayleigh}(b) show the ergodic rate under ideal LoS, Rician, and Rayleigh channels with different average NLoS power in a single-PA, single-UE scenario. Simulations are conducted in a $10 \times 10$ m$^2$ area with a single UE. The transmit power is set to $10$ dBm. The power of AWGN is configured as $-174 + 10\log_{10}(BW)$ dBm, where the bandwidth is set to $BW = 100$ MHz. The path loss exponent of NLoS scattering is set to $3$. The in-waveguide attenuation corresponds to about $1.47$ dB/m at $28$ GHz when PTFE is adopted as the dielectric waveguide~\cite{pozar2021microwave}. As illustrated in Fig.~\ref{fig:PASS_Rician_Rayleigh}(a), when the NLoS component is weak, the ergodic rate under Rician channel closely follows that under ideal LoS channel, which implies that the deterministic LoS component dominates the effective channel gain. As illustrated in Fig.~\ref{fig:PASS_Rician_Rayleigh}(b), when the average NLoS power increases, the NLoS component becomes more influential, and the ergodic rate gap between the Rician and ideal LoS channels becomes more pronounced. Moreover, the Rayleigh channel exhibits more distinct rate variations with the PA position due to the presence of in-waveguide attenuation. In addition, for all channels, the trade-off between waveguide loss and propagation loss yields a unique optimal PA position along the waveguide. The aforementioned results suggest that the deployment and performance evaluation of PASS should account for the strength of NLoS propagation, particularly in obstacle-rich environments where NLoS scattering can substantially impact the expected LoS channel performance.

Based on the channel model that incorporated both random LoS and NLoS components, Xu~\textit{et al.}~\cite{xu2025pinchingantennadesignrandomlos} investigated two optimization criteria: average SNR threshold coverage maximization and fairness-aware worst-grid average-SNR maximization, which target statistical average performance and weakest-link reliability under unfavorable NLoS conditions, respectively. Furthermore, the channel model that incorporated random LoS and NLoS components was applied in environment-aware PASS designs~\cite{xu2026environmentawarenetworkleveldesigngeneralizedI, xu2026environmentawarenetworkleveldesigngeneralizedII}. Overall, these studies show that incorporating NLoS propagation is essential for extending PASS analysis from idealized LoS conditions to more realistic environments.

\subsection{Discussion and Outlook}

\begin{itemize}
\item \textit{Uplink Signal Design and Scheduling for PASS:} uplink PASS faces a unique signal leakage problem. Specifically, a user signal coupled into the waveguide through one PA may propagate along uncertain directions and be re-radiated through other activated PAs. Current solutions activate one PA per uplink waveguide or segment and rely on waveguide-level scheduling, failing to adapt to user distribution or traffic demand. Therefore, future work could develop tractable uplink signal models and UE-aware scheduling that address leakage suppression, hardware complexity and deployment flexibility jointly.
\item \textit{NLoS Scattering-Induced Phase Perturbation in Multi-PA Deployments:} the potential benefits of multi-PA deployment depend on coherent signal combining. However, NLoS scattering introduces random phase perturbations, which may disrupt phase coherent combining and reduce the achievable array gain. Under Rician channels, the deterministic LoS component still provides a stable phase reference, so the random phase perturbations may only partially degrade coherent combining. In contrast, under obstacle-rich Rayleigh fading channels, the signal phase is primarily determined by random multipath scattering, which makes phase alignment fragile. Hence, the effect of NLoS-induced random phase perturbations on coherent combining in multi-PA deployments remains largely unexplored and requires further investigation.
\item \textit{Performance Analysis of Multiuser and Multi-waveguide Deployments:} practical downlink PASS may involve multiple waveguides, multiple PAs, and multiple users, where the performance gain is jointly determined by multi-waveguide deployment, PA activation, and power allocation schemes. Future work could characterize OP, achievable rate, spectral efficiency and EE under different SNR regimes and beamforming strategies. The main analytical difficulty arises from the in-waveguide phase shifts, free-space path loss, and multiuser interference, for which tractable approximations under physically consistent channel models would be valuable.
\item \textit{Blockage-aware PASS Design in Dynamic Environments:} existing blockage models either assume perfect knowledge of obstacle locations or prescribed statistical distributions, which is rarely available in practice. Furthermore, LoS or NLoS states change over time due to moving obstacles and UEs, as well as PA repositioning, making blockage prediction and environment tracking essential. AI-driven techniques offer a promising means of enabling such capability.
\end{itemize}


\section{Resource Allocation for PASS}
	\label{sec:optimization}
	
	PASS have recently emerged as a revolutionary flexible-antenna architecture, wherein PAs can be dynamically positioned along dielectric waveguides to establish strong LoS links and mitigate large-scale path loss~\cite{ding2025flexible, suzuki2022pinching}. Unlike conventional fixed-position antenna arrays that rely solely on digital beamforming to shape the radiation pattern, PASS introduces an additional spatial DoF by allowing the physical relocation of radiating elements at the meter scale~\cite{11169486, yang2025pinching}. Consequently, PA position becomes a critical design dimension that directly governs the channel gains, interference patterns and the achievable communication performance.
	
	The optimization of PASS encompasses a broad spectrum of technical challenges, ranging from the joint design of transmit beamforming and PA positions to resource allocation under practical hardware constraints. This section focuses on three technical thrusts: \emph{(i)}~PA position optimization, \emph{(ii)}~transmit beamforming optimization, and \emph{(iii)}~EE and QoS-aware optimization. This section presents a comprehensive review of these optimization methodologies, highlighting the evolution from idealized single-user models to practical multi-user deployments under realistic constraints.
	
	\subsection{PA Position Optimization}
	\label{subsec:placement}
	
	Optimizing PA positions along waveguides, often termed pinching beamforming \cite{wang2025passmodeling}, is the core optimization problem in PASS. The channel between the PA and a user is shaped by both free-space path loss and in-waveguide propagation, so the PA position directly controls the amplitude and phase of the equivalent baseband channel \cite{ding2025flexible}. Existing optimization frameworks fall into three categories, continuous position optimization, discrete antenna activation, and design under practical deployment constraints. 
	
	\subsubsection{Continuous Position Optimization}
	The continuous position optimization assumes that PAs can be positioned arbitrarily along the waveguide without constraints, thereby establishing the theoretical performance upper bound. The idealization serves as the analytical cornerstone for pinching beamforming, as it fully exploits the continuous spatial DoFs offered by the waveguide. For the single-user single-PA scenario, Xu et~al.~\cite{10896748} derived the closed-form optimal position that maximizes the downlink achievable rate. Since the channel gain is monotonic in the distance, the optimum is the projection of the user onto the waveguide, subject to aperture boundaries. In contrast, the multi-user multi-PA scenario gives rise to a highly non‑convex sum‑rate function that is plagued by numerous local optima. Wang~et~al.~\cite{wang2025passmodeling} proposed a penalty-based alternating optimization (AO) that decoupled the problem by employing a low-complexity one-dimensional search for the PA positions. The one-dimensional search helps escape poor local optima, achieving near-optimal performance with minimal overhead.
	
	Beyond standard linear waveguides, Zhong et al.~\cite{Zhong2026TCOM_2D} extended PA design to a two‑dimensional radiating plane, optimizing PA positions via particle swarm optimization (PSO) to maximize the minimum received SNR for fairer coverage. Xue et al.~\cite{Xue2026arXiv_multicell} addressed multi‑cell PASS with inter‑cell interference coupling PA positions across cells, adopting alternating PSO‑based PA placement to maximize the weighted sum rate through position reconfiguration. Both works highlight the importance of PA position optimization in harnessing system flexibility, resorting to heuristics when analytical solutions are intractable.

	\subsubsection{Discrete Antenna Activation}
	
	Practical PASS implementations often restrict PAs to a discrete set of candidate locations due to hardware limitations such as finite-resolution actuators and electromagnetic coupling, reducing the design to selecting an active subset from a grid. This transforms the problem into a combinatorial one whose complexity grows exponentially with the grid size. 
	
	In the context of NOMA-assisted PASS, Wang et al.~\cite{Wang2025WCL_NOMA} addressed the sum rate maximization problem under discrete activation constraints, formulating it as a one-sided one-to-one matching problem and developing a low-complexity algorithm to obtain near-optimal solutions.
	Complementary to algorithmic development, Tyrovolas et al.~\cite{Tyrovolas2025Ergodic} quantified the rate loss compared to continuous placement. The analysis revealed that denser grids asymptotically recover the continuous performance, establishing grid density as the key design parameter. Moreover, Xie et al.~\cite{xie2025lowcomplexityplacement} proposed a low-complexity design with closed-form PA positions for multi-user downlink sum rate maximization, avoiding iterative searching and supporting real-time operation.
	

	\subsubsection{Position Optimization Under Practical Constraints}
	
	The transition from theoretical models to practical deployment necessitates the incorporation of realistic hardware and environmental constraints into the PA position optimization framework. Hardware constraints include minimum PA spacing and in-waveguide attenuation. Location uncertainty introduces robustness requirements. These factors render PA placement a non‑convex problem that demands dedicated algorithms.
	
	At the hardware level, multi-PA deployments must enforce a minimum inter-PA spacing to mitigate coupling. Zhang et al.~\cite{Zhang2026arXiv_multiwave} proposed a projected gradient ascent framework with a projection operator that enforces the spacing constraint during updates without compromising convergence. Lossless waveguide assumptions become unrealistic for long-distance deployments under attenuation. Tyrovolas et al.~\cite{10976621} developed a framework that jointly accounts for free-space path loss and waveguide attenuation, characterizing optimal PA placement in the lossless case. Chen et al.~\cite{11540399} further incorporated waveguide power loss into the PA model, uncovering a trade-off between in-waveguide power loss and free-space path loss. Moreover, perfect user location information is often unavailable. Zeng et al.~\cite{Zeng2026arXiv} addressed location uncertainty by transforming the single-PA worst-case design into a convex program via S-Procedure, before extending to multi-PA via block coordinate descent (BCD) to ensure QoS over the uncertainty region.
	
	
	The optimization of PA positions in PASS has been investigated through three complementary paradigms. Continuous position optimization provides closed-form single-user solutions, but its extension to multi-user scenarios relies on heuristic approaches. Discrete antenna activation reformulates the problem as combinatorial selection via matching or low-complexity designs, with the performance gap to continuous case narrowing as grid density grows. Practical constraints including minimum spacing, in-waveguide attenuation, and location uncertainty are tackled via projection operator, attenuation trade-off, and  robust optimization, respectively. Collectively, these works reveal a clear evolution from idealized models toward deployable solutions. Future research must address the joint accommodation of multiple coupled constraints while balancing modeling fidelity, computational efficiency, and provable performance guarantees.
	
	\subsection{Transmit Beamforming Optimization}
	\label{subsec:beamforming}
	
	While PA placement determines the large-scale channel gains, transmit beamforming governs fine-grained spatial signal distribution and inter-user interference management through digital precoding weights. However, PA position optimization and beamforming design are inherently coupled rather than independent. A joint framework that simultaneously optimizes both PA positions and beamforming weights is therefore essential to unlock the full potential of PASS. The following exposition traces the development of beamforming optimization within such joint designs, organized into two parts, which are foundational multi-user precoding and beamforming for advanced PASS architectures.
	
	\subsubsection{Foundational Multi-User Precoding}
	
	
	Joint optimization of PA positions and transmit precoding faces a core challenge. The tight coupling between PA placement and beamforming weights renders the sum rate function highly non-convex, precluding straightforward decomposition and necessitating sophisticated algorithmic frameworks.
	
	Early work addressed this challenge through penalty-based AO frameworks that decouple the joint problem into tractable subproblems. Wang et al.~\cite{wang2025passmodeling} established the paradigm for multi-user multi-PA downlink systems, employing a penalty-based AO that iterates between beamforming optimization and a low-complexity one-dimensional search for PA positions. Zhao et al.~\cite{M14} generalized the design to multi-waveguide multi-user setups, jointly optimizing the transmission structure alongside the precoding vectors for more flexible resource allocation. Bereyhi et al.~\cite{11414143} further investigated MIMO-PASS architectures with multiple RF chains connected to multiple waveguides, proposing low-complexity hybrid precoding algorithms based on fractional programming and Gauss-Seidel iterations for the downlink. To mitigate computational overhead, Sun et al.~\cite{Sun2025TWC} proposed a low-complexity element-wise sequential optimization framework for sum rate maximization. The key innovation is a closed-form expression for the sum rate as a single-variable function of the pinching beamformer. The downlink adopts maximum ratio transmission (MRT), zero-forcing (ZF), and minimum mean square error (MMSE) as baseband beamformers, offering flexible performance-complexity tradeoffs.
	
	
	\subsubsection{Beamforming for Advanced PASS Architectures}
	
	Emerging PASS architectures require tailored beamforming strategies to exploit their structural advantages. Gan and Liu~\cite{gan2026centerfedmodeling} proposed center-fed PASS (C-PASS), where signals propagate bidirectionally from the waveguide center, yielding bidirectional effective channels. They developed an AO framework that updates precoding and power splitting in closed form, while optimizing PA positions and radiation coefficients via BCD. Jiang et al.~\cite{Jiang2026arXiv_SWAN} introduced segmented waveguide arrays with a tri-hybrid beamforming framework that integrates digital baseband processing for inter-segment interference cancellation, analog phase shifting for beamsteering, and pinching positioning for segment-level configuration. Wang et al.~\cite{Wang2026Tcom_HPASS} combined reconfigurable leaky-wave antennas with conventional PAs in hybrid PASS (H-PASS), developing a penalty-based two-loop algorithm for discrete position PAs and an alternating-minimization method for continuous position PAs. Beyond structural innovations, recent work has revisited the physical-layer fundamentals of PASS. Shao et al.~\cite{Shao2026arXiv} established the polarization-aware full-wave model for multi-mode PASS, revealing that polarization matching efficiency introduces additional design DoFs overlooked in prior channel models.
	
		
	In summary, transmit beamforming optimization in PASS follows two complementary trajectories. Foundational frameworks employ penalty-based AO to decouple beamforming from PA positioning, evolving toward multi-waveguide, MIMO, and element-wise designs that trade performance for efficiency. Advanced architectures introduce new DoFs and coupled constraints. The proliferation of architecture‑specific algorithms, though performance‑effective, fragments the methodological landscape, thereby impeding the emergence of a unified framework across diverse PASS deployments.
	\subsection{EE and QoS-Aware Optimization}
	\label{subsec:Energy}

	Unlike conventional fixed-antenna systems, PASS introduces an additional energy expenditure from the mechanical power consumed by PA repositioning. Motion-induced cost scales with movement frequency and distance, fundamentally altering the resource allocation landscape. The present subsection examines the joint management of these coupled factors for energy efficient and QoS-guaranteed communication.
	
	For EE maximization, the objective is defined as:
	\begin{equation}
		\eta_{\mathrm{EE}} = \frac{\sum_{k=1}^{K}R_k}{\sum_{k=1}^{K}P^{\mathrm{tx}}_k + P_{\mathrm{c}} + P_{\mathrm{m}}},
		\label{eq:EE}
	\end{equation}
	where $R_k$ denotes the achievable rate of the $k$-th user, $P^{\mathrm{tx}}_k$ is the transmit power allocated to the $k$-th user, $P_{\mathrm{c}}$ represents the fixed circuit power consumption, and $P_{\mathrm{m}}$ accounts for the mechanical power consumed by PA repositioning.
	
	Early works incorporated motion cost directly into optimization. Xu et al.~\cite{Xu2025arXiv_motion} jointly optimized radiation power, PA position, and beamforming. For continuous motion, they adopted an alternating direction method of multipliers (ADMM), which handles the coupled variables efficiently. For discrete motion scenarios, they used BCD to alternate between power allocation and position updates. Zhang et al.~\cite{Zeng2026TVT_EE} addressed EE maximization in a TDMA-based multi-user system under minimum rate constraints. Their approach first determines the optimal PA positions geometrically based on user distribution, then decomposes the remaining joint power and time allocation into subproblems, significantly reducing computational complexity. Beyond generic joint optimization, architectural advances can yield substantial efficiency gains. Zhang et al.~\cite{Zhang2026Tcom_DSD} introduced a dual-scale deployment framework that distinguishes coarse-range movement for global coverage from fine-scale local adjustments for precise service, with a penalty-based alternating algorithm that effectively reduces the energy overhead of PA repositioning. For wireless powered communication networks, Li \textit{et al.}~\cite{Li2026WCL_SWIPT} jointly optimized PA positions and resource allocation. Joint PA positioning and power allocation have also been used for downlink power minimization~\cite{Wang2025WCL_power} and uplink sum-rate and EE maximization~\cite{11215676,11131179}. Gan \textit{et al.}~\cite{Gan2025arXiv_WPT} studied wireless power transfer (WPT), jointly optimizing beamforming, PA positions, and power radiation ratios to improve power conversion efficiency while meeting communication rate requirements.
	
Building upon the SE-oriented frameworks, further studies have explored the interplay between SE and EE, along with application-specific QoS guarantees. Zhou et al.~\cite{Zhou2026ICC_SEEE} optimized the weighted product of SE and EE, proving that under maximum-ratio transmission, PA placement becomes power-independent, where PA positions are first determined without power allocation, followed by a closed-form power expression that ensures optimal trade-off. Xu et al.~\cite{11419117} demonstrated that PA-assisted downlink achieves significant power reduction for ultra-reliable and low-latency communication compared to fixed-antenna baselines, highlighting PASS's potential for mission-critical applications. Complementing this power-oriented design, Lin \textit{et al.}~\cite{Lin2025arXiv_URLLC} optimized finite-blocklength downlink rates through PA placement and phase alignment. For computation-oriented services, Liu \textit{et al.}~\cite{Liu2025arXiv_MEC} maximized computation capacity in wireless powered mobile edge computing (MEC) through joint PA positioning and resource allocation, while Gu \textit{et al.}~\cite{Gu2026arXiv_AirComp} minimized aggregation MSE in segmented-waveguide over-the-air computation (AirComp) through PA placement and segment-wise phase control.
	
	Parallel to EE and QoS designs, specialized frameworks exploit structural properties of specific PASS configurations. Zhang et al.~\cite{Zhang2025arXiv_DiPASS} introduced equal-quota power allocation for directional PASS. The closed-form decoupling strategy separates power control from geometric optimization. Wang et al.~\cite{M13} developed a comprehensive framework for multi-waveguide systems, integrating waveguide assignment, antenna activation, and power allocation. The assignment and activation subproblems are formulated as coalition-formation games for distributed decision-making, while power allocation is solved via monotonic optimization for global optimality. 
	
	The EE and QoS-aware PASS designs have progressively moved from idealized EE maximization toward comprehensive frameworks that accommodate motion energy, SE-EE trade-offs, and QoS constraints. Despite these advances, a unified framework that provides both EE optimality and QoS guarantees remains elusive. The coupling between hardware and algorithmic design, while yielding performance gains, fragments the optimization landscape and hinders generalizable methodologies.

	\subsection{PASS Optimization Techniques Outlook}
	\label{subsec:outlook}
	
	The optimization frameworks surveyed in Sections~\ref{subsec:placement}--\ref{subsec:Energy} reveal that PASS introduces a rich design space through position-dependent channel control. However, the flexibility fundamentally couples PA position with beamforming, resource allocation, and multiple access. Although substantial algorithmic advances have been made, several practical and theoretical challenges remain to be addressed:
	
	\begin{itemize}
		\item {Holistic optimization under multiple practical constraints:} existing works typically address one or two constraints at a time, such as minimum spacing, waveguide attenuation, or LoS blockage. A unified framework that jointly handles multiple hardware and environmental impairments while maintaining tractable complexity is still lacking. Future research should develop modular optimization architectures that can incorporate heterogeneous constraints without full re‑design, enabling a systematic trade‑off among modeling fidelity, algorithm efficiency, and performance guarantees.
		
		\item {Dynamic adaptation in mobile environments:} most current designs assume quasi‑static users and channels. In practical indoor deployments, user mobility, time‑varying blockages, and fluctuating traffic demands require continuous reconfiguration of PA positions, precoding, and resource assignments. Developing low‑latency algorithms that jointly schedule PA movements and communication parameters under delayed and imperfect CSI is an important direction.
		
		
	\end{itemize}


\section{Synergistic Integration of AI and PASS}

The flexibility of PASS introduces significant optimization challenges. Joint optimization of transmit beamforming, PA positions, power allocation, and user scheduling yields high-dimensional non-convex problems \cite{11048566, wang2025passmodeling, 11358830}. Conventional convex optimization and alternating optimization face three fundamental obstacles. First, the objective function is highly multimodal due to tight coupling between transmit beamforming and PA positioning \cite{gan2026centerfedmodeling}. Second, computational complexity scales poorly with the number of PAs, waveguides, and users, making large-scale deployment intractable \cite{xie2025lowcomplexityplacement, 10976621, 11315149}. Third, dynamic environments with user mobility and time-varying blockages require millisecond-level responses that iterative algorithms cannot deliver \cite{ai23}.

AI offers a paradigm shift for PASS design. AI models learn complex mappings from environmental states to near-optimal configurations, enabling offline training and online inference that substantially reduce computational latency compared with iterative solvers \cite{liu2026passtutorial}. Moreover, architectures with built-in structural priors can exploit the permutation equivariance and graph topology inherent in PASS, which facilitates generalization across varying numbers of users, waveguides, and PAs \cite{11219348}. A wide spectrum of AI techniques has been explored for PASS, including DL, RL, FL, meta-learning, and LLMs. As PASS evolves from theory to practice, AI will serve as a cornerstone for unlocking its full potential in next-generation wireless networks \cite{11434944, liu2026surveypinchingantennasystemspass}.

\subsection{DL for PASS}

DL is a class of machine learning techniques that learn hierarchical representations from data through neural networks with multiple layers. In the context of PASS, DL offers a paradigm shift from conventional convex optimization. By learning complex mappings from environmental parameters to near-optimal system configurations, deep neural networks (DNNs) enable offline training and online inference, avoiding repeated iterative solving at runtime. Furthermore, when the network architecture is deliberately aligned with the problem structure, as in GNNs whose message-passing rules are symmetric over nodes, the resulting models exhibit permutation-equivariant properties that facilitate deployment across systems with different numbers of PAs, waveguides, and users.

The core idea of employing DL in PASS is that neural networks are capable of directly learning the intricate coupling between PA placement, beamforming, and channel conditions. DL models can be trained in a supervised manner by using labels from offline optimization, or in an unsupervised manner that directly optimizes task objectives. The training process captures the underlying structure of PASS, enabling the network to produce high-quality solutions at inference time without iterative solving. Thus, DL has been increasingly explored for addressing the high-dimensional non-convex optimization problems inherent in PASS design.

\begin{figure}[t!]
\centering
\includegraphics[width=3 in]{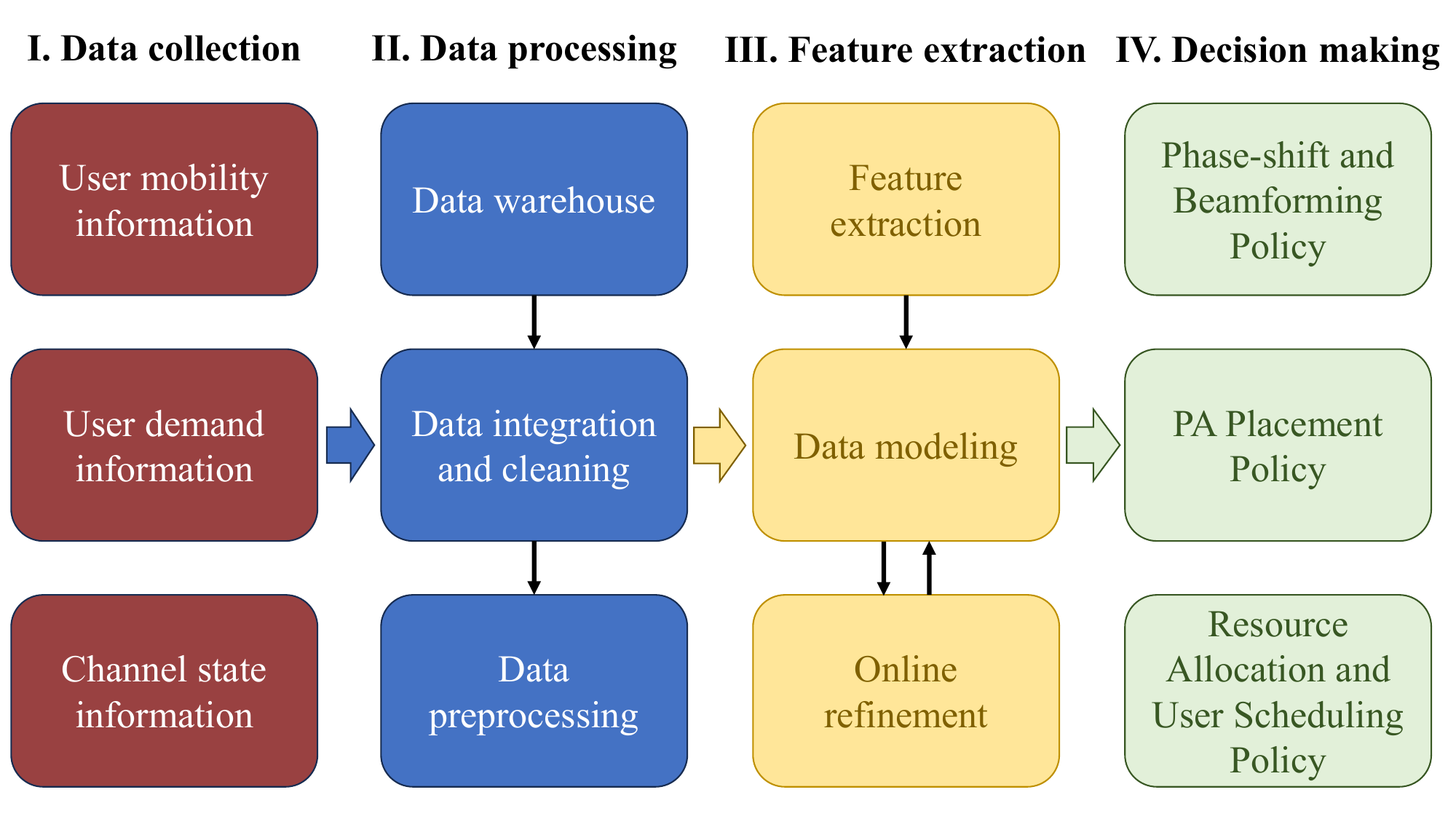}
\caption{Architecture of AI-enabled PASS networks.}
\label{fig:ai_pass_architecture}
\end{figure}

Fig.~\ref{fig:ai_pass_architecture} illustrates the architecture of AI-enabled PASS networks, in which environmental information collected from the physical layer is first processed through data processing and feature extraction modules before being fed into the AI engine. The AI engine then produces optimized control policies, including beamforming, PA placement, and resource allocation.

DL approaches can be organized along two dimensions. By network architecture, they include graph neural networks (GNNs), transformers, convolutional neural networks (CNNs), and generic DNNs. By training paradigm, the same architectures may be trained under supervised, unsupervised, or generative objectives depending on whether labeled optimal solutions are available. GNNs exploit the natural graph topology of PASS, where waveguides, PAs, and users form interconnected nodes \cite{11159498, 11063467, he2025risgnn, I7}. Transformers instead model global dependencies through attention mechanisms \cite{11303188, 11148833, xu2025jointtransmitpinching, 11018390}. 
CNNs extract spatial features from grid-structured channel information \cite{ai21}, DNNs approximate complex non-linear mappings for joint optimization \cite{ai08, I2, P6}, while unsupervised and generative paradigms enable label-free learning and channel estimation \cite{ai05, ai25, ai26, ai04, ai24}.
DL architectures have been applied to beamforming optimization, channel estimation, power allocation, ISAC, and PLS. The following subsections examine each category in detail.

\subsubsection{GNNs for PASS}

GNNs are a class of DL models designed to operate on graph-structured data, where nodes exchange information through message passing along edges. In PASS, the system naturally forms a graph in which waveguides, PAs, and users constitute nodes while their physical and electromagnetic relationships define edges. When the message-passing rules are shared and symmetric across nodes, the resulting aggregation is invariant to node permutation, so that a trained model can be evaluated on graphs with different numbers of nodes without redesigning the network. The general GNN update equation over a graph with $K$ vertices is expressed as:

\begin{equation}
\mathbf{e}_k^{(\ell+1)} = \sigma\left(\mathbf{M}\mathbf{e}_k^{(\ell)} + \sum_{i \in \mathcal{V}(k)} \mathbf{N}\mathbf{e}_i^{(\ell)}\right),
\label{eq:gnn_update}
\end{equation}
where $\mathbf{e}_k^{(\ell)}$ is the embedding of vertex $k$ at layer $\ell$, $\mathcal{V}(k)$ denotes its neighbors, $\mathbf{M}$ and $\mathbf{N}$ are learnable weight matrices, and $\sigma(\cdot)$ is a nonlinear activation function.

The core idea of employing GNNs in PASS is that they can learn the coupling between PA placement and beamforming by aggregating information from neighboring nodes, capturing how the configuration of one PA affects the channel quality of nearby users. GNNs can be trained in an unsupervised manner with the negative sum rate as the loss function, eliminating the need for labeled optimal solutions. Thus, GNNs are particularly effective for learning beamforming policies that adapt to the topology of PASS.

GNN architectures can be single-stage or multi-stage, with multi-stage designs reflecting the physical hierarchy of PASS optimization in which PA positions determine the channel, which then determines the optimal beamforming \cite{11159498}. 
Additionally, GNN-based PASS solutions can be categorized by how they handle the graph structure. 
Homogeneous GNNs treat all nodes identically, while heterogeneous GNNs distinguish between different node types such as PAs and users. Since PASS inherently exhibits a bipartite graph structure where edges exist exclusively between PA and user nodes, heterogeneous GNNs that exploit the bipartite asymmetry are particularly effective for capturing the functional dependencies between the two sides \cite{11063467}.
GNN architectures have been applied to joint PA placement and beamforming optimization, RIS-assisted systems, and ISAC \cite{he2025risgnn, I7}. 
More explicitly, 
Guo~\textit{et al.}~\cite{11159498} proposed the graphic pinching-antenna systems (GPASS) framework, which employs a two-stage GNN architecture to learn beamforming in PASS. A sub-GNN first learns optimal PA positions from graph-structured inputs encoding user locations and waveguide geometries, and a second sub-GNN then determines transmit beamformers based on the learned positions. The two-stage GPASS design mirrors the physical dependence of beamforming on PA placement and achieves performance close to fractional programming with significantly lower inference complexity. GPASS, however, treats all nodes homogeneously and does not explicitly distinguish the roles of PA and user vertices. To exploit the PA-user role asymmetry, Xie~\textit{et al.}~\cite{11063467} formulated downlink PASS as a bipartite graph and proposed the bipartite graph attention network (BGAT), whose attention mechanism identifies critical PA-user links and whose tailored readout guarantees feasible PA coordinates and power levels for EE maximization. BGAT was developed for single-waveguide scenarios, and the optimization space becomes substantially richer when passive reflecting elements are introduced. To extend BGAT beyond single-waveguide scenarios, He~\textit{et al.}~\cite{he2025risgnn} extended GNNs to RIS-assisted PASS with a three-stage heterogeneous graph architecture that successively determines PA positions from user locations, RIS phase shifts from the composite channel, and transmit beamformers, with customized activation functions enforcing the feasibility of each stage. The graph formulation becomes even richer in ISAC scenarios, where sensing targets constitute a third class of nodes. Guo~\textit{et al.}~\cite{I7} proposed SWISACGNN for PASS-enabled ISAC in low-altitude wireless networks, whose alternating update procedure is inspired by an alternative optimization algorithm and which jointly learns PA positions and beamforming to balance communication and sensing objectives. The above GNN studies demonstrate that GNNs offer a versatile and scalable framework for exploiting the topological structure of PASS across beamforming, antenna placement, RIS-assisted, and ISAC configurations. Their main limitation is that message passing aggregates information mainly from local neighborhoods, so capturing long-range interactions requires stacking multiple layers.

\subsubsection{Transformers for PASS}

Since message-passing GNNs reach distant nodes only through multiple aggregation hops, they may fail to capture long-range PA-user interactions within a shallow architecture, and stacking layers can lead to over-smoothing of node features. Transformers overcome the long-range modeling limitation through self-attention, which models dependencies among all elements in a sequence or set directly, enabling each node to attend to every other node in a single layer. The global attention capability is particularly valuable in large-scale PASS deployments. The transformer attention mechanism is defined as:

\begin{equation}
\text{Attention}(\mathbf{Q}, \mathbf{K}, \mathbf{V}) = \text{softmax}\left(\frac{\mathbf{Q}\mathbf{K}^T}{\sqrt{d_{\text{key}}}}\right)\mathbf{V},
\label{eq:attention}
\end{equation}
where $\mathbf{Q}$, $\mathbf{K}$, and $\mathbf{V}$ are the query, key, and value matrices, and $d_{\text{key}}$ is the dimension of the key vectors.

The core idea of employing transformers in PASS is that self-attention can model the global dependencies among all PAs and users simultaneously, capturing long-range interactions that message-passing GNNs may miss. Furthermore, the attention weights reveal which PA-user relationships most influence a given optimization objective, offering a degree of interpretability. Thus, transformers offer a powerful alternative for PASS optimization tasks that require global coordination.

\begin{figure}[t!]
\centering
\includegraphics[width=2.5 in]{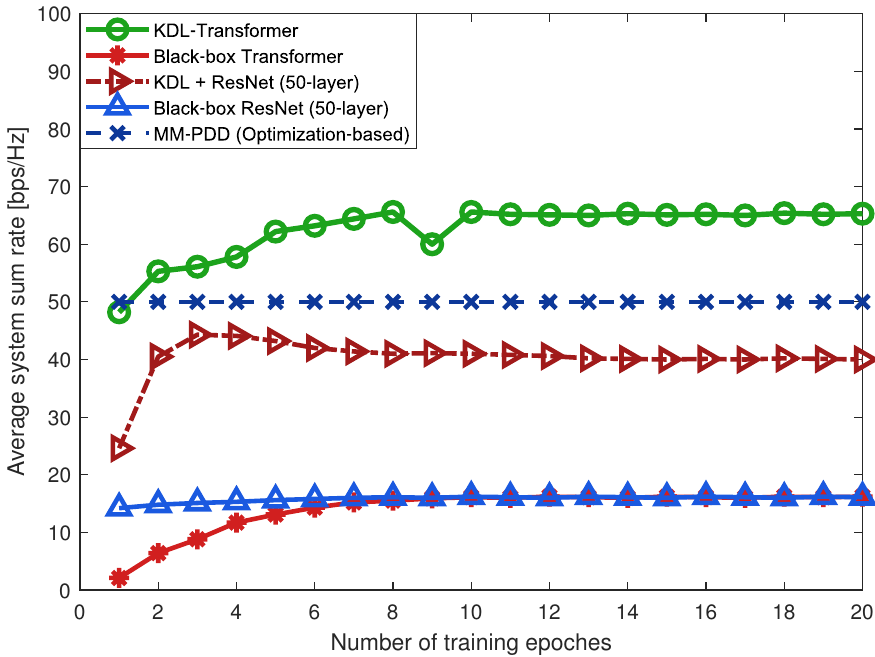}
\caption{Performance comparison of transformer with other methods in PASS \cite{11148833}.}
\label{fig:transformer_performance}
\end{figure}

Transformer-based PASS solutions differ in how much domain structure they embed. Guo~\textit{et al.}~\cite{11303188} proposed the PASSformer framework for tri-beamforming, which jointly optimizes digital, analog, and pinching beamforming to fully mitigate long-distance path loss. Rather than applying an unconstrained transformer, PASSformer restricts attention to connected nodes of a graph-structured input, combining permutation equivariance with the expressive power of global attention and achieving optimization-based performance at millisecond inference speeds. While PASSformer learns beamformers as a largely black-box mapping and provides no guarantee that its outputs satisfy optimality conditions, Xu~\textit{et al.}~\cite{11148833} addressed the lack of interpretability with the Karush-Kuhn-Tucker (KKT)-guided dual learning transformer (KDL-Transformer), which reconstructs Karush-Kuhn-Tucker solutions in an iteration-free manner by learning only the dual variables, while attention captures inter-PA, inter-user, and CSI-beamforming dependencies. Xu~\textit{et al.}~\cite{xu2025jointtransmitpinching} then systematically compared optimization-based and learning-based approaches for joint transmit and pinching beamforming, confirming that KKT-guided learning strikes the best balance between performance and computational speed. Whereas the above transformer designs target beamforming, channel estimation presents a different challenge because the information content of observations varies strongly across PAs. Xiao~\textit{et al.}~\cite{11018390} proposed a transformer-based channel estimator with a mixture-of-experts (MoE) architecture, whose self-attention learns the spatial relationships among PAs and routes observations to specialized experts. 
Fig.~\ref{fig:transformer_performance} compares the performance of transformer with other methods in PASS, showing that transformer-based methods outperform conventional black-box learning approaches.

\subsubsection{Other DL Models and Learning Paradigms}

Not every PASS problem exhibits the node-graph or global-attention structure that motivates GNNs and transformers. When the relevant data can be arranged on a grid, CNNs use convolutional filters to extract local spatial features from structured data such as channel matrices. Mohammadzadeh~\textit{et al.}~\cite{ai21} showed that a CNN can learn downlink power allocation for NOMA-assisted PASS, achieving performance close to weighted minimum mean square error (WMMSE) with substantially lower latency. For problems with moderate-dimensional inputs and no prominent spatial or graph structure, Karagiannidis~\textit{et al.}~ \cite{ai08} trained a DNN for two-state PASS optimization under user location uncertainty. Qin~\textit{et al.}~\cite{I2} formulated joint antenna position and transmit power optimization for PA-assisted ISAC as a DL problem. Then, Papanikolaou~\textit{et al.}~\cite{P6} developed a DNN-aided optimizer for PLS with artificial noise, jointly optimizing information beams, noise covariance, and PA positions under both perfect and imperfect CSI.

The architectures above typically rely on supervised labels generated by offline optimization, which are computationally expensive to obtain for PASS. Unsupervised learning removes the need for labeled data by optimizing task objectives directly. Kang~\textit{et al.}~\cite{ai05} introduced CaMPASS-Net, a dual-stream network with residual connections trained by a purpose-built unsupervised loss for capacity maximization in MIMO PASS. Perovi\'{c}~\textit{et al.}~ \cite{ai25} extended unsupervised DL to ISAC systems equipped with both PA and movable antennas, showing that combined antenna types yield larger sum rates. Channel estimation, in contrast, is an underdetermined inverse problem for which even unsupervised discriminative models are insufficient, since far fewer observations than channel coefficients are available. Generative models address the underdetermined nature of channel estimation by learning a prior distribution over channels and recovering the channel from limited observations through posterior inference. Chen and Wang \cite{ai26} formulated indoor near-field channel estimation as a reverse denoising process driven by a diffusion model, whose multipath-aware inference jointly captures deterministic waveguide-mediated propagation and stochastic scattering. For time-varying channels, Lv~\textit{et al.}~\cite{ai04} combined cross-attention with GRU decoders to track fading dynamics across estimation slots. Finally, beyond single-cell settings, Amhaz~\textit{et al.}~\cite{ai24} considered coordinated multipoint transmission with two PASS-equipped base stations serving users under SDMA, where a gradient-based meta-learning strategy tailored to large-scale optimization achieves a higher achievable-rate upper bound than conventional CoMP systems.

\subsection{RL for PASS}

The DL approaches discussed above learn a fixed mapping from environmental snapshots to configurations, and are therefore best suited to static or quasi-static deployments. In practice, however, users move and blockages vary over time, so that PA positions and beamforming must be continuously adjusted in a sequential decision-making process. RL is a machine learning paradigm in which an agent learns such policies by interacting with an environment and receiving rewards or penalties based on its actions. Unlike supervised learning, where labeled optimal solutions are required, RL enables the agent to discover near-optimal policies through trial-and-error exploration, which is particularly valuable under user mobility and time-varying channel conditions.

The core idea of employing RL in PASS is that the agent, typically implemented at the BS, learns a policy mapping observed environmental states to control actions including PA positions and beamforming coefficients. By maximizing cumulative rewards over long time horizons, RL incorporates farsighted system evolution rather than only instantaneous performance, and it can handle the distinct time scales of slow PA movement and fast beamforming adjustment.

RL frameworks for PASS differ in how they treat the separation between slow PA movement and fast beamforming adjustment. Zhao~\textit{et al.}~\cite{zhao2026dynamicplacementmobile} proposed a bilevel framework for mobile users in urban micro scenarios, where the learned policy proactively tracks users along their trajectories to maintain LoS links, creating a ``shadowing'' effect under mechanical energy constraints. Because the bilevel design still relies on an analytical inner-loop precoder, Amhaz~\textit{et al.}~\cite{ai22} instead developed an end-to-end deep deterministic policy gradient (DDPG) approach that maps states directly to both antenna placement and beamforming actions without requiring an explicit system model. The action space further expands when PASS is integrated with other system functions. Ma~\textit{et al.}~\cite{ai03} applied an SAC-based agent to PA-assisted NOMA MEC, where task offloading decisions join PA positions and power allocation in minimizing energy consumption. Pakravan~\textit{et al.}~\cite{ai19} considered wirelessly powered PASS, where the energy-harvesting time-switching ratio becomes an additional action optimized jointly with antenna positioning and power control. RL thus enables real-time adaptability across diverse PASS deployments, from basic beamforming to multi-resource co-optimization.

\subsection{FL for PASS}

The DL and RL approaches discussed above assume centralized training, in which raw data or experience must be aggregated at a single server. In geographically distributed PASS deployments, the resulting communication overhead and data governance constraints can make centralization infeasible. FL is a distributed machine learning paradigm in which each client trains a local model on its own data and transmits only model updates to a central server, which aggregates them into a global model. FL offers privacy advantages over centralized training because raw data never leave the clients. Nevertheless, model updates can still leak sensitive information, so mechanisms such as secure aggregation or differential privacy may still be required. In the context of PASS, the reconfigurability of PA positions provides a new physical-layer means of improving the communication efficiency of the training process itself.

The core idea of employing FL in PASS is that PA placement reshapes the channel quality between participating devices and the server, which directly governs the latency and energy cost of transmitting model updates. PAs can be activated near devices with poor links to compensate for stragglers and accelerate convergence. In other words, PASS introduces additional physical-layer design variables, namely PA positions and activation patterns, into the FL optimization problem, creating opportunities for joint design that conventional FL does not possess.

The literature has progressed from single-objective placement to heterogeneous-device handling, over-the-air aggregation, and finally theoretical characterization. Lin and Ding \cite{ai01} first observed that the duration of a synchronous FL round is dictated by its slowest clients, and proposed jointly optimizing PA placement and client participation to minimize the expected time-to-accuracy. Tail-latency optimization alone, however, does not explicitly account for heterogeneous device capabilities. Wu~\textit{et al.}~\cite{ai07} therefore considered a hybrid conventional-and-pinching network and employed a fuzzy-logic client classification together with a deep eeinforcement learning (DRL)-based algorithm to jointly optimize PA placement and resource allocation, explicitly mitigating stragglers. While the above FL designs assume orthogonal update transmissions, Asaad and Bereyhi \cite{ai14} exploited the signal-superposition property of over-the-air aggregation, jointly tuning PASS parameters and scheduling devices to minimize the energy consumption of over-the-air FL. Azimi-Abarghouyi and Brinton~\cite{azimiabarghouyi2026airpassovertheairfederatedlearning} developed AirPASS for over-the-air FL, jointly optimizing device selection, receive beamforming, and PA positions through homotopy-based methods. Optimizing latency alone can degrade learning quality, and vice versa. Asaad~\textit{et al.}~\cite{ai12} addressed the latency-quality trade-off in FedPASS, a two-tier iterative algorithm that jointly minimizes the end-to-end round latency and an upper bound on the FL optimality gap. Complementing the above algorithmic designs, Lin~\textit{et al.}~\cite{ai20} provided convergence analysis proving that PASS shortens the worst-link distance in synchronous FL and increases the on-time completion probability in asynchronous FL.

\subsection{Meta-Learning for PASS}

FL distributes training across devices, but a model trained for one deployment still cannot serve a new deployment with a different user distribution or blockage pattern without retraining from scratch, which is computationally expensive and data-hungry. Meta-learning, also known as learning-to-learn, addresses the need for deployment-specific retraining by training models to adapt quickly to new tasks with minimal data. Instead of learning a single task-specific mapping, meta-learning seeks an initialization that is close to optimal across a distribution of tasks, so that a new task can be handled with only a few gradient steps. In the context of PASS, rapid few-shot adaptation is particularly valuable when real-world deployments encounter user distributions, blockage patterns, and environmental conditions that differ from those seen during training.

The core idea is a bilevel structure in which an inner loop adapts the model to a specific task with a few gradient steps, while an outer loop updates the meta-parameters across tasks to minimize the sum of task-specific adapted losses. Properly trained, such an initialization helps mitigate domain shifts across deployments and reduces the performance drop from simulation to real environments, although it does not eliminate the sim-to-real gap by itself.

Zhou~\textit{et al.}~\cite{zhou2025gradientmetalearning} proposed the gradient meta-learning joint optimization (GML-JO) framework for multi-waveguide PASS, which maximizes the weighted sum rate by decomposing the non-convex problem into beamforming and antenna-position sub-problems and meta-learning an initialization from which each new user distribution can be adapted with minimal data, outperforming conventional joint optimization that is restarted for every scenario. GML-JO targets the communication objective under positioning uncertainty. When security constraints are introduced, the agent must additionally react to eavesdropper mobility. Musri~\textit{et al.}~ \cite{musri2026adaptivepinchingantennaoptimization} addressed eavesdropper mobility with a MAML/Reptile-based meta-learning controller that few-shot adapts antenna positions and transmit power under user-location uncertainty and physical-layer-security constraints, enabling real-time response to dynamic security threats. A related gradient-based meta-learning strategy was applied by Amhaz~\textit{et al.}~\cite{ai24} to the large-scale coordinated multipoint setting, where it operates without pre-training while attaining a higher achievable-rate bound than conventional CoMP schemes.

\subsection{LLMs for PASS}

Meta-learning still requires several gradient steps and a small set of task-specific data at each new deployment. LLMs offer a complementary route to generalization. Pre-trained on broad corpora and adapted through prompting or lightweight fine-tuning, they can perform zero-shot or few-shot reasoning over multimodal environmental context without task-specific training from scratch. In the context of PASS, LLMs leverage their pre-trained reasoning capabilities for context-aware beam training and system configuration, integrating information such as user location history, building layouts, and traffic patterns that is difficult to encode in a conventional neural network input.

LLM-enabled PASS solutions can be categorized by their application scope. Conventional beam training exhaustively probes candidate beams, whose overhead grows with the codebook size. Gan~\textit{et al.}~\cite{ai15} proposed an LLM-enabled beam training framework for downlink multi-user MIMO PASS, in which multimodal environmental information is encoded into high-dimensional tokens and the LLM generates a prioritized, context-aware probing sequence for both single-user and multi-user codebooks, substantially reducing training overhead while maintaining beam alignment accuracy. Beam training operates on conventional PASS, whereas SWANs additionally require segment selection and transmit/receive partitioning. Gao~\textit{et al.}~\cite{gao2026integratedsensingsegmented} first established an ISAC framework for SWAN with a hybrid segment selection and multiplexing protocol and developed a segment-hysteresis RL algorithm to jointly optimize segment selection and PA positioning. Such RL policies, however, must be retrained when the number of users changes. To enable cross-configuration transfer, Gao~\textit{et al.}~ \cite{ai13} constructed a CSI-induced self-graph capturing interactions between communication users and sensing targets and attached it to a LoRA-adapted LLM backbone. Then, Gao~\textit{et al.}~\cite{gao2026llmenabledpartitioning} extended the CSI-induced self-graph LLM design with two task-specific heads for antenna deployment and beamforming prediction, whose permutation-invariant graph representation allows the learned policy to transfer across varying user configurations with only lightweight adaptation rather than full retraining.

\subsection{Open Challenges and Future Directions}

Despite significant progress, several critical challenges remain.

\begin{itemize}
  \item {Data and Model Challenges} the integration of AI into PASS is hindered by data scarcity and the sim-to-real gap, as real-world measurement datasets for PASS remain scarce. Physics-based generative models and domain randomization offer promising directions for synthesizing training data and improving robustness. In parallel, incorporating physical constraints such as electromagnetic compatibility, hardware limitations, and regulatory requirements into learning frameworks remains an open challenge.

  \item {Generalization and Scalability} current AI models may suffer performance degradation under unseen scenarios due to domain shifts. Domain adaptation and meta-learning are promising directions to improve zero-shot and few-shot generalization. Moreover, PASS must operate reliably under user mobility, time-varying channels, and partial CSI, necessitating adaptive and uncertainty-aware models. Lightweight architectures and distributed inference are also essential for real-time large-scale deployments.

  \item {Theoretical and Methodological Challenges} the sample complexity, approximation error, and optimality gap of learning-based PASS remain uncharacterized, calling for a unified framework bridging information theory and statistical learning. Hybrid architectures that combine GNNs for graph representation, transformers for global attention, meta-learning for rapid adaptation, and DRL for sequential decisions represent a promising but largely uncharted direction.
\end{itemize}

\section{Conclusions}
\label{sec:conclusion}

This paper presented a comprehensive survey of PASS towards integrated positioning, communication, and navigation. First, the waveguide evolution and the electromagnetic foundations of PASS were revisited. Representative waveguide structures were compared in terms of the modal behavior, field confinement and attenuation, based on which PASS was distinguished from LCX under a unified channel model. Then, PASS-based positioning was surveyed from both the BS-side and the user-side perspectives, covering single-waveguide and multi-waveguide architectures under both LoS and multipath conditions. The RSSI-based, phase-aware and delay-based algorithms were compared in terms of their observables, advantages and limitations. Building upon these foundations, the performance of PASS was analyzed for both downlink and uplink transmissions, followed by a comparison of the probabilistic and geometry-aware blockage models as well as the mixed LoS/NLoS channel models. Furthermore, the resource allocation designs conceived for PASS were systematically reviewed, including PA position optimization, transmit beamforming, and EE and QoS-aware optimization. Finally, AI-based solutions were elaborated to tackle the high-dimensional non-convex problems encountered in dynamic and complex scenarios. 

Despite these advances, the research on PASS is still at an early stage, and ample opportunities remain for important contributions in this field. Electromagnetic-consistent models are needed to bridge full-wave analysis with tractable communication- and positioning-oriented designs. The mutual dependence between the PA positions and the CSI calls for treating positioning, communication and navigation as a single closed loop. The conflicting objectives governed by the same set of highly multimodal variables motivate near-real-time AI-aided optimization. Prototyping and field trials also remain essential for validating both the channel models and the algorithms surveyed in this paper. Overall, PASS extends antenna flexibility from the wavelength scale to the meter scale and, more importantly, unifies positioning, communication and navigation over the same guided-wave infrastructure. We hope that this survey provides a useful reference for the design and optimization of PASS-empowered 6G networks.

\bibliographystyle{IEEEtran}
\bibliography{IEEEabrv,PASSsurvey_20260422}

\begin{thebibliography}{100}
\providecommand{\url}[1]{#1}
\csname url@samestyle\endcsname
\providecommand{\newblock}{\relax}
\providecommand{\bibinfo}[2]{#2}
\providecommand{\BIBentrySTDinterwordspacing}{\spaceskip=0pt\relax}
\providecommand{\BIBentryALTinterwordstretchfactor}{4}
\providecommand{\BIBentryALTinterwordspacing}{\spaceskip=\fontdimen2\font plus
\BIBentryALTinterwordstretchfactor\fontdimen3\font minus
  \fontdimen4\font\relax}
\providecommand{\BIBforeignlanguage}[2]{{%
\expandafter\ifx\csname l@#1\endcsname\relax
\typeout{** WARNING: IEEEtran.bst: No hyphenation pattern has been}%
\typeout{** loaded for the language `#1'. Using the pattern for}%
\typeout{** the default language instead.}%
\else
\language=\csname l@#1\endcsname
\fi
#2}}
\providecommand{\BIBdecl}{\relax}
\BIBdecl

\bibitem{11433651}
Y.~Zhao \emph{et~al.}, ``Reconfigurable antennas for next-generation mobile
  communication networks: A comprehensive survey and tutorial,'' \emph{IEEE
  Commun. Surv. Tutorials}, vol.~28, pp. 5267--5306, 2026.

\bibitem{10858129}
C.~You \emph{et~al.}, ``Next generation advanced transceiver technologies for
  6{G} and beyond,'' \emph{IEEE J. Sel. Areas Commun.}, vol.~43, no.~3, pp.
  582--627, Mar. 2025.

\bibitem{Zhu2026MovableReconfigurable}
L.~Zhu \emph{et~al.}, ``Movable and reconfigurable antennas for 6{G}: Unlocking
  electromagnetic-domain design and optimization,'' \emph{npj Wireless
  Technol.}, vol.~2, no.~5, 2026.

\bibitem{zhang20236g}
H.~Zhang, N.~Shlezinger, F.~Guidi, D.~Dardari, and Y.~C. Eldar, ``6g wireless
  communications: From far-field beam steering to near-field beam focusing,''
  \emph{IEEE Commun. Mag.}, vol.~61, no.~4, pp. 72--77, Apr. 2023.

\bibitem{11219348}
X.~Yu, J.~Wang, R.~Q. Hu, D.~I. Kim, N.~Al-Dhahir, H.~Wymeersch, and N.~Zhao,
  ``Intelligent flexible position antenna systems for networking: A survey,''
  \emph{IEEE Trans. Netw. Sci. Eng}, vol.~13, pp. 3105--3126, 2026.

\bibitem{liu2026surveypinchingantennasystemspass}
Y.~Liu \emph{et~al.}, ``A survey of pinching-antenna systems ({PASS}),''
  \emph{arXiv}, vol. 2601.18927, 2026.

\bibitem{rappaport2002book}
T.~S. Rappaport, \emph{Wireless Communications: Principles and Practice},
  2nd~ed.\hskip 1em plus 0.5em minus 0.4em\relax Upper Saddle River, NJ:
  Prentice Hall, 2002.

\bibitem{mouly1992gsm}
M.~Mouly and M.-B. Pautet, \emph{The {GSM} System for Mobile
  Communications}.\hskip 1em plus 0.5em minus 0.4em\relax Washington, DC:
  Telecom Publishing, 1992.

\bibitem{liberti1999smart}
J.~C. Liberti and T.~S. Rappaport, \emph{Smart Antennas for Wireless
  Communications: {IS}-95 and Third Generation {CDMA} Applications}.\hskip 1em
  plus 0.5em minus 0.4em\relax Upper Saddle River, NJ: Prentice Hall, 1999.

\bibitem{6736761}
E.~G. Larsson, O.~Edfors, F.~Tufvesson, and T.~L. Marzetta, ``Massive {MIMO}
  for next generation wireless systems,'' \emph{IEEE Commun. Mag.}, vol.~52,
  no.~2, pp. 186--195, Feb. 2014.

\bibitem{MIMO_poi_1}
F.~Wen, H.~Wymeersch, B.~Peng, W.~P. Tay, H.~C. So, and D.~Yang, ``A survey on
  {5G} massive {MIMO} localization,'' \emph{Digit. Signal Process.}, vol.~94,
  pp. 21--28, Nov. 2019.

\bibitem{7593259}
J.~G. Andrews, T.~Bai, M.~N. Kulkarni, A.~Alkhateeb, A.~K. Gupta, and R.~W.
  Heath, ``Modeling and analyzing millimeter wave cellular systems,''
  \emph{IEEE Trans. Commun.}, vol.~65, no.~1, pp. 403--430, Jan. 2017.

\bibitem{7448962}
A.~Thornburg, T.~Bai, and R.~W. Heath, ``Performance analysis of outdoor
  {mmWave} ad hoc networks,'' \emph{IEEE Trans. Signal Process.}, vol.~64,
  no.~15, pp. 4065--4079, Aug. 2016.

\bibitem{cui2023nearfield}
M.~Cui, Z.~Wu, Y.~Lu, X.~Wei, and L.~Dai, ``Near-field {MIMO} communications
  for {6G}: Fundamentals, challenges, potentials, and future directions,''
  \emph{IEEE Commun. Mag.}, vol.~61, no.~1, pp. 40--46, Jan. 2023.

\bibitem{an2023beamfocusing}
J.~An, C.~Yuen, L.~Dai, M.~Di~Renzo, M.~Debbah, and L.~Hanzo, ``Toward
  beamfocusing-aided near-field communications: Research advances, potential,
  and challenges,'' \emph{arXiv}, vol. 2309.09242, 2023.

\bibitem{9264694}
K.-K. Wong, A.~Shojaeifard, K.-F. Tong, and Y.~Zhang, ``Fluid antenna
  systems,'' \emph{IEEE Trans. Wireless Commun.}, vol.~20, no.~3, pp.
  1950--1962, Mar. 2021.

\bibitem{wong2023faspart1}
K.-K. Wong, W.~K. New, X.~Hao, K.-F. Tong, and C.-B. Chae, ``Fluid antenna
  system---part {I}: Preliminaries,'' \emph{IEEE Commun. Lett.}, vol.~27,
  no.~8, pp. 1919--1923, Aug. 2023.

\bibitem{new2025fastutorial}
W.~K. New \emph{et~al.}, ``A tutorial on fluid antenna system for {6G}
  networks: Encompassing communication theory, optimization methods and machine
  learning,'' \emph{IEEE Commun. Surveys Tuts.}, vol.~27, no.~4, pp.
  2325--2377, 2025.

\bibitem{10286328}
L.~Zhu, W.~Ma, and R.~Zhang, ``Movable antennas for wireless communication:
  Opportunities and challenges,'' \emph{IEEE Commun. Mag.}, vol.~62, no.~6, pp.
  114--120, Jun. 2024.

\bibitem{Shao2026SixDMA}
X.~Shao \emph{et~al.}, ``A tutorial on six-dimensional movable antenna for 6{G}
  networks: Synergizing positionable and rotatable antennas,'' \emph{IEEE
  Commun. Surveys Tuts.}, pp. 1--1, 2026.

\bibitem{Zheng2026RATutorial}
B.~Zheng \emph{et~al.}, ``Rotatable antenna-empowered wireless networks: A
  tutorial,'' \emph{arXiv}, vol. 2603.25559, 2026.

\bibitem{southworth1950principles}
G.~C. Southworth, ``Principles and applications of waveguide transmission,''
  \emph{Bell Syst. Tech. J.}, vol.~29, no.~3, pp. 295--342, Jul. 1950.

\bibitem{southworth1936general}
------, ``Hyper-frequency wave guides---general considerations and experimental
  results,'' \emph{Bell Syst. Tech. J.}, vol.~15, no.~2, pp. 284--309, Apr.
  1936.

\bibitem{barrow1936hollow}
W.~L. Barrow, ``Transmission of electromagnetic waves in hollow tubes of
  metal,'' \emph{Proc. IRE}, vol.~24, no.~10, pp. 1298--1328, Oct. 1936.

\bibitem{wang2001periodicslots}
J.~H. Wang and K.~K. Mei, ``Theory and analysis of leaky coaxial cables with
  periodic slots,'' \emph{IEEE Trans. Antennas Propag.}, vol.~49, no.~12, pp.
  1723--1732, Dec. 2001.

\bibitem{delogne1975leakytheory}
P.~P. Delogne and M.~Safak, ``Electromagnetic theory of the leaky coaxial
  cable,'' \emph{Radio Electron. Eng.}, vol.~45, no.~5, pp. 233--240, May 1975.

\bibitem{zhang2001leakyfeeder}
Y.~P. Zhang, ``Indoor radiated-mode leaky feeder propagation at 2.0 {GHz},''
  \emph{IEEE Trans. Veh. Technol.}, vol.~50, no.~2, pp. 536--545, Mar. 2001.

\bibitem{11533561}
Y.~Xu, Z.~Ding, O.~A. Dobre, and T.-H. Chang, ``Pinching-antenna system design
  with {LoS} blockage: Does in-waveguide attenuation matter?'' \emph{IEEE
  Trans. Wireless Commun.}, vol.~25, pp. 17\,322--17\,337, 2026.

\bibitem{11178241}
K.~Wang, C.~Ouyang, Y.~Liu, and Z.~Ding, ``Pinching-antenna systems with {LoS}
  blockages,'' \emph{IEEE Wireless Commun. Lett.}, vol.~14, no.~12, pp.
  4122--4126, Dec. 2025.

\bibitem{wang2026blockageeffectpinchingantennasystems}
J.~Wang, J.~Wang, T.~Hou, X.~Sun, and A.~Nallanathan, ``On the blockage effect
  in pinching-antenna systems ({PASS}),'' \emph{arXiv}, vol. 2607.13582, 2026.

\bibitem{11169486}
Y.~Liu, Z.~Wang, X.~Mu, C.~Ouyang, X.~Xu, and Z.~Ding, ``Pinching-antenna
  systems: Architecture designs, opportunities, and outlook,'' \emph{IEEE
  Commun. Mag.}, vol.~64, no.~1, pp. 190--196, Jan. 2026.

\bibitem{liu2026passtutorial}
Y.~Liu \emph{et~al.}, ``Pinching-antenna systems ({PASS}): A tutorial,''
  \emph{IEEE Trans. Commun.}, vol.~74, p. 4881–4918, 2026.

\bibitem{11434944}
Y.~Xu \emph{et~al.}, ``Generalized pinching-antenna systems: A tutorial on
  principles, design strategies, and future directions,'' \emph{IEEE Commun.
  Surv. Tutorials}, vol.~28, pp. 5872--5908, 2026.

\bibitem{yang2025pinching}
Z.~Yang \emph{et~al.}, ``Pinching antennas: Principles, applications and
  challenges,'' \emph{IEEE Wireless Commun.}, Early Access, 2025.

\bibitem{R3}
M.~Zeng, J.~Wang, O.~A. Dobre, Z.~Ding, G.~K. Karagiannidis, R.~Schober, and
  H.~V. Poor, ``{Resource Allocation for Pinching-Antenna Systems:
  State-of-the-Art, Key Techniques and Open Issues},'' \emph{IEEE Wireless
  Commun.}, pp. 1--7, 2026.

\bibitem{R4}
K.~R. Wijewardhana, A.~Yadav, M.~Zeng, M.~Elsayed, O.~A. Dobre, and Z.~Ding,
  ``{Wireless-Fed Pinching-Antenna Systems (Wi-PASS) for NextG Wireless
  Networks},'' \emph{IEEE Commun. Mag.}, pp. 1--7, 2026.

\bibitem{illi2026pinchingantennasystemspass}
E.~Illi and M.~Qaraqe, ``Pinching antenna systems ({PASS}): Enabling
  reconfigurable and controllable wireless channels -- a comprehensive
  survey,'' \emph{arXiv}, vol. 2604.04521, 2026.

\bibitem{R9}
Y.~Lu, W.~Mao, X.~Xie, S.~Liao, Y.~Xu, Y.~Lin, R.~Zhang, B.~Ai, X.~Wang, and
  Z.~Ding, ``A survey on pinching-antenna systems: Optimization techniques,
  intelligent designs, and multifunctional applications,'' \emph{IEEE Commun.
  Surv. Tutorials}, vol.~28, pp. 6592--6627, 2026.

\bibitem{rayleigh1897passage}
L.~Rayleigh, ``On the passage of electric waves through tubes, or the
  vibrations of dielectric cylinders,'' \emph{The London, Edinburgh, and Dublin
  Philosophical Magazine and Journal of Science}, vol.~43, no. 262, pp.
  125--132, Feb. 1897.

\bibitem{packard1984origin}
K.~S. Packard, ``The origin of waveguides: A case of multiple rediscovery,''
  \emph{IEEE Trans. Microw. Theory Techn.}, vol.~32, no.~9, pp. 961--969, Sep.
  1984.

\bibitem{carson1936mathematical}
J.~R. Carson, S.~P. Mead, and S.~A. Schelkunoff, ``Hyper-frequency wave
  guides---mathematical theory,'' \emph{Bell Syst. Tech. J.}, vol.~15, no.~2,
  pp. 310--333, Apr. 1936.

\bibitem{monticone2015leakywave}
F.~Monticone and A.~Al{\`u}, ``Leaky-wave theory, techniques, and applications:
  From microwaves to visible frequencies,'' \emph{Proc. IEEE}, vol. 103, no.~5,
  pp. 793--821, May 2015.

\bibitem{chandler1949rod}
C.~H. Chandler, ``An investigation of dielectric rod as wave guide,'' \emph{J.
  Appl. Phys.}, vol.~20, no.~12, pp. 1188--1192, Dec. 1949.

\bibitem{elsasser1949attenuation}
W.~M. Elsasser, ``Attenuation in a dielectric circular rod,'' \emph{J. Appl.
  Phys.}, vol.~20, no.~12, pp. 1193--1196, Dec. 1949.

\bibitem{snitzer1961cylindrical}
E.~Snitzer, ``Cylindrical dielectric waveguide modes,'' \emph{J. Opt. Soc.
  Am.}, vol.~51, no.~5, pp. 491--498, May 1961.

\bibitem{king1957losses}
D.~D. King and S.~P. Schlesinger, ``Losses in dielectric image lines,''
  \emph{IRE Trans. Microw. Theory Techn.}, vol.~5, no.~1, pp. 31--35, Jan.
  1957.

\bibitem{kelebekler2021leaky}
E.~Kelebekler, ``Investigation of the leaky-wave characteristics of a
  cylindrical dielectric rod using the coefficient matrix of the system of
  characteristic equations,'' \emph{J. Electromagn. Eng. Sci.}, vol.~21, no.~3,
  pp. 189--200, Jul. 2021.

\bibitem{kao1966fibre}
K.~C. Kao and G.~A. Hockham, ``Dielectric-fibre surface waveguides for optical
  frequencies,'' \emph{Proc. IEE}, vol. 113, no.~7, pp. 1151--1158, Jul. 1966.

\bibitem{marcatili1969rectangular}
E.~A.~J. Marcatili, ``Dielectric rectangular waveguide and directional coupler
  for integrated optics,'' \emph{Bell Syst. Tech. J.}, vol.~48, no.~7, pp.
  2071--2102, Sep. 1969.

\bibitem{jablonski1978circular}
D.~Jablonski, ``Attenuation characteristics of circular dielectric waveguide at
  millimeter wavelengths,'' \emph{IEEE Trans. Microw. Theory Techn.}, vol.~26,
  no.~9, pp. 667--671, Sep. 1978.

\bibitem{yoneyama1983nrd}
T.~Yoneyama and S.~Nishida, ``Nonradiative dielectric waveguide circuit
  components,'' \emph{Int. J. Infrared Millim. Waves}, vol.~4, no.~3, pp.
  439--449, May 1983.

\bibitem{malekabadi2014ribbon}
S.~A. Malekabadi, S.~A. Charlebois, D.~Deslandes, and F.~Boone,
  ``High-resistivity silicon dielectric ribbon waveguide for single-mode
  low-loss propagation at f/g-bands,'' \emph{IEEE Trans. Terahertz Sci.
  Technol.}, vol.~4, no.~4, pp. 447--453, Jul. 2014.

\bibitem{atakaramians2013terahertz}
S.~Atakaramians, S.~Afshar~V., T.~M. Monro, and D.~Abbott, ``Terahertz
  dielectric waveguides,'' \emph{Adv. Opt. Photonics}, vol.~5, no.~2, pp.
  169--215, Jun. 2013.

\bibitem{james1967cylindrical}
J.~R. James, ``Theoretical investigation of cylindrical dielectric-rod
  antennas,'' \emph{Proc. IEE}, vol. 114, no.~3, pp. 309--319, Mar. 1967.

\bibitem{callahan1968moths}
P.~S. Callahan, ``A high frequency dielectric waveguide on the antennae of
  night-flying moths (saturnidae),'' \emph{Appl. Opt.}, vol.~7, no.~8, pp.
  1425--1430, Aug. 1968.

\bibitem{mangalgiri2017nanorod}
G.~M. Mangalgiri, P.~Manley, W.~Riedel, and M.~Schmid, ``Dielectric nanorod
  scattering and its influence on material interfaces,'' \emph{Sci. Rep.},
  vol.~7, no.~1, p. 4311, Jun. 2017.

\bibitem{smith2017waveguidefed}
D.~R. Smith, O.~Yurduseven, L.~Pulido-Mancera, P.~Bowen, and N.~B. Kundtz,
  ``Analysis of a waveguide-fed metasurface antenna,'' \emph{Phys. Rev. Appl.},
  vol.~8, no.~5, p. 054048, Nov. 2017.

\bibitem{faenzi2019metasurface}
M.~Faenzi, G.~Minatti, D.~Gonzalez-Ovejero, F.~Caminita, E.~Martini, and
  S.~Maci, ``Metasurface antennas: New models, applications and realizations,''
  \emph{Sci. Rep.}, vol.~9, no.~1, p. 10178, Jul. 2019.

\bibitem{suzuki2022pinching}
A.~Fukuda, H.~Yamamoto, H.~Okazaki, Y.~Suzuki, and K.~Kawai, ``Pinching
  antenna: Using a dielectric waveguide as an antenna,'' \emph{NTT DOCOMO
  Technical J.}, vol.~23, no.~3, pp. 5--12, Jan. 2022.

\bibitem{marcuvitz1951waveguide}
N.~Marcuvitz, \emph{Waveguide Handbook}, ser. MIT Radiation Laboratory
  Series.\hskip 1em plus 0.5em minus 0.4em\relax New York, NY, USA:
  McGraw-Hill, 1951, vol.~10.

\bibitem{pozar2021microwave}
D.~M. Pozar, \emph{Microwave Engineering}, 4th~ed.\hskip 1em plus 0.5em minus
  0.4em\relax Hoboken, NJ, USA: John Wiley \& Sons, 2011.

\bibitem{goell1969circular}
J.~E. Goell, ``A circular-harmonic computer analysis of rectangular dielectric
  waveguides,'' \emph{Bell Syst. Tech. J.}, vol.~48, no.~7, pp. 2133--2160,
  Sep. 1969.

\bibitem{hou2019rectangularrod}
Y.~Hou, Y.~Li, Z.~Zhang, and Q.~Xue, ``Rectangular dielectric rod antenna fed
  by air-substrate parallel strip line,'' \emph{IEEE Trans. Antennas Propag.},
  vol.~67, no.~10, pp. 6308--6316, Oct. 2019.

\bibitem{huang2019reconfigurable}
C.~Huang, A.~Zappone, G.~C. Alexandropoulos, M.~Debbah, and C.~Yuen,
  ``Reconfigurable intelligent surfaces for energy efficiency in wireless
  communication,'' \emph{IEEE Trans. Wireless Commun.}, vol.~18, no.~8, pp.
  4157--4170, Aug. 2019.

\bibitem{wang2025passmodeling}
Z.~Wang, C.~Ouyang, X.~Mu, Y.~Liu, and Z.~Ding, ``Modeling and beamforming
  optimization for pinching-antenna systems,'' \emph{IEEE Trans. Commun.},
  vol.~73, no.~12, pp. 13\,904--13\,919, Dec. 2025.

\bibitem{xu2026powerradiation}
X.~Xu, X.~Mu, Z.~Wang, Y.~Liu, and A.~Nallanathan, ``Pinching-antenna systems
  ({PASS}): Power radiation model and optimal beamforming design,'' \emph{IEEE
  Trans. Commun.}, vol.~74, pp. 2160--2175, 2026.

\bibitem{hou2017capacitylcx}
Y.~Hou, S.~Tsukamoto, S.~Li, T.~Higashino, K.~Kobayashi, and M.~Okada,
  ``Capacity evaluation of {MIMO} channel with one leaky coaxial cable used as
  two antennas over linear-cell environments,'' \emph{IEEE Trans. Veh.
  Technol.}, vol.~66, no.~6, pp. 4636--4646, Jun. 2017.

\bibitem{wang2001adjustablelcx}
J.~H. Wang, ``Leaky coaxial cable with adjustable coupling loss for mobile
  communications in complex environments,'' \emph{IEEE Microw. Wireless Compon.
  Lett.}, vol.~11, no.~8, pp. 346--348, Aug. 2001.

\bibitem{wang2016modelinglcx}
H.~Wang, F.~R. Yu, and L.~Jiang, ``Modeling of radio channels with leaky
  coaxial cable for {LTE}-m based cbtc systems,'' \emph{IEEE Commun. Lett.},
  vol.~20, no.~5, pp. 1038--1041, May 2016.

\bibitem{wang2025generalizedlcxpass}
K.~Wang, Z.~Ding, and L.~Hanzo, ``Generalized pinching-antenna systems: A
  leaky-coaxial-cable perspective,'' \emph{arXiv}, vol. 2512.04979, 2025.

\bibitem{ding2025flexible}
Z.~Ding, R.~Schober, and H.~Vincent~Poor, ``Flexible-antenna systems: A
  pinching-antenna perspective,'' \emph{IEEE Trans. Commun.}, vol.~73, no.~10,
  pp. 9236--9253, Oct. 2025.

\bibitem{zhang2025pinchingantennasystemspassbasedindoor}
Y.~Zhang, X.~Sun, J.~Wang, T.~Hou, A.~Li, Y.~Liu, and A.~Nallanathan,
  ``Pinching-antenna systems ({PASS})-based indoor positioning,'' \emph{arXiv},
  vol. 2508.08185, 2025.

\bibitem{PASS_poi_4}
H.~Feng, E.~Bedeer, M.~Zeng, X.~Li, S.~Gong, and Q.-V. Pham, ``Phase-aware
  localization in pinching antenna systems: {CRLB} analysis and {ML}
  estimation,'' \emph{IEEE Commun. Lett.}, vol.~30, pp. 2412--2416, 2026.

\bibitem{PASS_poi_2}
X.~Xu, X.~Mu, Y.~Liu, H.~Xing, and A.~Nallanathan, ``User localization and
  channel estimation for pinching-antenna systems ({PASS}),'' \emph{arXiv},
  vol. 2512.14351, 2025.

\bibitem{PASS_poi_40}
Y.~Zhang, X.~Sun, J.~Wang, T.~Hou, A.~Li, Y.~Liu, and A.~Nallanathan,
  ``Integrated positioning and communications for {PASS}: A robust approach,''
  \emph{IEEE Transactions on Wireless Communications}, 2026.

\bibitem{PASS_poi_3}
M.~Liu, Y.~Xiao, S.~Yang, G.~Wu, X.~Lei, and W.~Xiang, ``Joint channel
  estimation and localization in pinching-antenna {OFDM} systems: The blessing
  of multipath,'' \emph{arXiv}, vol. 2512.09432, 2025.

\bibitem{PASS_poi_7}
Y.~Zhang, X.~Sun, T.~Hou, A.~Li, and Y.~Liu, ``Uplink positioning for {PASS} in
  multipath environments,'' \emph{arXiv}, vol. 2607.20069, 2026.

\bibitem{trodition_indoor_1}
F.~Zafari, A.~Gkelias, and K.~K. Leung, ``A survey of indoor localization
  systems and technologies,'' \emph{IEEE Commun. Surv. Tutor.}, vol.~21, no.~3,
  pp. 2568--2599, Apr. 2019.

\bibitem{trodition_indoor_2}
A.~Yassin, Y.~Nasser, M.~Awad, A.~Al-Dubai, R.~Liu, C.~Yuen, R.~Raulefs, and
  E.~Aboutanios, ``Recent advances in indoor localization: A survey on
  theoretical approaches and applications,'' \emph{IEEE Commun. Surv. Tutor.},
  vol.~19, no.~2, pp. 1327--1346, Nov. 2016.

\bibitem{wifi_1}
S.~Yiu, M.~Dashti, H.~Claussen, and F.~Perez-Cruz, ``Wireless {RSSI}
  fingerprinting localization,'' \emph{Signal Process.}, vol. 131, pp.
  235--244, Feb. 2017.

\bibitem{wifi_2}
S.~He and S.-H.~G. Chan, ``{Wi-Fi} fingerprint-based indoor positioning: Recent
  advances and comparisons,'' \emph{IEEE Commun. Surv. Tutor.}, vol.~18, no.~1,
  pp. 466--490, Aug. 2015.

\bibitem{BLE}
R.~Faragher and R.~Harle, ``Location fingerprinting with {Bluetooth} low energy
  beacons,'' \emph{IEEE J. Sel. Areas Commun.}, vol.~33, no.~11, pp.
  2418--2428, May. 2015.

\bibitem{UWB_1}
J.-Y. Lee and R.~A. Scholtz, ``Ranging in a dense multipath environment using
  an {UWB} radio link,'' \emph{IEEE J. Sel. Areas Commun.}, vol.~20, no.~9, pp.
  1677--1683, Dec. 2002.

\bibitem{UWB_2}
D.~Dardari, A.~Conti, U.~Ferner, A.~Giorgetti, and M.~Z. Win, ``Ranging with
  ultrawide bandwidth signals in multipath environments,'' \emph{Proc. IEEE},
  vol.~97, no.~2, pp. 404--426, Mar. 2009.

\bibitem{li2026pinching}
Z.~Li, J.~Wang, T.~Hou, and A.~Li, ``Pinching-antenna systems ({PASS})-based
  user-side navigation: An anchor-line-based approach,'' \emph{arXiv}, vol.
  2607.13485, 2026.

\bibitem{PASS_poi_5}
J.~He, X.~Mu, H.~Q. Ngo, and M.~Matthaiou, ``Pinching-antenna system-assisted
  localization: A stochastic geometry perspective,'' \emph{IEEE Wireless
  Commun. Lett.}, vol.~15, pp. 1737--1741, 2026.

\bibitem{example2025}
\BIBentryALTinterwordspacing
{NTT DOCOMO INC.} (2022) Pinching antenna. [Online]. Available:
  \url{https://www.docomo.ne.jp/english/info/media_center/event/mwc21/pdf/06_MWC2021_docomo_Pinching_Antenna_en.pdf}
\BIBentrySTDinterwordspacing

\bibitem{10976621}
D.~Tyrovolas, S.~A. Tegos, P.~D. Diamantoulakis, S.~Ioannidis, C.~K. Liaskos,
  and G.~K. Karagiannidis, ``Performance analysis of pinching-antenna
  systems,'' \emph{IEEE Trans. Cognit. Commun. Networking}, vol.~12, pp.
  590--601, 2026.

\bibitem{11540399}
S.~Chen, W.~Duan, J.~Gu, S.~Dang, M.~Wen, and P.-H. Ho, ``Pinching-antenna
  systems: Waveguide-power loss and free-space path loss trade-off,''
  \emph{IEEE Trans. Wireless Commun.}, vol.~25, pp. 17\,942--17\,957, 2026.

\bibitem{10896748}
Y.~Xu, Z.~Ding, and G.~K. Karagiannidis, ``Rate maximization for downlink
  pinching-antenna systems,'' \emph{IEEE Wireless Commun. Lett.}, vol.~14,
  no.~5, pp. 1431--1435, May. 2025.

\bibitem{11434513}
T.~Hou, J.~Wang, W.~Gong, A.~Li, and A.~Nallanathan, ``On the performance of
  pinching-antenna array,'' \emph{IEEE Wireless Commun. Lett.}, vol.~15, pp.
  2403--2407, 2026.

\bibitem{11457366}
Y.~Xu, Z.~Ding, R.~Schober, and T.-H. Chang, ``Pinching-antenna systems with
  in-waveguide attenuation: Performance analysis and algorithm design,''
  \emph{IEEE Trans. Wireless Commun.}, vol.~25, pp. 14\,564--14\,580, 2026.

\bibitem{M4}
Y.~Cheng, C.~Ouyang, Y.~Liu, and G.~K. Karagiannidis, ``On the performance of
  pinching-antenna systems ({PASS}) with orthogonal and non-orthogonal multiple
  access,'' \emph{arXiv}, vol. 2506.02420, 2025.

\bibitem{cheng2026singlewaveguidemultiplepinchingantennasystemsoma}
Y.~Cheng, H.~Li, Z.~Wang, and Z.~Ding, ``Single-waveguide
  multiple-pinching-antenna systems: {OMA} versus {NOMA},'' \emph{arXiv}, vol.
  2604.00588, 2026.

\bibitem{yue2026performanceanalysispinchingantenna}
X.~Yue, X.~Tao, J.~Zhao, X.~Lei, Y.~Liu, and Z.~Ding, ``Performance analysis of
  pinching antenna systems enabled {NOMA} communications,'' \emph{arXiv}, vol.
  2604.25285, 2026.

\bibitem{10981775}
C.~Ouyang, Z.~Wang, Y.~Liu, and Z.~Ding, ``Array gain for pinching-antenna
  systems ({PASS}),'' \emph{IEEE Commun. Lett.}, vol.~29, no.~6, pp.
  1471--1475, Jun. 2025.

\bibitem{11358830}
C.~Ouyang, Z.~Wang, Y.~Liu, H.~Shin, and Z.~Ding, ``Capacity characterization
  of pinching-antenna systems,'' \emph{IEEE Trans. Wireless Commun.}, vol.~25,
  pp. 10\,387--10\,404, 2026.

\bibitem{11391537}
M.~Qian, X.~Mu, L.~You, and M.~Matthaiou, ``Pinching-antenna-based
  communications: Spectral efficiency analysis and deployment strategies,''
  \emph{IEEE Trans. Commun.}, vol.~74, pp. 4758--4771, 2026.

\bibitem{tegos2025minimumdatarateuplink}
S.~A. Tegos, P.~D. Diamantoulakis, Z.~Ding, and G.~K. Karagiannidis, ``Minimum
  data rate maximization for uplink pinching-antenna systems,'' \emph{IEEE
  Wireless Commun. Lett.}, vol.~14, no.~5, pp. 1516--1520, May. 2025.

\bibitem{11195810}
T.~Hou, Y.~Liu, and A.~Nallanathan, ``On the performance of uplink pinching
  antenna systems ({PASS}),'' \emph{IEEE Trans. Commun.}, vol.~74, pp. 92--105,
  2026.

\bibitem{11175694}
J.~Zhang, H.~Xu, C.~Ouyang, Q.~Zou, and H.~Yang, ``Uplink sum-rate maximization
  for pinching antenna-assisted multiuser {MISO} communications,'' \emph{IEEE
  Commun. Lett.}, vol.~29, no.~12, pp. 2795--2799, Sep. 2025.

\bibitem{11215676}
M.~Zeng, J.~Wang, X.~Li, G.~Wang, O.~A. Dobre, and Z.~Ding, ``Sum rate
  maximization for {NOMA}-assisted uplink pinching-antenna systems,''
  \emph{IEEE Wireless Commun. Lett.}, vol.~15, pp. 280--284, 2026.

\bibitem{11131179}
M.~Zeng, X.~Li, J.~Wang, G.~Huang, O.~A. Dobre, and Z.~Ding, ``Energy-efficient
  resource allocation for {NOMA}-assisted uplink pinching-antenna systems,''
  \emph{IEEE Wireless Commun. Lett.}, vol.~14, no.~11, pp. 3695--3699, Nov.
  2025.

\bibitem{11414143}
A.~Bereyhi, C.~Ouyang, S.~Asaad, Z.~Ding, and H.~Vincent~Poor, ``{MIMO-PASS}:
  Uplink and downlink transmission via {MIMO} pinching-antenna systems,''
  \emph{IEEE Trans. Commun.}, vol.~74, pp. 5701--5716, 2026.

\bibitem{11348983}
C.~Ouyang, H.~Jiang, Z.~Wang, Y.~Liu, and Z.~Ding, ``Uplink and downlink
  communications in segmented waveguide-enabled pinching-antenna systems
  ({SWANs}),'' \emph{IEEE Trans. Commun.}, vol.~74, pp. 3688--3703, 2026.

\bibitem{9512383}
G.~Charan, M.~Alrabeiah, and A.~Alkhateeb, ``Vision-aided 6{G} wireless
  communications: Blockage prediction and proactive handoff,'' \emph{IEEE
  Trans. Veh. Technol.}, vol.~70, no.~10, pp. 10\,193--10\,208, Oct. 2021.

\bibitem{3gpp.38.901}
3rd Generation Partnership Project~(3GPP), ``Study on channel model for
  frequencies from 0.5 to 100 {GHz} ({Release} 19),'' Tech. Rep. 38.901, Jul.
  2025.

\bibitem{11036558}
Z.~Ding and H.~Vincent~Poor, ``{LoS} blockage in pinching-antenna systems:
  Curse or blessing?'' \emph{IEEE Wireless Commun. Lett.}, vol.~14, no.~9, pp.
  2798--2802, Sep. 2025.

\bibitem{11516156}
L.~Li, Y.~Xu, T.~Cai, and T.-H. Chang, ``Power minimization in pinching-antenna
  systems under probabilistic {LoS} blockage,'' \emph{IEEE Wireless Commun.
  Lett.}, vol.~15, pp. 3139--3143, 2026.

\bibitem{11475477}
M.~F. Hanif and Y.~Liu, ``Multicasting pinching-antenna systems with {LoS}
  blockage,'' \emph{IEEE Wireless Commun. Lett.}, vol.~15, pp. 2498--2502,
  2026.

\bibitem{11315149}
Y.~Sun, Z.~Ding, and G.~K. Karagiannidis, ``A stochastic geometric analysis on
  multi-cell pinching-antenna systems under blockage effect,'' \emph{IEEE
  Wireless Commun. Lett.}, vol.~15, pp. 1085--1089, 2026.

\bibitem{ding2025environmentdivisionmultipleaccess}
Z.~Ding, R.~Schober, and H.~V. Poor, ``Environment division multiple access
  ({EDMA}): A feasibility study via pinching antennas,'' \emph{arXiv}, vol.
  2511.03820, 2025.

\bibitem{xie2026pinchingantennasblockageawareenvironments}
X.~Xie, F.~Fang, Z.~Ding, and X.~Wang, ``Pinching antennas in blockage-aware
  environments: Modeling, design, and optimization,'' \emph{arXiv}, vol.
  2601.01277, 2026.

\bibitem{zhao2026robustsecureblockageawarepinching}
R.~Zhao, S.~Hu, D.~Mishra, and D.~W.~K. Ng, ``Robust and secure blockage-aware
  pinching antenna-assisted wireless communication,'' \emph{arXiv}, vol.
  2601.06430, 2026.

\bibitem{xu2026environmentawarenetworkleveldesigngeneralizedII}
Y.~Xu, Z.~Ding, X.~Y. Zhang, T.~Q. Duong, and T.-H. Chang, ``Environment-aware
  network-level design of generalized pinching-antenna systems--part {II}:
  Geometry-aware case,'' \emph{arXiv}, vol. 2602.17032, 2026.

\bibitem{11526989}
Z.~Sun, X.~Mu, S.~Han, S.~Xu, and M.~Matthaiou, ``Pinching-antenna-enabled
  cognitive radio networks,'' \emph{IEEE Trans. Commun.}, vol.~74, pp.
  9185--9199, 2026.

\bibitem{wang2026performancepinchingantennasystemspass}
J.~Wang, J.~Wang, T.~Hou, A.~Li, Y.~Liu, and A.~Nallanathan, ``On the
  performance of pinching-antenna systems ({PASS}) under dynamic channels with
  blockages,'' \emph{arXiv}, vol. 2607.13581, 2026.

\bibitem{xu2025pinchingantennadesignrandomlos}
Y.~Xu, Y.~Lu, Z.~Ding, and T.-H. Chang, ``Pinching-antenna system design under
  random {LoS} and {NLoS} channels,'' \emph{arXiv}, vol. 2512.04719, 2025.

\bibitem{xu2026environmentawarenetworkleveldesigngeneralizedI}
Y.~Xu, Z.~Ding, X.~Y. Zhang, T.~Q. Duong, and T.-H. Chang, ``Environment-aware
  network-level design of generalized pinching-antenna systems--part {I}:
  Traffic-aware case,'' \emph{arXiv}, vol. 2602.17023, 2026.

\bibitem{Zhong2026TCOM_2D}
Y.~Zhong, Y.~Xiao, Y.~Li, H.~Chen, X.~Lei, and P.~Fan, ``2-{D} pinching-antenna
  systems: Modeling and beamforming design,'' \emph{IEEE Trans. Commun.},
  vol.~74, pp. 8254--8266, 2026.

\bibitem{Xue2026arXiv_multicell}
Q.~Xue, C.~Ouyang, Y.~Liu, and Z.~Ding, ``Pinching-antenna enabled multicell
  wireless systems,'' \emph{arXiv}, vol. 2606.12888, 2026.

\bibitem{Wang2025WCL_NOMA}
K.~Wang, Z.~Ding, and R.~Schober, ``Antenna activation for {NOMA} assisted
  pinching-antenna systems,'' \emph{IEEE Wireless Commun. Lett.}, vol.~14,
  no.~5, pp. 1528--1530, May 2025.

\bibitem{Tyrovolas2025Ergodic}
D.~Tyrovolas, S.~A. Tegos, Y.~Xiao, P.~D. Diamantoulakis, S.~Ioannidis,
  C.~Liaskos, and G.~K. Karagiannidis, ``Ergodic rate analysis of two-state
  pinching-antenna systems,'' \emph{arXiv}, vol. 2511.01798, 2025.

\bibitem{xie2025lowcomplexityplacement}
X.~Xie, F.~Fang, Z.~Ding, and X.~Wang, ``A low-complexity placement design of
  pinching-antenna systems,'' \emph{IEEE Commun. Lett.}, vol.~29, no.~8, pp.
  1784--1788, Aug. 2025.

\bibitem{Zhang2026arXiv_multiwave}
Y.~Zhang, X.~Sun, J.~Wang, T.~Hou, A.~Li, Y.~Liu, and A.~Nallanathan,
  ``Multi-waveguide pinching antenna placement optimization for rate
  maximization,'' \emph{arXiv}, vol. 2512.18711, 2026.

\bibitem{Zeng2026arXiv}
M.~Zeng, X.~Mu, Y.~Liu, and Z.~Ding, ``Robust single- and multi-pinching
  antenna systems under user location uncertainty,'' \emph{arXiv}, vol.
  2604.09774, 2026.

\bibitem{M14}
J.~Zhao, H.~Song, X.~Mu, K.~Cai, Y.~Zhu, and Y.~Liu, ``Pinching-antenna
  systems-enabled multi-user communications: Transmission structures and
  beamforming optimization,'' \emph{IEEE Trans. Commun.}, vol.~74, pp.
  2316--2330, 2026.

\bibitem{Sun2025TWC}
M.~Sun, C.~Ouyang, S.~Wu, and Y.~Liu, ``Multiuser beamforming for
  pinching-antenna systems: An element-wise optimization framework,''
  \emph{IEEE Trans. Wireless Commun.}, vol.~25, pp. 6538--6552, Oct. 2025.

\bibitem{gan2026centerfedmodeling}
X.~Gan and Y.~Liu, ``Center-fed pinching antenna system ({C}-{PASS}): Modeling,
  analysis, and beamforming design,'' \emph{arXiv}, vol. 2602.14805, 2026.

\bibitem{Jiang2026arXiv_SWAN}
H.~Jiang, Z.~Wang, Y.~Liu, A.~Nallanathan, and Z.~Ding, ``{SWAN} for uplink
  tri-hybrid beamforming: Riemannian optimization and interleaved topologies,''
  \emph{arXiv}, vol. 2604.25285, 2026.

\bibitem{Wang2026Tcom_HPASS}
K.~Wang, Z.~Ding, and D.~K.~C. So, ``Hybrid pinching antenna systems:
  Architecture and beamforming design,'' \emph{IEEE Trans. Commun.}, vol.~74,
  pp. 12\,129--12\,144, 2026.

\bibitem{Shao2026arXiv}
Y.~Shao, C.~Ouyang, Y.~Liu, and Z.~Ding, ``Multi-mode pinching-antenna systems:
  Polarization-aware full-wave modeling and optimization,'' \emph{arXiv}, vol.
  2604.01778, 2026.

\bibitem{Xu2025arXiv_motion}
Y.~Xu, D.~Xu, X.~Yu, S.~Song, Z.~Ding, and R.~Schober, ``Joint radiation power,
  antenna position, and beamforming optimization for pinching-antenna systems
  with motion power consumption,'' \emph{IEEE Trans. Wireless Commun.},
  vol.~25, pp. 7825--7841, 2026.

\bibitem{Zeng2026TVT_EE}
M.~Zeng, J.~Wang, G.~Zhou, F.~Fang, and X.~Wang, ``Energy-efficient design for
  downlink pinching-antenna systems with {QoS} guarantee,'' \emph{IEEE Trans.
  Veh. Technol.}, vol.~75, no.~2, pp. 3503--3507, 2026.

\bibitem{Zhang2026Tcom_DSD}
X.~Zhang, Y.~Lu, Z.~Ding, and T.-H. Chang, ``Dual-scale antenna deployment for
  pinching-antenna systems,'' \emph{IEEE Trans. Commun.}, vol.~74, pp.
  9772--9786, 2026.

\bibitem{Li2026WCL_SWIPT}
Y.~Li, H.~Xu, M.~Zeng, and Y.~Liu, ``Pinching antenna-aided wireless powered
  communication networks,'' \emph{IEEE Wireless Commun. Lett.}, vol.~15, pp.
  255--259, 2026.

\bibitem{Wang2025WCL_power}
S.~Mohammadzadeh, K.~Cumanan, C.~Li, and Z.~Ding, ``Efficient downlink power
  allocation for {NOMA}-based pinching-antenna systems,'' \emph{IEEE Wireless
  Commun. Lett.}, vol.~14, no.~12, pp. 4187--4191, Dec. 2025.

\bibitem{Gan2025arXiv_WPT}
W.~Gan and Y.~Liu, ``Enabling wireless power transfer in pinching-antenna
  systems,'' \emph{arXiv}, vol. 2511.11465, 2025.

\bibitem{Zhou2026ICC_SEEE}
Z.~Zhou, Y.~Liu, and Z.~Ding, ``Spectral and energy efficiency trade-off: A
  pinching antenna perspective,'' in \emph{Proc. IEEE ICC}, 2026.

\bibitem{11419117}
Y.~Xu, Y.~Lu, D.~Cai, R.~Jiang, and F.~Fang, ``Pinching antenna-enabled {URLLC}
  downlink,'' \emph{IEEE Wireless Commun. Lett.}, vol.~15, pp. 2075--2079,
  2026.

\bibitem{Lin2025arXiv_URLLC}
X.~Lin, Y.~Liu, and Z.~Ding, ``Rate optimization for downlink {URLLC} via
  pinching antenna arrays,'' \emph{arXiv}, vol. 2509.01222, 2025.

\bibitem{Liu2025arXiv_MEC}
P.~Liu, M.~Hua, G.~Chen, X.~Wang, and Z.~Fei, ``Computation capacity
  maximization for {PASS}-assisted wireless powered {MEC},'' \emph{arXiv}, vol.
  2506.07598, 2025.

\bibitem{Gu2026arXiv_AirComp}
Y.~Gu, C.~Ouyang, and Y.~Liu, ``Over-the-air computation via segmented
  waveguide-enabled {PASS} array,'' \emph{arXiv}, vol. 2602.14805, 2026.

\bibitem{Zhang2025arXiv_DiPASS}
R.~Zhang, Y.~Shao, and Y.~Liu, ``Directional pinching-antenna systems,''
  \emph{arXiv}, vol. 2511.19133, 2025.

\bibitem{M13}
K.~Wang, Z.~Ding, and G.~K. Karagiannidis, ``Antenna activation and resource
  allocation in multi-waveguide pinching-antenna systems,'' \emph{IEEE Trans.
  Wireless Commun.}, vol.~25, pp. 4070--4082, 2026.

\bibitem{11048566}
L.~Zhang, X.~Mu, A.~Liu, and Y.~Liu, ``Two-timescale joint transmit and
  pinching beamforming for pinching-antenna systems,'' \emph{IEEE Wireless
  Commun. Lett.}, vol.~14, no.~9, pp. 2897--2901, Sept. 2025.

\bibitem{ai23}
F.~Fang, Z.~Ding, V.~C.~M. Leung, and L.~Hanzo, ``Pinching antennas meet {AI}
  in next-generation wireless networks,'' \emph{arXiv}, vol. 2511.07442, 2025.

\bibitem{11159498}
J.~Guo, Y.~Liu, and A.~Nallanathan, ``A graph neural network for learning
  beamforming in pinching antenna systems ({PASS}),'' \emph{IEEE Wireless
  Commun. Lett}, vol.~14, no.~12, pp. 3962--3966, Dec. 2025.

\bibitem{11063467}
X.~Xie, Y.~Lu, and Z.~Ding, ``Graph neural network enabled pinching antennas,''
  \emph{IEEE Wireless Commun. Lett.}, vol.~14, no.~9, pp. 2982--2986, Sept.
  2025.

\bibitem{he2025risgnn}
C.~He, Y.~Lu, Y.~Xu, C.-Y. Chi, B.~Ai, and A.~Nallanathan, ``{RIS}-assisted
  downlink pinching-antenna systems: {GNN}-enabled optimization approaches,''
  \emph{arXiv}, vol. 2511.20305, 2025.

\bibitem{I7}
J.~Guo, Y.~Liu, and A.~Nallanathan, ``Learning beamforming for pinching antenna
  system-enabled {ISAC} in low-altitude wireless networks,'' \emph{arXiv}, vol.
  2512.04293, 2025.

\bibitem{11303188}
------, ``Graph transformer for tri-beamforming in pinching antenna systems
  ({PASS}),'' \emph{IEEE J. Sel. Areas Commun.}, vol.~44, pp. 2872--2888, 2026.

\bibitem{11148833}
X.~Xu, Y.~Liu, X.~Mu, D.~Gan, and A.~Nallanathan, ``Beamforming for pinching
  antenna systems ({PASS}): {KKT}-guided large model learning,'' in \emph{2025
  IEEE/CIC Int. Conf. Commun. China (ICCC)}, Aug. 2025, pp. 1--6.

\bibitem{xu2025jointtransmitpinching}
X.~Xu, X.~Mu, Y.~Liu, and A.~Nallanathan, ``Joint transmit and pinching
  beamforming for pinching antenna system ({PASS}): Optimization-based or
  learning-based?'' \emph{IEEE Trans. Wireless Commun.}, vol.~25, pp.
  11\,449--11\,464, 2026.

\bibitem{11018390}
J.~Xiao, J.~Wang, and Y.~Liu, ``Channel estimation for pinching-antenna systems
  ({PASS}),'' \emph{IEEE Commun. Lett.}, vol.~29, no.~8, pp. 1789--1793, Aug.
  2025.

\bibitem{ai21}
S.~Mohammadzadeh, K.~Cumanan, and Z.~Ding, ``{NOMA} assisted downlink power
  allocation in pinching antenna systems using convolutional neural network,''
  \emph{IEEE Internet Things J.}, vol.~13, no.~12, pp. 26\,299--26\,314, Mar.
  2026.

\bibitem{ai08}
O.~G. Karagiannidis, V.~E. Galanopoulou, P.~D. Diamantoulakis, Z.~Ding, and
  O.~Dobre, ``Deep learning optimization of two-state pinching antennas
  systems,'' \emph{arXiv}, vol. 2507.06222, 2025.

\bibitem{I2}
Y.~Qin, Y.~Fu, and H.~Zhang, ``Joint antenna position and transmit power
  optimization for pinching antenna-assisted {ISAC} systems,'' \emph{IEEE
  Wireless Commun. Lett.}, vol.~14, no.~11, pp. 3535--3539, Nov. 2025.

\bibitem{P6}
P.~P. Papanikolaou, D.~Bozanis, S.~A. Tegos, P.~D. Diamantoulakis,
  P.~Sarigiannidis, and G.~K. Karagiannidis, ``Physical layer security with
  artificial noise in {MIMO} pinching-antenna systems,'' \emph{arXiv}, vol.
  2511.23079, 2025.

\bibitem{ai05}
J.-M. Kang, S.~Yun, and I.-M. Kim, ``{CaMPASS-Net}: A deep learning framework
  on capacity maximization for {MIMO} pinching antenna systems in {IoT},''
  \emph{IEEE Internet Things J.}, vol.~12, no.~21, pp. 45\,917--45\,920, Nov.
  2025.

\bibitem{ai25}
N.~S. Perović, K.~Singh, and C.-P. Li, ``{DL}-driven optimization for {ISAC}
  system equipped with pinching and movable antennas,'' \emph{arXiv}, vol.
  2605.17629, 2026.

\bibitem{ai26}
Y.~Chen and Q.~Wang, ``Generative diffusion-based channel estimation for
  pinching antenna-assisted indoor {NFC},'' \emph{Electronics}, vol.~15, no.~4,
  p. 730, Feb. 2026.

\bibitem{ai04}
S.~Lv, T.~Liu, R.~Yang, M.~Li, and Y.~Liu, ``Deep learning-driven time-varying
  fading channel estimation for pinching antenna systems,'' \emph{IEEE Wireless
  Commun. Lett.}, vol.~15, pp. 2569--2573, Nov. 2026.

\bibitem{ai24}
A.~Amhaz, S.~Khisa, M.~Elhattab, C.~Assi, and S.~Sharafeddine, ``Coordinated
  multipoint transmission in pinching antenna systems,'' \emph{arXiv}, vol.
  2510.23837, 2025.

\bibitem{zhao2026dynamicplacementmobile}
Q.~Zhao, Z.~Feng, X.~Xie, H.~Qin, Y.~Liu, and X.~Zhang, ``Dynamic antenna
  placement for mobile users in urban micro pinching-antenna systems,''
  \emph{arXiv}, vol. 2603.00619, 2026.

\bibitem{ai22}
A.~Amhaz, M.~Elhattab, C.~Assi, and S.~Sharafeddine, ``Joint beamforming and
  antenna placement optimization in pinching antenna systems with user
  mobility: A deep reinforcement learning approach,'' \emph{arXiv}, vol.
  2605.08039, 2026.

\bibitem{ai03}
Y.~Ma, W.~Huang, Z.~Ding, D.~K.~C. So, and V.~Bhatia, ``Joint optimization in
  pinching-antenna-assisted {NOMA-MEC} systems via {DRL},'' \emph{IEEE Wireless
  Commun. Lett.}, vol.~15, pp. 2493--2497, 2026.

\bibitem{ai19}
S.~Pakravan, M.~Ahmadzadeh, M.~Zeng, X.~Li, and F.~Fang, ``{AI}-empowered
  resource allocation for wirelessly powered pinching-antenna systems,''
  \emph{IEEE Trans. Veh. Technol.}, pp. 1--6, Early Access, 2026.

\bibitem{ai01}
Y.~Lin and Z.~Ding, ``Tail-latency-aware federated learning with pinching
  antenna: Latency, participation, and placement,'' \emph{Entropy}, vol.~28,
  no.~3, p. 341, 2026.

\bibitem{ai07}
B.~Wu, F.~Fang, M.~Zeng, and X.~Wang, ``Straggler-resilient federated learning
  over a hybrid conventional and pinching antenna network,'' \emph{IEEE Trans.
  Veh. Technol.}, vol.~74, no.~7, pp. 15\,404--15\,409, Jul. 2026.

\bibitem{ai14}
S.~Asaad and A.~Bereyhi, ``Energy-efficient over-the-air federated learning via
  pinching antenna systems,'' in \emph{ICC 2026 - IEEE Int. Conf. Commun.},
  Jul. 2026, pp. 1--6.

\bibitem{azimiabarghouyi2026airpassovertheairfederatedlearning}
S.~M. Azimi-Abarghouyi and C.~G. Brinton, ``Airpass: Over-the-air federated
  learning via pinching antenna systems,'' \emph{arXiv}, vol. 2607.06768, 2026.

\bibitem{ai12}
S.~Asaad, H.~Tabassum, and P.~Wang, ``Pinching antennas-assisted low-latency
  federated learning over multi-user wireless networks,'' \emph{arXiv}, vol.
  2603.08595, 2026.

\bibitem{ai20}
Y.~Lin, Z.~Chen, and Z.~Ding, ``Pinching-antenna-enabled federated learning:
  Tail latency, participation, and convergence analysis,'' \emph{arXiv}, vol.
  2510.23315, 2025.

\bibitem{zhou2025gradientmetalearning}
K.~Zhou, W.~Zhou, D.~Cai, X.~Lei, Y.~Xu, Z.~Ding, and P.~Fan, ``A gradient
  meta-learning joint optimization for beamforming and antenna position in
  pinching-antenna systems,'' \emph{IEEE Trans. Commun.}, vol.~74, pp.
  1099--1112, 2026.

\bibitem{musri2026adaptivepinchingantennaoptimization}
K.~T. Musri, A.~Y. Sarhan, O.~A. Abdullah, and H.~Al-Hraishawi, ``Adaptive
  pinching antenna optimization via meta-learning for physical-layer security
  in dynamic wireless networks,'' \emph{arXiv}, vol. 2601.00115, 2026.

\bibitem{ai15}
D.~Gan, X.~Xu, X.~Ge, and Y.~Liu, ``{LLM} enabled beam training for pinching
  antenna systems ({PASS}),'' \emph{arXiv}, vol. 2511.09453, 2025.

\bibitem{gao2026integratedsensingsegmented}
Q.~Gao, R.~Zhong, H.~Shin, and Y.~Liu, ``Integrated sensing and communication
  for segmented waveguide-enabled pinching antenna systems,'' \emph{arXiv},
  vol. 2601.20658, 2026.

\bibitem{ai13}
Q.~Gao, R.~Zhong, and Y.~Liu, ``Graph-enhanced {LLM} for {SWAN-ISAC},''
  \emph{arXiv}, vol. 2604.10256, 2026.

\bibitem{gao2026llmenabledpartitioning}
Q.~Gao, R.~Zhong, H.~Shin, and Y.~Liu, ``{LLM}-enabled antenna partitioning and
  beamforming optimization for segmented pinching antenna assisted {ISAC}
  systems,'' \emph{arXiv}, vol. 2604.10372, 2026.

\end{thebibliography}

\end{document}